\documentclass{aa}  
\usepackage{natbib,epsfig,txfonts,graphicx}
\bibpunct{(}{)}{;}{a}{}{,}

\newcommand{\rsun}{\ensuremath{R_{\odot}}}

\newcommand{\Teff}{\ensuremath{T_{\rm eff}}}
\newcommand{\vinf}{\ensuremath{v_{\infty}}}
\newcommand{\mdot}{\ensuremath{\dot{M}}}

\newcommand{\msunyr}{\ensuremath{M_{\odot} {\rm yr}^{-1}}}

\newcommand{\mdu}{\ensuremath{10^{-6}\,M_{\odot} {\rm yr}^{-1}}}

\newcommand{\beq}{\begin{equation}}
\newcommand{\eeq}{\end{equation}}
\newcommand{\beqa}{\begin{eqnarray}}
\newcommand{\eeqa}{\end{eqnarray}}

\newcommand{\nbeq}{\begin{equation*}}
\newcommand{\neeq}{\end{equation*}}

\newcommand{\kms}{\ensuremath{{\rm km}\,{\rm s}^{-1}}}

\newcommand{\rarrow}{\rightarrow}
\newcommand{\dd}{{\rm d}}

\newcommand{\HII} {H\,{\sc ii}}
\newcommand{\HeI} {He\,{\sc i}}
\newcommand{\HeII}{He\,{\sc ii}}

\newcommand{\CIII}{C\,{\sc iii}}
\newcommand{\CIV}{C\,{\sc iv}}

\newcommand\NII{N\,{\sc ii}}
\newcommand\NIII{N\,{\sc iii}}
\newcommand\NIV{N\,{\sc iv}}
\newcommand\NV{N\,{\sc v}}

\newcommand\OII{O\,{\sc ii}}
\newcommand\OIII{O\,{\sc iii}}
\newcommand\OIV{O\,{\sc iv}}

\newcommand\NeII{Ne\,{\sc ii}}

\newcommand\SiIII{Si\,{\sc iii}}
\newcommand\SiIV{Si\,{\sc iv}}

\newcommand{\Hd} {H$_{\rm \delta}$}

\newcommand{\Bra}{Br$_{\rm \alpha}$}

\newcommand{\Rstar}{\ensuremath{R_{\ast}}}

\newcommand{\logg}{\ensuremath{\log g}}

\newcommand{\vturb}{\ensuremath{v_{\rm turb}}}

\newcommand{\taur}{\ensuremath{\tau_{\rm Ross}}}

\newcommand{\trip}{\ensuremath{\lambda\lambda4634-4640-4642}}
\newcommand{\qua}{\ensuremath{\lambda\lambda4510-4514-4518}}

\newcommand{\Jbar}{\ensuremath{\bar J}}

\begin{document}
\title{Nitrogen line spectroscopy of O-stars}
\subtitle{I. Nitrogen III emission line formation revisited
\thanks{Appendix A, B, and C are only available in electronic form at
http://www.edpsciences.org}}

\author{J.G. Rivero Gonz\'alez\inst{1}, J. Puls\inst{1} \and 
F. Najarro\inst{2}}

\institute{Universit\"atssternwarte M\"unchen, Scheinerstr. 1, 81679 M\"unchen, 
           Germany, \email{jorge@usm.uni-muenchen.de} 
	   \and
	   Centro de Astrobiolog\'{\i}a, (CSIC-INTA), 
	   Ctra. Torrej\'on a Ajalvir km 4,
           28850 Torrej\'on de Ardoz, Spain}

\date{Received; Accepted}

\abstract {Evolutionary models of massive stars predict a surface
enrichment of Nitrogen, due to rotational mixing.  Recent studies
within the {\it VLT-FLAMES survey of massive stars}\, have challenged
(part of) these predictions. Such systematic determinations of
Nitrogen abundances, however, have been mostly performed only for cooler
(B-type) objects. 
}
{This is the first paper in a series dealing with optical Nitrogen
spectroscopy of O-type stars, aiming at the analysis of
Nitrogen abundances for stellar samples of significant size, to place
further constraints on the early evolution of massive stars.
Here we concentrate on the formation of the optical \NIII\ lines at
\trip\ that are fundamental for the definition of the different
morphological `f'-classes.} 
{We implement a new Nitrogen model atom into the NLTE
atmosphere/spectrum synthesis code {\sc fastwind}, and compare the
resulting optical N{\sc iii} spectra with other predictions,
mostly from the seminal work by Mihalas \& Hummer (1973, ApJ 179, 827,
`MH'), and from the alternative code {\sc cmfgen}.}
{Using similar model atmospheres as MH (not blanketed and wind-free),
we are able to reproduce their results, in particular the optical
triplet emission lines. According to MH, these should be strongly
related to dielectronic recombination and the drain by certain
two-electron transitions. However, using realistic, fully
line-blanketed atmospheres at solar abundances, the key role of the
dielectronic recombinations controlling these emission features is 
superseded -- for O-star conditions -- by the strength of the stellar
wind and metallicity. Thus, in the case of wind-free (weak wind)
models, the resulting lower ionizing EUV-fluxes severely suppress the
emission. As the mass loss rate is increased, pumping through the
\NIII\ resonance line(s) in the presence of a near-photospheric
velocity field (i.e., the Swings-mechanism) results in a net optical
triplet line emission. A comparison with results from {\sc cmfgen}\,
is mostly satisfactory, except for the range 30,000~K $\leq \Teff\
\leq$ 35,000~K, where {\sc cmfgen}\, triggers the triplet emission at
lower \Teff\ than {\sc fastwind}. This effect could be traced down to
line overlap effects between the \NIII\ and \OIII\ resonance lines
that cannot be simulated by {\sc fastwind}\, so far, due to the lack
of a detailed \OIII\ model atom.}
{Since the efficiency of dielectronic recombination and `two electron
drain' strongly depends on the degree of line-blanketing/-blocking, we
predict the emission to become stronger in a metal-poor environment, 
though lower wind-strengths and Nitrogen abundances might counteract
this effect. Weak winded stars (if existent in the decisive parameter
range) should display less triplet emission than their counterparts
with `normal' winds.}


\titlerunning{\NIII\ emission line formation revisited}
\authorrunning{J.G. Rivero Gonz\'alez, J. Puls, \& F. Najarro}

\maketitle
%

\section{Introduction}
\label{Introduction}

One of the key aspects of massive star evolution is rotational mixing
and its impact. Evolutionary models including rotation (e.g.,
\citealt{Heger00, MeynetMaeder00, Brott11a}) predict a surface
enrichment of Nitrogen with an associated Carbon depletion during the
main sequence evolution.\footnote{due to a rapid achievement of the CN
equilibrium, whilst the Oxygen depletion implied by the full CNO
equilibrium is only found in rapidly rotating and more massive stars
at later stages, e.g., \citealt{Brott11a}.} The faster a star rotates,
the more mixing will occur, and the larger the Nitrogen surface
abundance that should be observed.

Several studies \citep{hunter08, hunter09, Brott11b} have challenged
the predicted effects of rotational mixing on the basis of
observations performed within the VLT-FLAMES survey of massive stars
\citep{evans06}. These studies provide the first statistically
significant abundance measurements of Galactic, LMC, and SMC B-type
stars, 
covering a broad range of rotational velocities.

For the Galactic case, the mean B-star Nitrogen abundance as derived by
\citet{hunter09b} is in quite good agreement with the corresponding baseline
abundance. For the Magellanic Clouds stars, however, the derived Nitrogen
abundances show clearly the presence of an enrichment, where this enrichment
is more extreme in the SMC than in the LMC.\footnote{Baseline abundances
for all three environments from \HII\ regions and unevolved B-stars, see
\citet{hunter07}.} Theoretical considerations have major difficulties in
explaining several aspects of the accumulated results: Within the population
of (LMC) core-Hydrogen burning objects, both unenriched fast rotators and
highly enriched slow rotators have been found, in contradiction to standard
theory,
as well as slowly rotating, highly enriched B-supergiants (see below).
Taken together, these results imply that standard rotational mixing
might be not as dominant as usually quoted, and/or that other
enrichment processes might be present as well \citep{Brott11b}.

Interestingly, there exist only few rapidly rotating B-supergiants,
since there is a steep drop of rotation rates below \Teff\ $\approx$ 20~kK
(e.g., \citealt{Howarth97}). Recently, \citet{Vink10} tried to explain this
finding based on two alternative scenarios. In the first scenario, the low
rotation rates of B-supergiants are suggested to be caused by braking due to
an increased mass loss for \Teff\ $<$ 25~kK, related to the so-called
bi-stability jump \citep{PP90, vink00}. Since the reality of such an
increased mass loss is still debated \citep{MP08, Puls10}, \citet{Vink10}
discuss an alternative scenario where the slowly rotating B-supergiants
might form an entirely separate, non core hydrogen-burning population. E.g.,
they might be products of binary evolution (though this is not generally
expected to lead to slowly rotating stars), or they might be post-RSG or
blue-loop stars. 

Support of this second scenario is the finding that the majority of the
cooler (LMC) objects is strongly Nitrogen-enriched (see above), and Vink et
al. argue that ``although rotating models can in principle account for large
N abundances, the fact that such a large number of the cooler objects is
found to be N enriched suggests an evolved nature for these stars.'' 
A careful Nitrogen analysis of their (early) progenitors, the O-type
stars, will certainly help to further constrain these ideas and
present massive star evolution in general. Note that one of the
scientific drivers of the current VLT-FLAMES Tarantula survey
\citep{Evans11} is just such an analysis of an unprecedented sample of
`normal' O-stars and emission-line stars. 

Until to date, however, Nitrogen abundances have been {\it
systematically} derived only for the cooler subset of the previous
VLT-FLAMES survey, by means of analyzing \NII\ alone, whereas
corresponding results are missing for the most massive and hottest
stars. More generally, when inspecting the available literature
for massive stars, one realizes that metallic abundances, in
particular of the key element Nitrogen, have been derived only
for a small number of O-type stars (e.g., \citealt{bouret03, Hillier03,
walborn04, Heap06}). The simple reason is that they are difficult to
determine, since the formation of \NIII/\NIV\ lines (and lines from
similar ions of C and O) is problematic due to the impact of various
processes that are absent or negligible at cooler spectral types.

One might argue that the determination of Nitrogen and other metallic
abundances of hotter stars could or even should be performed via UV
wind-lines, since these are clearly visible as long as the wind-strength is
not too low, and the line-formation is less complex and less dependent on
accurate atomic models than for photospheric lines connecting intermediate
or even high-lying levels. Note, however, that the results of such analyses
strongly depend on the assumptions regarding and the treatment of wind X-ray
emission and wind clumping (\citealt{puls08b} and references therein). A
careful photospheric analysis, on the other hand, remains rather unaffected
by such problems as long as X-ray emission and clumping do not start (very)
close to the photosphere, and we will follow the latter approach,
concentrating on optical lines. 

This paper is the first in a series of upcoming publications dealing with
Nitrogen spectroscopy of O-type stars. The major objective of this project
is the analysis of optical spectra from stellar samples of significant size 
in different environments, to derive the corresponding Nitrogen abundances
which are key to our understanding of the early evolution of massive stars.
In the present study, we will concentrate on the formation of the optical
\NIII\ emission lines at \trip, which are fundamental for the definition of the
different morphological `f'-classes. During our implementation of Nitrogen
into the NLTE-atmosphere/line formation code {\sc fastwind} \citep{puls05}
it turned out that the canonical explanation in terms of dielectronic
recombination \citep{mihalas73} no longer or only partly applies when modern
atmosphere codes including line-blocking/blanketing and winds are used to
synthesize the \NIII\ spectrum. Since the f-features are observed in the 
majority of O-stars and strongly dependent on the Nitrogen abundance, a
thorough re-investigation of their formation process is required, in order
to avoid wrong conclusions.  

This paper is organized as follows: At first
(Sect.~\ref{intro-niii}) we recapitulate previous explanations
for the \NIII\ triplet emission, in particular the standard picture as
provided by \citet{mihalas73}. 
In Sect.~\ref{dr} we discuss the dielectronic recombination process
and its implementation into {\sc fastwind}. Sect.~\ref{atom niii}
presents our new \NIII\ model ion, and in Sect.~\ref{test-dr} we
investigate the dependence of the triplet emission on various
processes. We compare our results with corresponding ones from the
alternative code {\sc cmfgen} \citep{hilliermiller98} in 
Sect~\ref{comp-cmfgen}, and discuss the impact of coupling with \OIII\
via corresponding resonance lines in Sect.~\ref{Ocoup}. The dependence
of the emission strength on specific parameters is discussed in
Sect.~\ref{varpar}, and Sect.~\ref{conclusions} provides our summary
and conclusions.

In the next paper of this series (`Paper~II'), we will present our
complete Nitrogen model atom, and perform a Nitrogen abundance
analysis for the LMC O-stars from the previous VLT-FLAMES survey of
massive stars.

\section{\NIII\ emission lines from O-stars - status quo}
\label{intro-niii} 
The presence of emission in the \NIII\ triplet at \trip~\AA, in
combination with the behaviour of \HeII~$\lambda4686$, is used for
classification purposes (`f'-features, see \citealt{walborn71b}, 
\citealt{sota11}), and to discriminate O-stars with such line emission
from pure absorption-line objects.  
%
%
As stressed by \citet[hereafter BM71]{brucato71},
these emission lines
originate in the stellar photosphere and not in an ``exterior shell'' 
(see also \citealt{Heap06} for more recent work).
%
%
In this case, the most plausible explanation is by invoking NLTE
effects. In NLTE, line emission occurs when the corresponding source
function at formation depths is larger than the continuum intensity at
transition frequency $\nu_0$. Such a large source function becomes
possible if the upper level of the transition is (considerably)
overpopulated with respect to the lower one, i.e., if $b_{\rm u} >
b_{\rm l}$ (with $b_{\rm u}$ and $b_{\rm l}$ the NLTE
departure coefficients of the upper and lower level, respectively). 
Note that both coefficients can lie below unity.

\begin{figure}
\resizebox{\hsize}{!}
  {\includegraphics{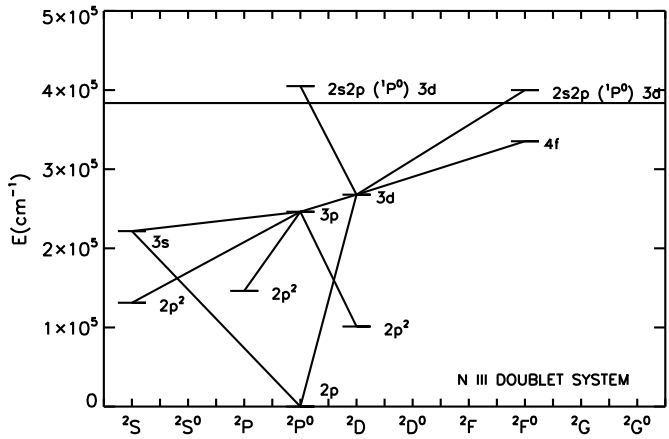}} 
\caption{
Grotrian diagram displaying the transitions involved in the \NIII\
emission lines problem. The horizontal line marks the \NIII\
ionization threshold. The \NIII\ \trip\ triplet is formed by the
transitions 3d $\rightarrow$ 3p, while the absorption lines at
$\lambda\lambda$4097-4103 are due to the transitions 3p $\rightarrow$
3s. An effective drain of 3p is provided by the `two electron
transitions' 3p $\rightarrow$ 2p$^2 (^2$S, $^2$P, $^2$D).  The levels
above the ionization limit are the autoionizing levels that feed level
3d via dielectronic recombination (Sect.~\ref{niii-case}). Cascade
processes (4f $\rightarrow$ 3d) can overpopulate
the 3d state as well. The Swings mechanism involves the resonance transitions
2p $\rightarrow$ 3s and 2p $\rightarrow$ 3d. Note that the (energetic)
positions of the autoionizing niveaus have been shifted upwards for
clarity.} 
\label{dr-mihalas}
\end{figure}

For the \NIII\ triplet produced by the 3p - 3d transitions (see
Fig.~\ref{dr-mihalas}), such a mechanism should result in a (relative)
overpopulation of the upper level, 3d. In the early work about
Of stars, the fluorescence mechanism developed by \cite{bowen35} has
been suggested to trigger such an overpopulation.
%
Many authors (\citealt{swings40,swings48,oke54}) argued against the
relevance of the Bowen mechanism in Of stars, because of the lack of \OIII\
emission lines at $\lambda\lambda$3340, 3444, 3759 \AA\ which are connected
to the involved levels. 

An alternative suggestion is due to \cite{swings48} and relies on an intense
continuum that may directly 
pump the resonance transition 2p $\rarrow$ 3d (thus producing the required
overpopulation, $b_{{\rm 3d}} > b_{{\rm 3p}}$) as well as 2p $\rarrow$ 3s, 
while the transition 2p - 3p is radiatively forbidden. The implied
overpopulation of level 3s due to pumping may then explain why the
transitions 3s - 3p at $\lambda\lambda$4097-4103 are always in
absorption ($b_{{\rm 3s}} > b_{{\rm 3p}}$). 

Without such pumping of the 3s level, an auxiliary draining mechanism
for the 3p level is needed, since otherwise an overpopulation,
$b_{{\rm 3p}} > b_{{\rm 3s}}$, might occur due to cascade processes,
implying the presence of emission at $\lambda\lambda$4097-4103, which
is not observed. Indeed, the anomalous `two electron
transitions' 2p$^2$ ($^2$S, $^2$P, $^2$D) - 3p with transition
probabilities comparable to the 3p $\rarrow$ 3s one electron transition have
been identified by \cite{nikitin63} as potentially important draining
processes. Calculations by BM71 show that the presence
of these draining processes is sufficient both to ensure the
overpopulation of 3d (relative to 3p) and to prevent the
overpopulation of 3p relative to 3s.
Thus, these `two electron transitions' play a key role.

BM71 also noted a problem for the Swings mechanism when applied to
realistic conditions. The 2p $\rightarrow$ 3s and 2p $\rightarrow$ 3d
resonance lines are expected to be much more opaque than 3p
$\rightarrow$ 3s and 3d $\rightarrow$ 3p, and should be consequently
in detailed balance in the line forming region. That would mean no
pumping and no overpopulation of 3d relative to 3p (but see
Sect.~\ref{wind-effects}).
The problem became (preliminary) solved when BM71 suggested a
third potential mechanism. They realized the existence of a large number of
autoionizing levels that either connect directly to the 3d state or to
levels that can cascade downwards to 3d. Hence, the latter state may
become strongly overpopulated by {\it dielectronic recombination} (`DR', see
Sect.~\ref{dr}). 

The most influential analysis of the \NIII\ emission lines problem until now 
has been carried out by \citet[`MH']{mihalas73}, building on the work by
BM71. They used static, plane-parallel models trying to explain
the effect for O((f)) and O(f) stars. As a final result, they were able to
reproduce the \NIII\ triplet emission at the observed temperatures and
gravities in parallel with absorption at $\lambda\lambda4097-4103$,
%
by overpopulating level 3d primarily via dielectronic
recombination. The subsequent 3d~$\rarrow$~3p cascade produces the
emission. The strong drain 3p $\rarrow$ 2p$^2$ via `two electron
transitions' enhances the overpopulation of 3d relative to 3p by
depopulating 3p and prevents emission in the 3p~$\rarrow$~3s lines.
Until to-date, dielectronic recombination is the canonical explanation
for the formation of the f-features. 

\section{Dielectronic recombination}
\label{dr}

If two electrons are excited within a complex atom/ion with several
electrons, they can give rise to states with energies both below and
above the ionization potential. States above the ionization limit, under
certain selection rules, may preferentially autoionize to the ground state
of the ion plus a free electron. 

Thus the ionization from an initial bound
state A(i) to an ionized final state A$^+$(f) can occur either directly,
or by a (different) transition from the initial bound state A(i) to an
intermediate doubly excited state $\bar{\rm A}$ above the ionization potential
that finally autoionizes to A$^+$(f).
%
%
In this respect, Photo-Excitation of the Core (PEC) resonances \citep{yu87} are
particularly strong, because they correspond to a single electron
transition in which the outer electron is a spectator and does not
change. 

The inverse process is also possible, if an ion 
collides with an electron of sufficient energy, leading to a doubly
excited state. Generally, the compound state will immediately
autoionize again (large autoionization probabilities, $A^{\rm a} \sim
10^{13}-10^{14}$~s$^{-1}$). In some cases, however, a stabilizing
transition occurs, in which one of the excited electrons, usually the
one in the lower state, radiatively decays to the lowest available
quantum state. This process is the dielectronic
recombination,
%
and can be summarized as the capture of an
electron by the target leading to an intermediate doubly excited state 
that stabilises by emitting a photon rather than an electron.

\subsection{The \NIII\ emission triplet at \trip} 
\label{niii-case} 

As pointed out by \cite{mihalas71}, there are two important autoionizing
series in the \NIII\ ion, of the form 2s2p($^1$P$^0$)$nl$ and 2s2p($^3$P$^0$)$nl$,
along with few bound double excitation states with similar configuration. 
Since states of the form 2s2p$nl$ are directly coupled to 2s$^2nl$ states,
they are of great importance. In the following, we consider only the
states from the singlet series, because the transitions from the
triplet series to 2s$^2nl$ states are not electric dipole allowed transitions
in LS coupling.

The singlet series comprises only two bound configurations, 2s2p3s and
2s2p3p, whilst the 2s2p3d configuration lies only 1.6~eV
above the ionization potential and is of major importance, since the 
low position in the continuum produces strong dielectronic recombination,
\beq
{\rm 2s}^2(^1{\rm S}) \mbox{ (\NIV) }+ \mathrm{e}^- \rarrow {\rm 2s2p}(^1{\rm
P}^0){\rm
3d} (^2{\rm P}^0\,,\ ^2{\rm F}^0).
\label{stab}
\eeq 
Note that the core electron transition (see above),
2s$^2$($^1$S)~$\rarrow$~2s2p($^1$P$^0$), is
equivalent to the resonance transition in \NIV. Thus, any 2s$^2$($^1$S)$nl$
$\rarrow$ 2s2p($^1$P$^0$)$nl$ transition gives rise to strong, broad resonances
(see Fig.~\ref{smooth-op}), unless the 2s2p($^1$P$^0$)$nl$ state is truly
bound. 

The autoionizing states (Eq.~\ref{stab}) can stabilize via two
alternative routes, either
\beq
{\rm 2s2p}(^1{\rm P}^0){\rm 3d} (^2{\rm P}^0\,,\ ^2{\rm F}^0) \rarrow
{\rm 2s2p}(^1{\rm P}^0){\rm 3p} (^2{\rm S}\,,\ ^2{\rm P}\,,\ ^2{\rm D}) +
h\nu,
\label{stab_1}
\eeq
that end in the doubly excited bound configuration mentioned above, or the
one that might overpopulate the upper level of the transition 3d $\rarrow$ 3p,
\beq
{\rm 2s2p}(^1{\rm P}^0){\rm 3d} (^2{\rm P}^0\,,\ ^2{\rm F}^0) \rarrow
{\rm 2s}^2{\rm 3d} (^2{\rm D}) + h\nu,
\label{stab_2}
\eeq
thus playing a (potential) key role in the \NIII\ triplet emission.
The latter route is displayed in the upper part of
Fig.~\ref{dr-mihalas}.

\subsection{Implementation into {\sc fastwind}}
\label{fw-dr}

Dielectronic recombination and its inverse process have been implemented 
into {\sc fastwind} in two different ways, to allow us to use different data
sets. Specifically, we implemented
\begin{itemize}
\item[(i)] an \textit{implicit method} where the
contribution of the dielectronic recombination is already included in the
photoionization cross-sections. This method needs to be used for data that
has been calculated, e.g., within the OPACITY project \citep{cunto92} and 
\item[(ii)] an \textit{explicit method} using resonance-free
photoionization cross-sections in combination with explicitly included 
stabilizing transitions from autoionizing levels. This method will be used
when we have information regarding transition frequencies and strengths of
the stabilizing transitions. 
\end{itemize}
Later on we discuss the advantages and
disadvantages of both methods, whereas in Appendix~\ref{exp-dr} we provide
some details on the explicit method, and in Appendix~\ref{imp-dr} we show
that the implicit and explicit method are consistent as long as the
autoionizing states are in LTE, as often the case (or frequently assumed). 

\begin{figure}
\resizebox{\hsize}{!}
 {\includegraphics{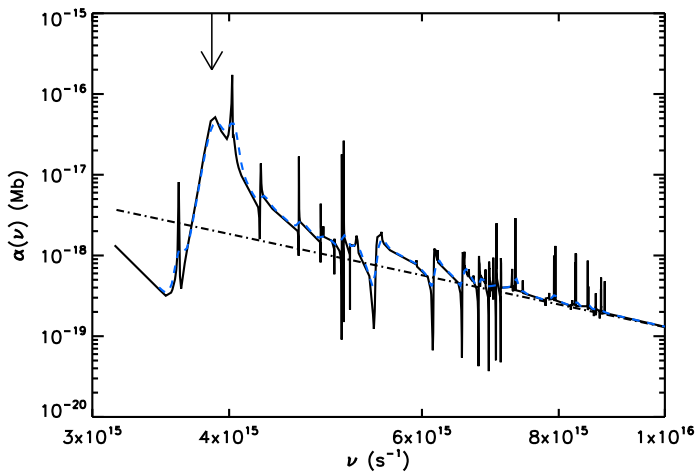}}
\caption{
Comparison of the `raw' cross-section from the OPACITY project (black), and
the smoothed one (grey/blue), for the 3d $^2$D state of \NIII. Note the
numerous complex resonances. An example for a PEC resonance, being much
broader than the usual Rydberg resonances, is marked by the vertical
arrow.
Dashed: corresponding resonance-free data in terms of the~\citet{seaton58}
approximation from {\sc wm}-basic.}
\label{smooth-op}
\end{figure}

Photoionization cross-sections from the OPACITY
project data include the contributions of dielectronic recombination and
will be used within the implicit method. All these cross-sections display
complex resonances (where the largest and widest ones are the PEC
resonances), which somewhat complicate the implementation of this
method. Since some of the resonances are quite narrow, care must be taken
when sampling the cross-sections. If performing a straightforward
calculation, the radiative transfer would need to be solved at each point of
the fine-frequency grid required by the resonances in the `raw' data, which
would increase the computational effort considerably. To circumvent the
problem we implemented a method that is also used in {\sc wm}-basic
\citep{pauldrach01} and in {\sc cmfgen} \citep{hilliermiller98} (see
also \citealt{Bautista98}). The raw
OPACITY project cross-sections are smoothed (Fig.~\ref{smooth-op}) to adapt
them to the standard continuum frequency grid used within the code,
which has a typical resolution of a couple of hundred \kms. To this end, the
data are convolved with a Gaussian profile of typically 3,000~\kms width,
via a Fast Fourier Transform. This convolution ensures that the area
under the original data remains conserved, 
and that all resonances are treated with sufficient accuracy. By
means of this approach, the ionization and recombination rates should
be accurately represented, except for recombination rates at very low
temperatures, where the recombination coefficient is quite sensitive
to the exact location of the resonance (see~\citealt{hilliermiller98}).

\subsection{Implicit vs. explicit method}
\label{imp_vs_exp}

Though under similar assumptions both methods achieve similar
results, there are certain advantages and disadvantages that are
summarized in the following (see also \citealt{hilliermiller98}).

\paragraph{The implicit method} has the advantage that for states that can
autoionize in LS coupling their contribution to dielectronic recombination
(and the inverse process) is already included within the photoionization
data. 
There is no need to look for both the important autoionizing series 
to each level and the oscillator strengths of the 
stabilizing transitions. On the other hand, there are also disadvantages. 
Narrow resonances are not always well resolved, and, if the resonance
is strong, the dielectronic recombination rate from such a resonance could become
erroneous. Besides, the positions of the resonances are only approximate.
If line coincidences are important, this can affect the transfer and the 
corresponding rates. \citet{hilliermiller98} suggested to avoid the
implicit approach for those transitions where dielectronic recombination is
an important mechanism. In our new \NIII\ model ion (Sect.~\ref{atom
niii}) we have followed their advice.
Finally, dielectronic recombination rates calculated by the implicit method
are inevitably based on the assumption that the autoionizing levels are in 
LTE with respect to the ground state of the next higher ion.\footnote{This will
almost always be the case, otherwise the states are more or less bound.}
In the rare case that this were no
longer true, the implicit method cannot be used. Then, the autoionizing
levels need to be included into the model atom, and all transitions need
to be treated explicitly.

\paragraph{The explicit method} has the theoretical advantage that
resonances can be inserted at the correct wavelength if known. (Though
one has to admit that resonance positions will never be very accurate.
For the PECs this does not matter though.) Often, however, the profile
functions for the resonances are difficult to obtain (which, on the
other hand, are included in the implicit approach). The width of these
Fano profiles is set by the autoionization coefficients, which,
frequently, are not available. To overcome this problem, we
follow the approach by MH and assume the resonance to be
wide (which is true for the most important PEC-resonances), such that
we can use the mean intensity instead of the scattering integral when
calculating the rates, and become independent of the specific profile.
When line coincidences play a role, this might lead to certain errors
though.

\section{The \NIII\ model ion}
\label{atom niii}
We implemented a new Nitrogen model atom into the {\sc fastwind}
database, consisting of \NII\ to \NV. In the following, we provide
some details of the \NIII\ model ion, whilst the remaining ions will
be described in Paper~II.
Our \NIII\ model consists of 41 levels, quite similar to the \NIII\
model as used within {\sc wm}-basic. LS-coupled terms up to principal
quantum number $n = 6$ and angular momentum $l = 4$ have been
considered. Table~\ref{atom_lev_niii} provides detailed information
about the selected levels. 
All fine-structure sub-levels have been packed into one LS-coupled
term. \footnote{When calculating the final synthetic profiles we
use, when necessary, the un-packed levels by assuming that $n_i/g_i$
-- with occupation number $n_i$ and statistical weight $g_i$ -- is
similar for each sub-level within a packed level, due to strong
collisional coupling.} Two spin systems (doublet and quartet) are
present and treated simultaneously (see
Fig.~\ref{niii-doublet-quartet} for Grotrian diagrams).

We account for all (193) allowed electric dipole radiative transitions
between the 41 levels, as well as for (164) radiative intercombination
transitions between the two spin system. Oscillator strengths have
been taken from NIST\footnote{
http://www.nist.gov/physlab/data/asd.cfm, firstly described in
\citet{nist}} when possible, and else from the {\sc wm}-basic
database.\footnote{see \citet{pauldrach94c}. In brief, the atomic
structure code {\sc superstructure} (\citealt{eissner69, eissner91})
has been used to calculate all bound state energies in LS and
intermediate coupling as well as related atomic data, particularly
oscillator strengths including those for stabilizing transitions.}
NIST \NIII\ data are mostly from \citet{Bell95}, and from OPACITY
project calculations by \citet{Fernley99}.

Roughly one thousand bound-bound collisional transitions between all
levels are accounted for.
(i) For all collisions between the 11 lowest levels, 2s$^2$ 2p,
2s 2p$^2$, 2p$^3$, and 2s$^2$ 3$l$ ($l$~=~s, p, d), comprising doublet
and quartet terms,
we use the collision strengths as
calculated by \citet{stafford94}, from the ab initio R-matrix method 
\citep{berrington87}. 
(ii) For most of the optically allowed transitions between higher levels (i.e.,
from level 11 as the lower one on), the \citet{vanregemorter62}
approximation is applied.
(iii) For the optically forbidden transitions and the remainder of optically
allowed ones (transitions involving the highest level), the semi-empirical
formula from \citet{allen73} (with $\Omega$~=~1) is used.

Photoionization cross-sections have been taken from the OPACITY
Project on-line atomic database,
TOPbase\footnote{http://cdsweb.u-strasbg.fr/topbase/topbase.html}
\citep{cunto92}. These cross-sections have been computed by
\citet{Fernley99} by the R-matrix method using the close-coupling
approximation and contain, as pointed out previously, complex
resonance structures.

For excited \NIII\ levels with no OPACITY Project data available (5g $^2$G
and 6g $^2$G) and in those cases where we
apply the explicit method for dielectronic recombination,
resonance-free cross-sections are used, provided in terms of the
\citet{seaton58} approximation,
%
\beq
\alpha(\nu) = \alpha_{0}[\beta(\nu_{0}/\nu)^{s} +
(1-\beta)(\nu_{0}/\nu)^{s+1}],
\label{Seaton_cross}
\eeq
with $\alpha_{0}$ the cross-section at threshold $\nu_0$, and
$\beta$ and $s$ fit parameters from the {\sc wm}-basic atomic
database. For most cross-sections, a reasonable consistency between
these and the OPACITY project data is found, if one compares the
resonance-free contribution only, see Fig.~\ref{smooth-op}.

The most important dielectronic recombination and reverse ionization processes
are treated by the explicit method (in particular, recombination to the
strategic 3d level). Corresponding atomic data (wavelengths and
oscillator strengths of the stabilizing transitions) are from the 
{\sc wm}-basic atomic database as well.
Finally, the cross-sections for collisional ionization are derived 
following the \citet{seaton62} formula, with threshold
cross-sections from {\sc wm}-basic.

\section{\NIII\ (emission) line formation} 
\label{test-dr} 
In the following, we describe the results of an extensive test series
regarding our newly developed \NIII\ model ion. In particular,
we check if the triplet appears in emission in the observed
parameter range, and if the $\lambda4097$ line remains always in absorption.
Note that these lines should be strongly correlated, i.e., the stronger
the emission in the triplet, the weaker the absorption at $\lambda4097$,
since both transitions share the level 3p (Fig.~\ref{dr-mihalas}).
A change in the corresponding departure coefficient leads to
a change in both lines. For example, if $b_{{\rm 3p}}$ becomes
diminished due to a more efficient drain by the `two electron
transitions' (see Sect.~\ref{intro-niii}), this leads to more emission at
\trip\ and to less absorption at $\lambda4097$, due to less cascading.


For all tests, we calculated model-grids that cover the
same stellar parameters (O-type dwarfs and (super-) giants) as used by
MH, listed in Table~\ref{models-mih}. All tests have been performed by
means of {\sc fastwind}, using our {\it complete} Nitrogen model atom.

\begin{table}
\caption{Model grid used by \cite{mihalas73} and in our test series.}
\label{models-mih}
\tabcolsep0.8mm
\begin{center}
\begin{tabular}{c|ccccccccc} 
\hline
\hline
Model T& 3233 &  3540 & 3533 & 3740 & 3735 & 4040 & 4035 & 4540 &5040 \\
\hline
\Teff (kK) & 32.5 & 35.0 & 35.0 & 37.5 & 37.5 & 40.0 & 40.0 & 45.0 & 50.0 \\
\logg      & 3.3  & 4.0  & 3.3  & 4.0  & 3.5  & 4.0  & 3.5  & 4.0  &4.0\\  
\hline
\mdot & 1.0 & 0.35 & 1.82 & 0.58 & 3.16 & 0.93 & 5.3 & 8.05 & 17.5\\  
\hline
\end{tabular}
\end{center}
\tablefoot{
The mass-loss rates provided in the last row (in units of \mdu) refer to our
tests of wind effects (Sect.~\ref{wind-effects}) alone. All other tests have
been performed with negligible \mdot.
}
\end{table}

\subsection{Comparison with the results from MH}
\label{mih-hum} 

\begin{figure*}
\resizebox{\hsize}{!}
 {\includegraphics{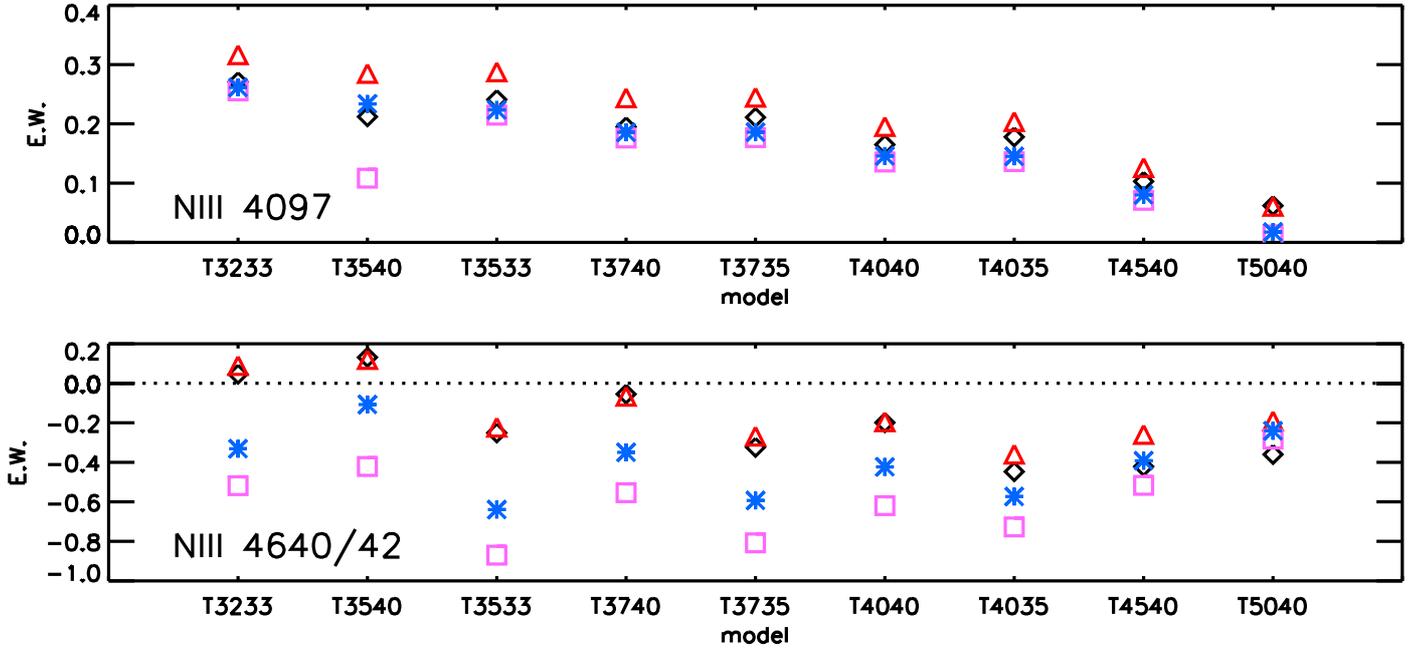}}
\caption{Comparison of equivalent widths (EW) for \NIII\ $\lambda4097$
and $\lambda\lambda4640/42$. All atmospheric models calculated with
`pseudo line-blanketing' (see text). Black diamonds: results from MH;
red triangles: new {\sc fastwind} calculations with `mixed' \NIII\
ionic model (new level structure, but important transition data from
MH); purple squares: new {\sc fastwind} calculations with new \NIII\
model; blue asterisks: as squares, but DR-contribution to 3d level
diminished by a factor of two. Here and in the following,
positive and negative EWs refer to absorption and emission lines,
respectively.}
\label{mh_vs_mhcalc}
\end{figure*}

First, we test if we are able to reproduce the MH-results for O((f))
and O(f) stars. To this end, we need to invoke (almost) identical
conditions, regarding both atmospheric and atomic models. Thus, we
modified our \NIII\ ionic model, replacing part of our new data with
those used by MH (`mixed' ionic model).
In particular, we replaced data for dielectronic recombination (to the
three draining levels, 2p$^2$ ($^2$D, $^2$S,$^2$P), to level 3d,
and to few higher important levels -- \#14, 17, 20 and 21, see
Table~\ref{atom_lev_niii}), the oscillator strengths for the `two
electron transitions' (Table~\ref{fval-mih}), and the photoionization
cross-sections for all levels below 3d (the cross-sections for
the latter did already agree). 

\begin{table}
\caption{Oscillator strengths for the `two electron transition' as used by
\cite{mihalas73} and within our new atomic model.}
\label{fval-mih}
\begin{center}
\begin{tabular}{ccc}
\hline 
\hline
\multicolumn{1}{c}{Transition}
&\multicolumn{1}{c}{$f_{\rm lu}$ (MH)}
&\multicolumn{1}{c}{$f_{\rm lu}$ (used in this work)}
\\
\hline
2p$^2$ $^2$D - 3p $^2$P$^0$ & $2.50 \cdot 10^{-4}$ & $4.38 \cdot 10^{-3}$ \\
2p$^2$ $^2$S - 3p $^2$P$^0$ & $2.00 \cdot 10^{-4}$ & $1.31 \cdot 10^{-2}$ \\
2p$^2$ $^2$P - 3p $^2$P$^0$ & $2.40 \cdot 10^{-2}$ & $4.02 \cdot 10^{-4}$ \\
\hline
\end{tabular}
\end{center}
\end{table}

Consistent with the MH-models, a Nitrogen abundance of [N]
\footnote{[A] = $\log$ A/H + 12, with A/H the number density of
element A with respect to Hydrogen} = 8.18 was adopted.  This is a
factor of 2.5 larger than the solar one, [N$_\odot$] = 7.78
\citep{asplund05}.\footnote{\citet{asplund09} provide a slightly
larger value, [N$_\odot$] = 7.83~$\pm$~0.05, where this difference is
irrelevant in the following context.} 
To account for the absence of a wind in their
models, a negligible mass-loss rate of $\mdot = 10^{-9} \msunyr$ was
used.

\begin{figure}
\resizebox{\hsize}{!}
   {\includegraphics{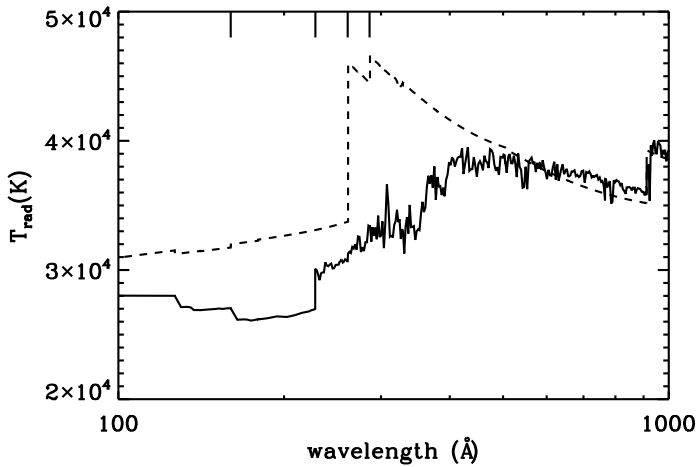}}
\caption {EUV radiation temperatures for model `T3735'
with pseudo (dashed) and full (solid) line-blanketing. Important ionization
edges are indicated as vertical markers. From left to right:
\NIV, \HeII, \NIII\ (ground-state), and \NIII\ 2p$^2$ $^2$D.}
\label{T3735_trad}
\end{figure}

Though line-blocking/blanketing could not be included into the
atmospheric models in 1973, MH realized its significance and
tried to incorporate some important aspect by means of a
`pseudo-blanketing' treatment. The \NIII\ photoionization edge (at
261~\AA) lies very close to the \HeII\ ground state edge (at
228~\AA) in a region of low continuum opacity and high emergent
flux (if no line-blocking due to the numerous EUV metal-lines is
present), and would cause severe underpopulation of the \NIII\ ground
state if the blanketing effect is neglected, as shown by BM71. MH
argued that heavy line-blanketing and much lower fluxes are to be
expected in this region. To simulate these effects, they extrapolated
the \HeII\ ground-state photoionization cross-section beyond the
Lyman-edge up to the \NIII\ edge, using a $\nu^{-3}$ extra\-polation.
To be consistent with their approach, we proceed in the same way by
including this treatment into {\sc fastwind} (see Fig.~\ref{T3735_trad}).

\begin{figure}
\resizebox{\hsize}{!}
  {\includegraphics{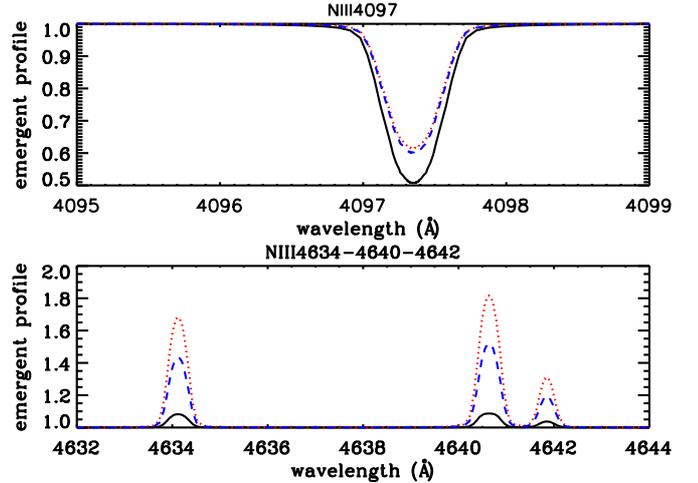}}
\caption{Comparison of {\sc fastwind} \NIII\ line profiles from model `T3740'
with `pseudo line-blanketing', using different atomic data sets.
Solid (black): `mixed' \NIII\ model (see text); dotted (red): new \NIII\ model;
dashed (blue): new \NIII\ model, but DR-contribution to 3d level
diminished by a factor of two.} 
\label{cas}
\end{figure}

Figure~\ref{mh_vs_mhcalc} displays the comparison between the
resulting equivalent widths from the MH models (black
symbols) and our models using the conditions as outlined above (red
triangles). Overall, the agreement for \trip\ is
satisfactory,\footnote{Here and in the following, we only display the
total equivalent widths of the $\lambda\lambda$4640/4642 components;
the behavior of $\lambda$4634 is analogous.} and slight differences
are present only for the hottest models. In agreement with the MH
results, our profiles turn from absorption into emission around $\Teff
\sim 37,000$~K for dwarfs and at $\Teff \sim 33,000$~K for
(super-)giants. As well, the $\lambda4097$ line is always in
absorption throughout the grid, though our calculations predict
moderately more absorption in this line. All lines show the same trend
in both sets, and the remaining differences might be attributed to
still somewhat different atomic data.\footnote{number of levels,
collisional data and LTE assumption concerning the quartet system
levels by MH.}

\begin{figure*}
\resizebox{\hsize}{!} {\includegraphics{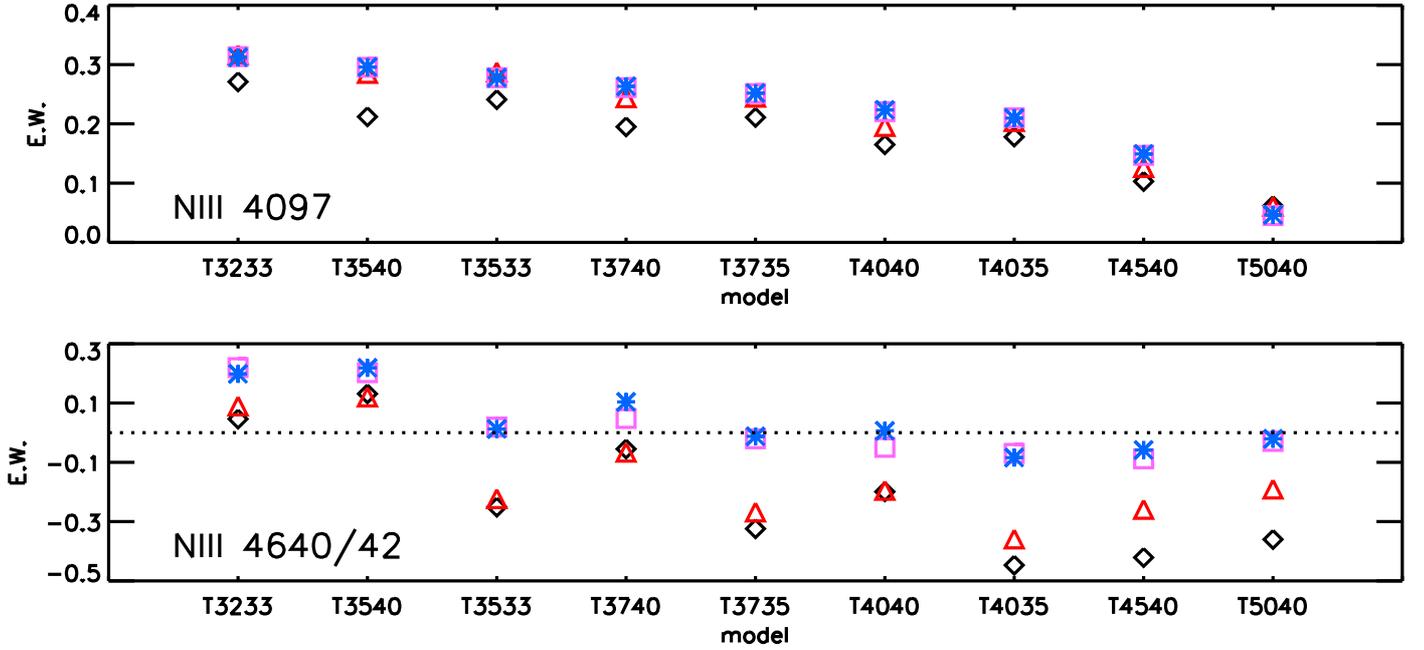}}
\caption{Comparison of equivalent widths (EW) for \NIII\ $\lambda4097$
and $\lambda\lambda4640/42$. Diamonds and triangles as
Fig.~\ref{mh_vs_mhcalc}. Purple squares: new results using new atomic
data and realistic line-blanketing; blue asterisks: as squares,
but DR-contribution to 3d level set to zero.} 
\label{full-blank-grid}
\end{figure*}


We note already here that in all cases the emission is more pronounced in
low-gravity objects. In high-gravity objects (dwarfs) part of the emission
is suppressed because of higher collisional rates ($\propto n_{\rm
e}$), driving the relative populations towards LTE.  

After demonstrating that we can (almost) reproduce the profiles
calculated by MH when similar conditions are applied, the next step is
to investigate the effect of the new \NIII\ atomic data implemented
during the present work. In fact, this leads to much more triplet
emission, see Figs.~\ref{mh_vs_mhcalc} (purple squares) and~\ref{cas}
(dotted profiles). Even for the coolest models, where MH still obtain
weak absorption, our calculations result in strong emission. This big
difference is produced by larger DR rates into 3d and a larger drain
of 3p by the `two electron transitions', due to 
larger oscillator strengths (see
Table~\ref{fval-mih}).\footnote{A similar effect has already been
noted by MH when they increased the corresponding oscillator strengths
in their atomic model.} Note also that the $\lambda4097$ line becomes
weaker, in accordance with the correlation predicted above.

The reaction of the emission strength on the dielectronic data also allows
us to check the validity of the implementation of dielectronic recombination
by the explicit method, described in Sect.~\ref{exp-dr}, particularly the
dependence on the oscillator strength of the stabilizing transition(s). 
Figures~\ref{mh_vs_mhcalc} (blue asterisks) and~\ref{cas} (dashed) show
the reaction of the triplet lines if we diminish the oscillator strength of the
strongest stabilizing transition to 3d by a factor of two. This reduction
leads to a significantly weaker emission throughout the whole grid, and a
slight increase in the absorption strength of $\lambda4097$. For a
further test on the consistency between implicit and explicit method,
see Appendix~\ref{consistency}.

\subsection{Models with full line-blanketing}
\label{full-blank}

In their study, MH could not consider the problem for realistic
atmospheres accounting for a consistent description of
line-blocking/blanketing, simply because such atmospheric models did not
exist at that time. To investigate the differential effect on the \NIII\
emission lines, we calculated the same grid of models, now including full
line-blanketing\footnote{Using background metallicities corresponding to the
`older' solar abundances from \citet{grevesse98}, but keeping [N]~=~8.18.} as
incorporated to {\sc fastwind}, and compare with the MH calculations
(Fig.~\ref{full-blank-grid}, purple squares vs. black diamonds,
respectively). 

Astonishingly, the triplet emission almost vanishes throughout the
whole grid. This dramatic result points to the importance of including
a realistic treatment of line-blanketing when investigating the
emission line problem in Of stars. It further implies, of course, that
the new mechanism preventing emission in line-blanketed models needs
to be understood, and that an alternative explanation/modeling for the
observed emission must be found.

Let us first investigate the `inhibition effect', by considering model
`T3735' that displays one of the largest reactions. The most
important consequence of the inclusion of line-blocking/blanketing is
the decrease of the ionizing fluxes in the EUV.
Figure~\ref{T3735_trad} displays corresponding radiation temperatures
as a function of wavelength, for the `pseudo line-blanketed', simple
model constructed in analogy to MH (dashed) and the fully
line-blanketed model (solid).

Indeed, the presence of substantially different ionizing fluxes around
and longward of the \NIII\ edge is the origin of the different triplet
emission, via two alternative routes. For part of the cooler models
(not the one displayed here), significantly higher radiation
temperatures in the \NIII\ continuum ($\lambda < 261$~\AA) of the
`pseudo line-blanketed' models lead to a strong ground-state
depopulation.  Moreover, the ground-state and the 2p$^2$ levels are
strongly collisionally coupled, from the deepest atmosphere to the
line-forming region of the photospheric lines. Thus, the ground-state
and the 2p$^2$ levels react in a coupled way. If the ground-state
becomes depopulated because of a strong radiation field, the 2p$^2$
levels become underpopulated as well, giving rise to a strong drain
from level 3p and thus a large source-function for the triplet
lines. Corresponding models including a realistic line-blocking (with
lower EUV-fluxes) cause much less depopulation of the ground-state and
the 2p$^2$ levels, thus suppressing any efficient drain and preventing
strong emission in the triplet lines.

For hotter models, e.g., model `T3735' as considered here, the
operating effect is vice versa, but leads to the same result. In the
simple, MH-like model, the 2p$^2$ levels become strongly depopulated
in a direct way, because (i) the ionizing fluxes at the corresponding
edges ($\lambda > 261$ \AA) are extremely high (the `pseudo
line-blanketing extends `only' until 261~\AA), and (ii) because
also a strong ionization via doubly excited levels is present, again
due to a strong radiation field at the corresponding transition
wavelengths around 290-415\AA. Due to the depopulation of the 2p$^2$
levels, the coupled ground-state becomes depopulated as
well. 

On the other hand, the depopulation via direct and `indirect'
ionization from 2p$^2$ does not work for consistently blocked models,
because of the much lower ionizing fluxes. Insofar, the extension of
the \HeII\ ground-state ionization cross section by MH, to mimic the
presence of line-blocking, was not sufficient. They should have
extrapolated this cross-section at least to the edge of the lowest
2p$^2$ level. In this case, however, they would have found much lower
\NIII\ emission, too low to be consistent with observations.

\begin{figure}
\resizebox{\hsize}{!} {\includegraphics{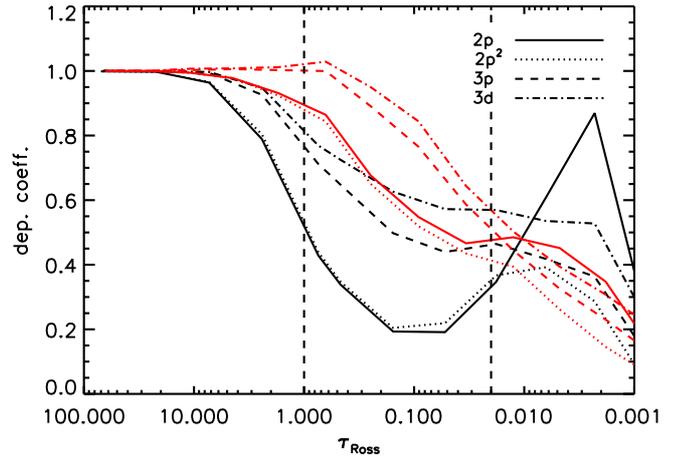}}
\caption{Departure coefficients for important levels regarding the
triplet emission, for model `T3735' with pseudo (black) and full (red)
line-blanketing (level 2p is the ground-state).
The formation region of the triplet lines is indicated by vertical dashed lines. See text.} 
\label{T3735-deps}
\end{figure}

In Fig.~\ref{T3735-deps}, we compare the corresponding departure
coefficients. Black curves refer to the simple, 'pseudo line-blanketed' model,
and red curves to the corresponding model with full line-blocking/-blanketing. 
Obviously, both models display strongly coupled ground and 2p$^2$ states 
in the formation region of photospheric lines (here: $\taur \ge
0.02$), where these states are much more depopulated in the simple model.

For the conditions discussed so far, a significant depopulation of
2p$^2$ is a prerequisite for obtaining strong emission lines in the
optical: only in this case, an efficient drain 3p $\rarrow$ 2p$^2$
due to cascading processes can be produced. The consequence of
different draining efficiencies becomes obvious if we investigate the
run of $b_{{\rm 3p}}$ (dashed). In the pseudo-blanketed model, this level
becomes much more depopulated relative to 3d (dashed-dotted),
leading to much stronger triplet emission than in the full
blanketed model.

\begin{figure}
\begin{minipage}{8cm}
\resizebox{\hsize}{!}
   {\includegraphics{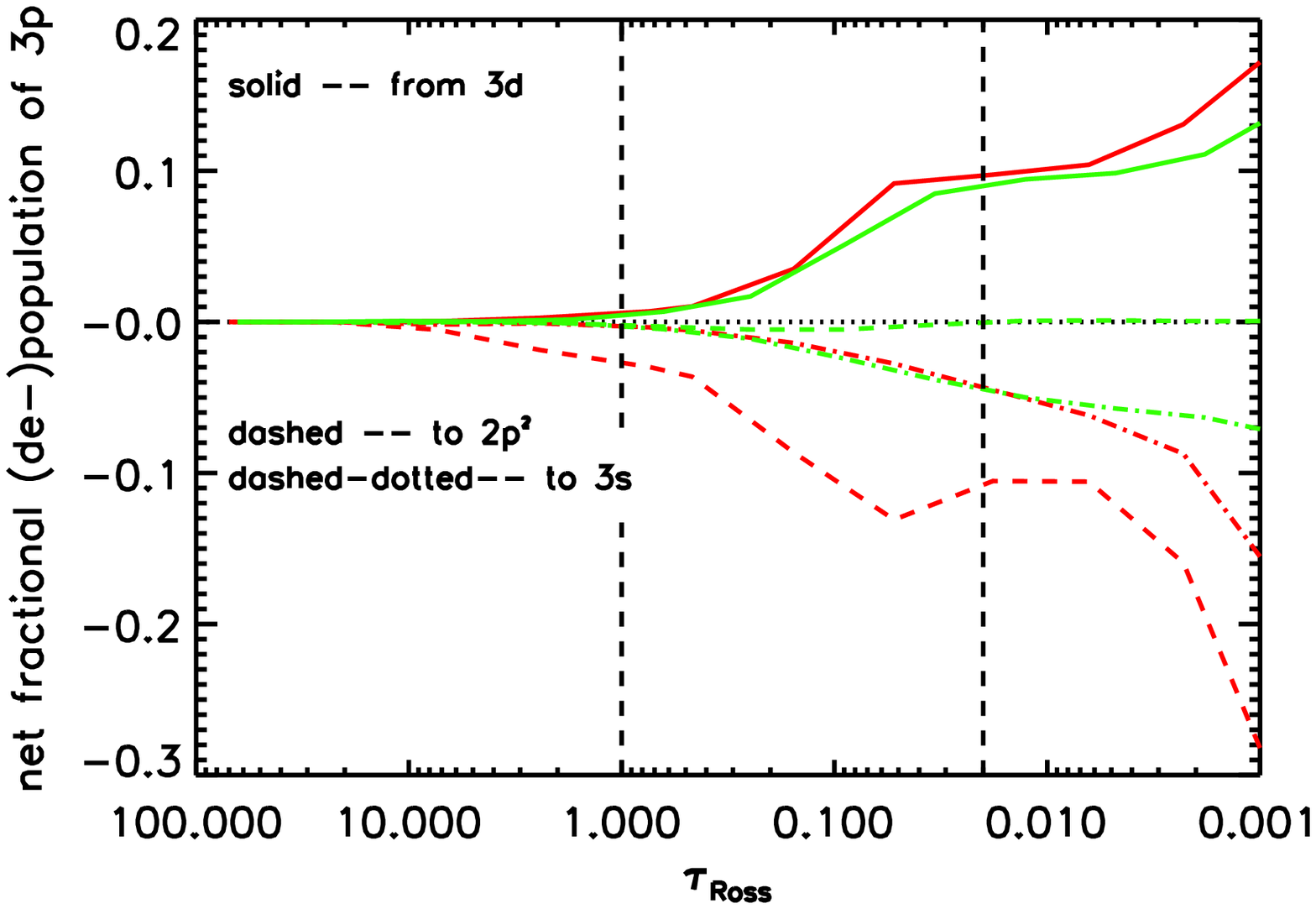}}
\end{minipage}
\hspace{-.5cm}
\begin{minipage}{8cm}
   \resizebox{\hsize}{!}
   {\includegraphics{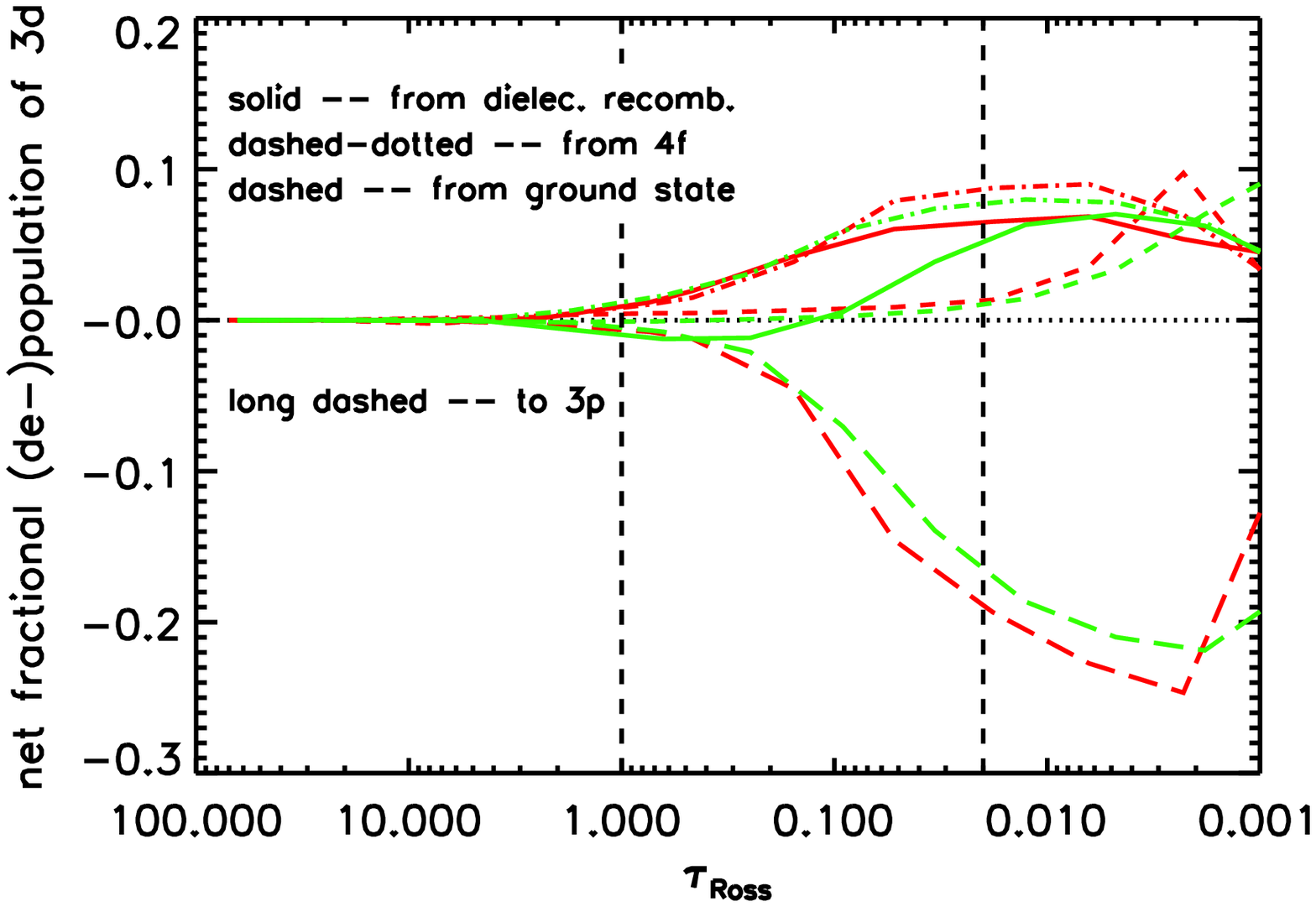}}
\end{minipage}
\caption {Fractional net rates to and from level 3p (upper panel) and level
3d (lower panel), for model `T3735' with pseudo (red) and full (green)
line-blanketing. The formation region of the triplet lines is
indicated by vertical dashed lines.} 
\label{pop-depop-T3735-10-11}
\end{figure}

The impact of the different processes can be examined in detail if we
investigate the corresponding net rates responsible for the population
and depopulation of level 3p (Fig.~\ref{pop-depop-T3735-10-11}, upper
panel). In this and the following similar plots, we display the
dominating individual net rates (i.e., $n_j R_{ji} - n_i R_{ij} > 0$
for population, with index $i$ the considered level) as a fraction of
the total population rate.\footnote{which in statistical equilibrium
is identical to the total depopulation rate.}

\begin{figure*}
\resizebox{\hsize}{!}
  {\includegraphics{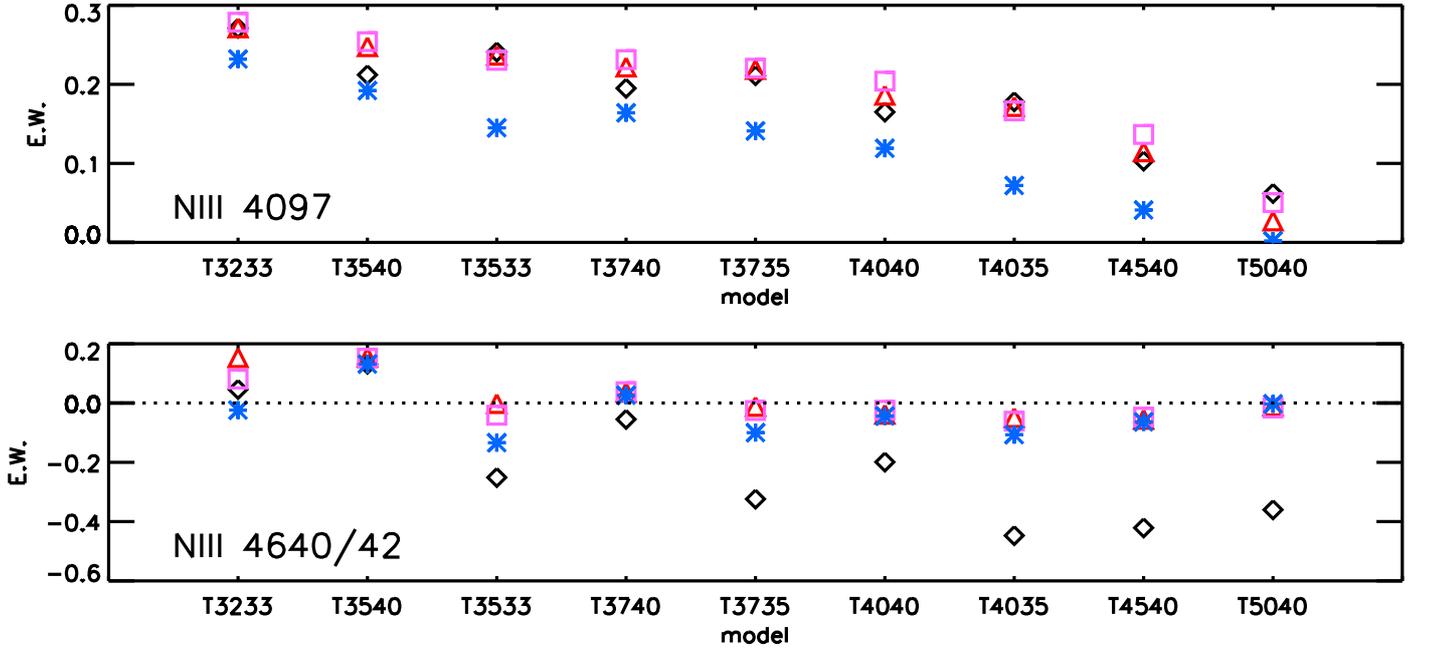}}
\caption{Comparison of equivalent widths for \NIII\ $\lambda4097$ and
$\lambda\lambda4640/42$. Black diamonds as in Fig.~\ref{mh_vs_mhcalc}
(original results from MH). Other models with [N]=7.92. Blue asterisks from {\sc
tlusty} (OSTAR2002), red triangles from {\sc fastwind}, using new atomic
data and realistic line-blanketing; purple squares: as triangles, but
with temperature and electron stratification from {\sc tlusty}. For
details, see text.} 
\label{grid-tl-blank}
\end{figure*}

Indeed, there is a dramatic difference in the net rates that
depopulate level 3p via the `two electron' drain (dashed). Whereas for
the `pseudo line-blanketed' model the net rate into 2p$^2$ is the
dominating one (resulting from the strong depopulation of 2p$^2$),
this rate almost vanishes for the model with full line-blanketing. In
contrast, the other two important processes, the population by 3d
(solid) and the depopulation into 3s (dashed-dotted) are rather
similar. Consequently, 3p becomes less depopulated compared to 3d in
the fully blanketed models. Moreover, since the (relative)
depopulation into 3s remains unaffected, and 3p has a larger
population (see Fig.~\ref{T3735-deps}), level 3s becomes stronger
populated as well.  Thus, the blanketed models produce more absorption
in $\lambda$4097 (cf. Fig.~\ref{full-blank-grid} with
Fig.~\ref{mh_vs_mhcalc}, squares in upper panels).

 Since in the older MH-models the triplet emission is also due to the 
strong population of 3d via DR (Fig.~\ref{pop-depop-T3735-10-11}, lower
panel, red solid), we have to check how the presence of full
line-blocking affects this process. In this case, the net-rate from DR
(green solid) becomes even negative in part of the line forming
region, i.e., the ionization via intermediate doubly excited states
partly outweighs the dielectronic recombination. This difference
originates from two effects: (i) the photospheric electron densities
in the older models are larger, because of missing radiation pressure
from metallic lines.
Higher electron densities imply higher recombination rates. (ii) In the fully
blanketed models, the radiation field at the frequency of the main
stabilizing transition to 3d (777~\AA) is slightly higher (see
Fig.~\ref {T3735_trad}), 
which leads to more ionization. Consequently, DR plays only a minor (or even
opposite) role in the fully blocked models, contrasted to the MH case.
In the blocked models, the major population is via the 4f level
(dashed-dotted). Note also that the Swings mechanism, i.e., a
population via the resonance line at 374~\AA\ (dashed), does not play
any role in photospheric regions, as already argued by BM71 and shown
by MH. 

\begin{figure*}
\resizebox{\hsize}{!}
  {\includegraphics{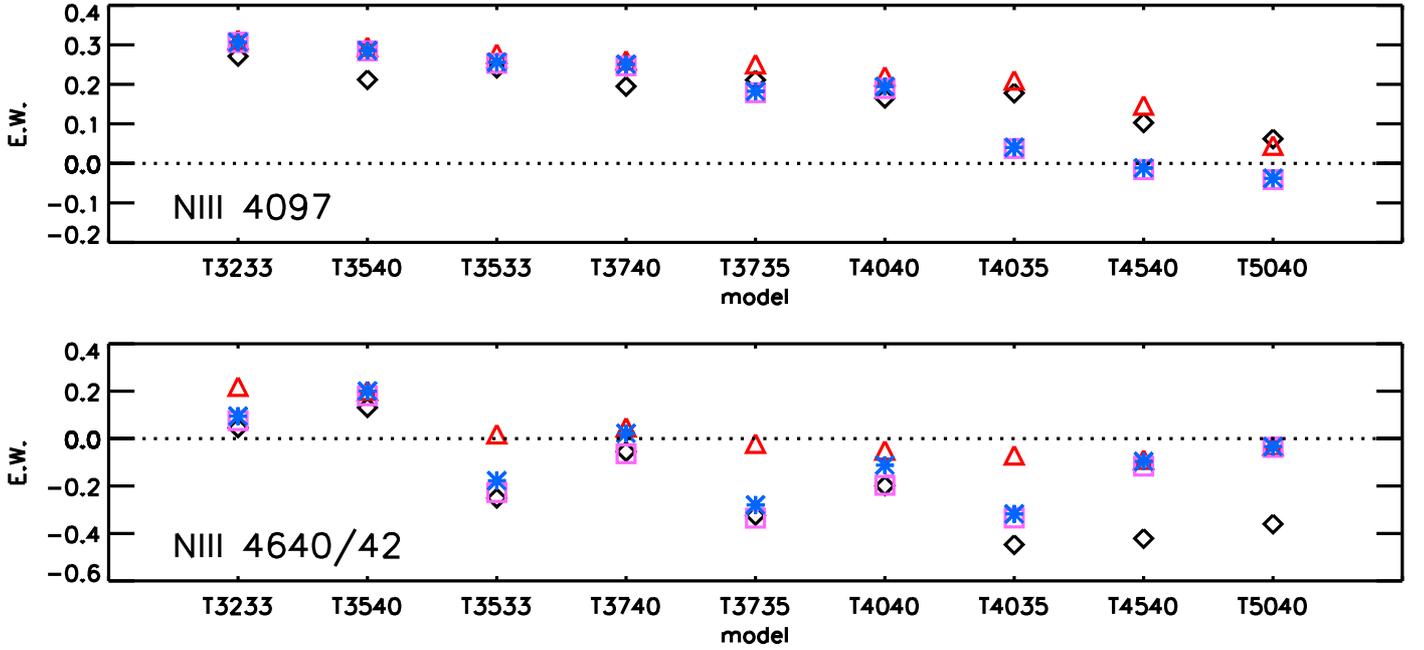}}
\caption{Comparison of equivalent widths (EW) for \NIII\ $\lambda4097$ 
and $\lambda\lambda4640/42$. Black diamonds as in
Fig.~\ref{mh_vs_mhcalc} (results from MH). Red triangles: new results using new
atomic data, realistic line-blanketing, 
no winds; purple rectangles: new results using new
atomic data, realistic line-blanketing and winds with a mass-loss rate
according to the wind-momentum luminosity relation provided by 
\cite{repolust04}; blue asterisks: as purple, but with DR
contribution to 3d level set to zero.} 
\label{grid-wind}
\end{figure*}

For a final check about the importance of DR to level 3d in models
with full line-blanketing, we tested its impact by switching off the
main stabilizing transition. From Fig.~\ref{full-blank-grid} (blue
asterisks), it is obvious that the effect indeed is marginal
throughout the whole grid.  {\it Thus, a correct treatment of
line-blocking seems to suppress both the efficiency of the draining
transitions and the dielectronic recombination}, and we have to ask
ourselves how the observed triplet emission is produced, since the
presence of line-blocking cannot be argued away.

Before tackling this problem, in Fig.~\ref{grid-tl-blank} we compare
our results (red triangles) with those from {\sc tlusty} 
(\citealt{hubeny88, hubeny95})\footnote{a code that assumes
plane-parallel geometry, hydrostatic and radiative equilibrium, and
calculates line-blanketed NLTE model atmospheres and corresponding
synthetic profiles. Due to its restrictions, only objects with
negligible winds can be analyzed.}
(blue asterisks), as provided by the OSTAR2002 grid\footnote{Equivalent
widths from own integration.} (\citealt{LanzHubeny03}; see also
\citealt{Heap06}). Contrasted to all
other similar plots, we used an abundance of [N] = 7.92, 
to be consistent with the 
{\sc tlusty} grid. Obviously, our {\sc fastwind}
results for the \NIII\ emission lines in dwarfs match exactly those
from {\sc tlusty}, whereas our results for supergiants display
somewhat less emission (or more absorption). On the other hand, the
absorption line at $\lambda$4097 is systematically stronger in all our
models, which points to different oscillator strengths. To rule out
potential differences in the atmospheric stratification, 
we calculated an additional grid with temperature and electron
structure taken from {\sc tlusty}, smoothly connected to the wind
structure as calculated by {\sc fastwind}.
Corresponding results are displayed by
purple squares, and differ hardly from those based on the original
{\sc fastwind} structure, except for model T3233 where the differences
in $T$ and $n_{\rm e}$ are larger than elsewhere. The remaining
discrepancies for supergiants can be explained by somewhat higher EUV
fluxes in the {\sc tlusty} models, favouring more \NIII\ triplet
emission. Nevertheless, the disagreement between {\sc
fastwind} and {\sc tlusty} is usually much weaker than between these
two codes and the results by MH,
which underpins our finding that the triplet emission
becomes strongly suppressed in fully blanketed models. 

\subsection{The impact of wind effects}
\label{wind-effects}

So far, one process has been neglected, namely the (general) presence
of winds in OB-stars. Note that MH had not the resources to reproduce
O-stars with truly ``extended'' atmospheres. Regarding their O(f) and
O((f)) objects, on the other hand, there was no need to consider wind
effects, because the observed \NIII\ triplet emission could be
simulated by accounting for DR and `two-electron' drain alone.  
Actually, MH pointed out that in Of-supergiants (with denser winds) the Swings mechanism
could play a crucial role in the overpopulation of 3d:
Velocity fields are able to shift the resonance lines into the
continuum, allowing them to become locally more transparent, and
deviations from detailed balance and strong pumping might occur. 

With the advent of new atmospheric codes, we are now able to
investigate the general role of winds, and to explain how the emission
lines are formed within such a scenario. To this end, we have
re-calculated the model grid from Table~\ref{models-mih}, now
including the presence of a wind. Prototypical values for terminal
velocity and velocity-field exponent have been used,
\vinf~=~2,000~\kms, $\beta~=~0.9$, and mass-loss rates were inferred
from the wind-momentum luminosity relationship (WLR) provided by
\cite{repolust04}, which differentiates between supergiants and other
luminosity classes. Clumping effects have been neglected, but will be
discussed in Sect.~\ref{varpar}. For a summary of the adopted
mass-loss rates, see Table~\ref{models-mih}.

With the inclusion of winds now, we almost recover the emission
predicted in the original MH calculations (Fig.~\ref{grid-wind}). Less
emission is produced only for the two hottest models, where the inclusion
of a wind has no effect. For all other models, however, the wind effect is large, both
for the supergiants and for the dwarfs with a rather low wind-density.

\begin{figure}
\begin{minipage}{8.5cm}
\resizebox{\hsize}{!}
   {\includegraphics{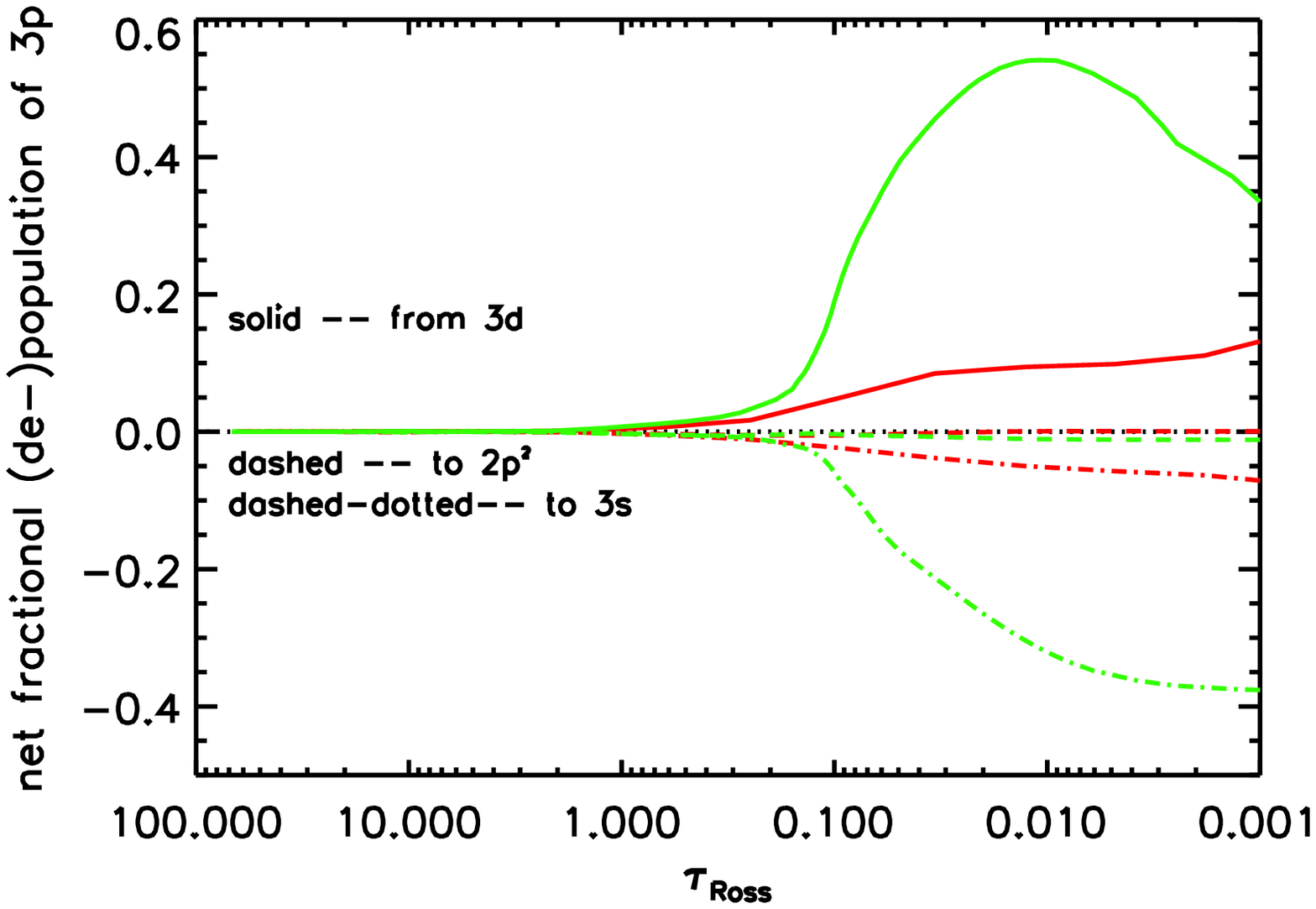}}
\end{minipage}
\hspace{-.5cm}
\begin{minipage}{8.5cm}
   \resizebox{\hsize}{!}
   {\includegraphics{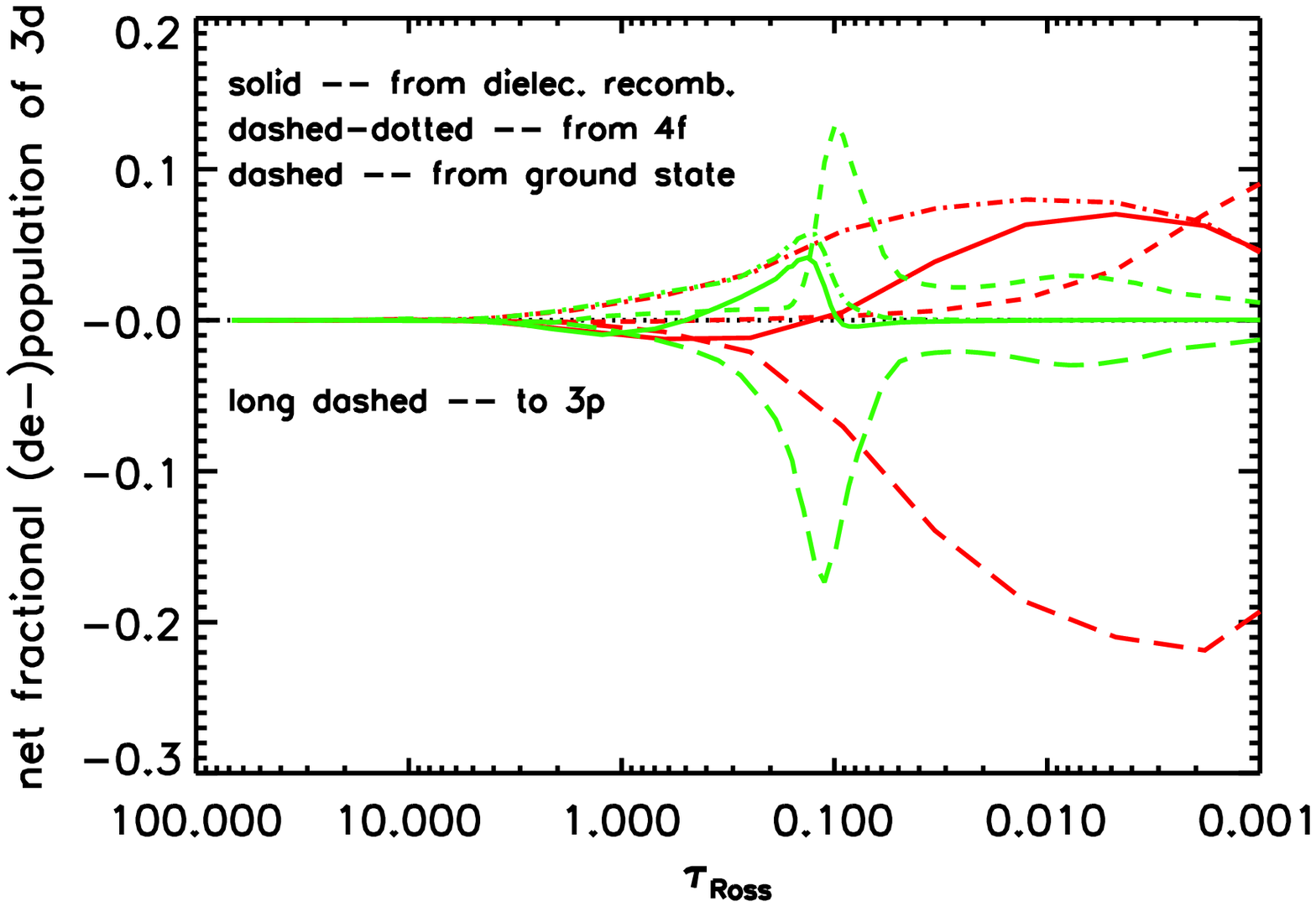}}
\end{minipage}
\caption{Fractional net rates to and from level 3p (top) and level 3d
(bottom), for model `T3735' (full line-blanketing) with no wind (red) and
with wind (green).} 
\label{pop-depop-T3735-10-11-wind}
\end{figure}

To understand the underlying mechanism, we inspect again the
fractional net rates, now for the wind-model `T3735'
(Fig.~\ref{pop-depop-T3735-10-11-wind}). Obviously, the wind induces a
significant overpopulation of the 3d level via the ground state
rather than the dielectronic recombination, just in the way as
indicated above. Due to the velocity field induced Doppler-shifts, the
resonance line(s) become desaturated, the rates are no longer in
detailed balance, and considerable pumping occurs because of the still
quite large radiation field at 374~\AA\ and the significant
ground-state population (larger than in the MH-like models).
Interestingly, this does not only happen in supergiants, as speculated
by MH, but also in dwarfs (at least those with non-negligible
mass-loss rate), because the velocity field sets in just in those
regions where the emission lines are formed.\footnote{An analogous
desaturation through velocity field induced Doppler shifts in stars
with low mass loss rates was found by \cite{najarro96} in the \HeI\
resonance lines of B giants.} 

The impact of the wind on the population of the 3p level is not as
extreme, and the `two electron transition' drain remains as weak
as for the wind-free model. Thus, the presence of emission
relies mostly on the overpopulation of the 3d level.

Again, we test the (remaining) influence of dielectronic recombination by
switching off the stabilizing transition to 3d. Though there is a certain
effect, the change is not extreme. Note also that the net rate from
dielectronic recombination is larger in the wind model than in the wind-free
model, see Fig.~\ref{pop-depop-T3735-10-11-wind} (green solid
line). Nevertheless, we conclude that dielectronic recombination plays, if
at all, only a secondary role in the overpopulation of the 3d level when
consistent atmospheric models are considered. {\it The crucial process is
pumping by the resonance lines.}

The reaction of $\lambda4097$ on velocity field effects is more
complex. On the one side, there is still the cascade from 3p to 3s,
giving rise to a certain correlation. On the other, the resonance line
to 3s (at 452~\AA) becomes efficient now, and can either feed (as
argued in Sect.~\ref{intro-niii}) or drain level 3s, in dependence of
its optical depth. Under the conditions discussed here, the resonance
line is pumping at cooler temperatures, with a zero net effect on the
strength of $\lambda4097$ (since the cascade from 3p becomes somewhat
decreased, compared to wind-free models). At higher temperatures, the
resonance line becomes optically thin,\footnote{Note that the
oscillator strength is more than a factor of 10 lower than that of the
resonance line feeding 3d.} because of a lower ground-state
population, and level 3s can cascade to the ground-state. Thus, the
absorption strength of $\lambda4097$ might become significantly
reduced, which explains, e.g., the strong deviation of model `T4035'
from the MH predictions. For this and the hotter models, $\lambda4097$
is very weak, and even appears in (very weak) emission for model `T5040'.

\section{Comparison with results from {\sc cmfgen}}
\label{comp-cmfgen} 

In this section, we compare the results from our {\sc fastwind} models
with corresponding ones from {\sc cmfgen}, a code that is considered to
produce highly reliable synthetic spectra, due to its approach of
calculating all lines (within its atomic database) in the comoving
frame. 
For this purpose, we used a grid of models for dwarfs and supergiants
in the O and early B-star range, which has already been used in previous
comparisons \citep{lenorzer04, puls05, Repo05}. 
{\sc fastwind} models have been calculated with three
`explicit' atoms, H, He and our new N atom. 
Stellar and wind parameters of the grid models are listed in
Table~\ref{grid-cmfgen}, with wind parameters following roughly the
WLR for Galactic stars. The model designations correspond only
coarsely to spectral types and are, with respect to recent
calibrations \citep{repolust04, martins05a}, somewhat too early. All
calculations have been performed with the `old' solar Nitrogen
abundance, [N] = 7.92 \citep{grevesse98}, and a micro-turbulence
$\vturb = 15\ \kms$. 

\begin{table}
\caption{Stellar and wind parameters of our model grid used to compare
synthetic \NII/\NIII\ profiles from {\sc fastwind} and {\sc cmfgen}. The grid
is a subset of the grid presented by \citet{lenorzer04}.}
\label{grid-cmfgen}
\tabcolsep1.8mm
\begin{center}
\begin{tabular}{ccrcccc}
\hline 
\hline
\multicolumn{7}{c}{Luminosity class V}\\
\hline
\multicolumn{1}{c}{Model}
&\multicolumn{1}{c}{\Teff}
&\multicolumn{1}{c}{\Rstar}
&\multicolumn{1}{c}{\logg}
&\multicolumn{1}{c}{\mdot}
&\multicolumn{1}{c}{\vinf}
&\multicolumn{1}{c}{$\beta$}\\
\multicolumn{1}{c}{}
&\multicolumn{1}{c}{(K)}
&\multicolumn{1}{c}{(\rsun)}
&\multicolumn{1}{c}{(cgs)}
&\multicolumn{1}{c}{(\mdu)}
&\multicolumn{1}{c}{(\kms)}
&\multicolumn{1}{c}{}
\\
\hline
d2v & 46100 & 11.4 & 4.01 & 2.52   & 3140 & 0.8\\
d4v & 41010 & 10.0 & 4.01 & 0.847  & 2850 & 0.8\\
d6v & 35900 & 8.8  & 3.95 & 0.210  & 2570 & 0.8\\
d8v & 32000 & 8.0  & 3.90 & 0.056  & 2400 & 0.8\\
d10v& 28000 & 7.4  & 3.87 & 0.0122 & 2210 & 0.8\\
\hline
\multicolumn{7}{c}{Luminosity class I}\\
\hline
\multicolumn{1}{c}{Model}
&\multicolumn{1}{c}{\Teff}
&\multicolumn{1}{c}{\Rstar}
&\multicolumn{1}{c}{\logg}
&\multicolumn{1}{c}{\mdot}
&\multicolumn{1}{c}{\vinf}
&\multicolumn{1}{c}{$\beta$}\\
\multicolumn{1}{c}{}
&\multicolumn{1}{c}{(K)}
&\multicolumn{1}{c}{(\rsun)}
&\multicolumn{1}{c}{(cgs)}
&\multicolumn{1}{c}{(\mdu)}
&\multicolumn{1}{c}{(\kms)}
&\multicolumn{1}{c}{}
\\
\hline
s2a & 44700  & 19.6 & 3.79 & 12.0  & 2620 & 1.0\\
s4a & 38700  & 21.8 & 3.57 & 7.35  & 2190 & 1.0\\
s6a & 32740  & 24.6 & 3.33 & 3.10  & 1810 & 1.0\\ 
s8a & 29760  & 26.2 & 3.21 & 1.53  & 1690 & 1.0\\
s10a & 23780 & 30.5 & 2.98 & 3.90  & 740  & 1.0\\
\hline
\end{tabular}
\end{center}
\end{table}

For the comparison between {\sc fastwind} and {\sc cmfgen} results,
we consider useful diagnostic \NII\ (to check the cooler models)
and \NIII\ lines in the blue part of the visual spectrum. 
Table~\ref{tab_lines_niii} lists important \NIII\ lines in the range
between 4,000 to 6,500~\AA, together with multiplet numbers according to
\citet{Moore75}, to provide an impression of how many independent lines
are present and which lines belong to the same multiplet.\footnote{
A similar table regarding \NII, \NIV\ and \NV\ lines will be presented
in Paper~II.} Additionally, we provide information about adjacent
lines present in the spectra of (early) B and O stars, where the major
source of contamination arises from \OII\ and \CIII. 

This set of lines can be split into five different groups.
Lines that belong to the first group (\#2,5,9-11 in
Table~\ref{tab_lines_niii}) are produced by cascade processes of the
(doublet) series 1s$^2$ 2s$^2$ $nl$ with $n = 3,4,5$ and $l$ = s, p,
d, f, g, and additional (over-) population of the 3d level. As
outlined in the previous sections, the behavior of these lines is
strongly coupled.

The second group (lines \#6-8), \qua, results from transitions
within the quartet system. 
We consider only the three strongest components of the corresponding
multiplet, that are also the least blended ones.

The third group (lines \#1,3,4) represents three lines, at
$\lambda\lambda$4003, 4195, and 4200 \AA, that are formed
between higher lying levels within the doublet system. These lines are
weaker than the ones from the previous two groups, but still worth to
use them within a comparison of codes and also within a final
abundance analysis. Note that \NIII$\lambda4195$ and
\NIII$\lambda4200$ are located within the Stark-wing and the core of
\HeII$\lambda4200$, respectively, which requires a consistent analysis
of the total line complex.

Lines at $\lambda\lambda$5320, 5327 \AA\ (\#12-13) and
$\lambda\lambda$6445, 6450, 6454, 6467 \AA\ (\#14-17) comprise the
fourth and fifth group, respectively. The former set of lines is
located in a spectral region that is rarely observed, and the latter
comprises a multiplet from the quartet system, in the red part of the
visual spectrum.

Although most of the lines listed in Table~\ref{tab_lines_niii} are
(usually) visible in not too early O-type spectra, their diagnostic
potential for abundance determinations is different. The lines at
\qua\ (from the quartet system) are certainly the best candidates to
infer abundances, since they are quite strong and their formation is
rather simple. Also the \NIII\ triplet itself provides valuable
information. Due to its complex formation -- when in emission -- and
additional problems (see below and Sect~\ref{Ocoup}), these lines
should be used only as secondary diagnostics, whenever possible. The 
remaining lines in the blue part of the visual
spectrum are rather weak and/or strongly blended with adjacent lines
(see Table~\ref{tab_lines_niii}). In particular, the (theoretically)
very interesting transition 3p $\rarrow$ 3s, \NIII$\lambda$4097, is
located within the Stark-wing of \Hd. Thus, these lines should be used
preferentially as a consistency check, and employed as a direct
abundance indicator only at low rotation. Lines in the yellow
part of the optical (group four) have not been considered by us so
far, since our observational material does not cover this spectral
range, and we are not able to judge their diagnostic value for O-type
stars. Finally, lines from group five in the red are usually rather weak,
and might be used only in high S/N spectra of slowly rotating stars.

\begin{table}
\caption{Diagnostic \NIII\ lines in the optical, 
together with adjacent blends.}
\label{tab_lines_niii}
\tabcolsep0.9mm
\begin{tabular}{rcrrl}
\hline 
\hline
\# & Wavelength & M\# & low-up & Blends \\
   & (\AA)  &     &        &  \\
\hline
1  &   4003.58 &17& 20-33 & \OII\ $\lambda$4007.46 \\ 
2  &   4097.33 & 1&  8-10 & \OII\ $\lambda$4097.26, 4098.24, \Hd\ $\lambda$4101.74 \\ 
3  &   4195.76 & 6& 13-22 & \OII\ $\lambda$4192.52, 4196.26, \SiIII$\lambda$4195.59,\\
   &           &  &       &\HeII\ $\lambda$4200.00  \\
4  &   4200.07 & 6& 13-22 & \HeII\ $\lambda$4200.00 \\
5  &   4379.11 &18& 21-34 & \OII\ $\lambda$4378.03, 4378.43, \CIII$\lambda$4379.47,\\
   &           &  &       & \NII\ $\lambda$4379.59 \\ 
6  &   4510.88 & 3& 12-16 & \NIII\ $\lambda$4510.92, \NeII$\lambda$4511.42 \\
7  &   4514.86 & 3& 12-16 & \OIII\ $\lambda$4513.83, \NeII $\lambda$4514.88,\\
   &           &  &       & \CIII$\lambda$4515.81, 4516.77 \\
8  &   4518.14 & 3& 12-16 & \NeII\ $\lambda$4518.14, \OIII$\lambda$4519.62 \\
9  &   4634.14 & 2& 10-11 & \SiIV\ $\lambda$4631.24, \OIV$\lambda$4632 \\
10 &   4640.64 & 2& 10-11 & \OII\ $\lambda$4638.86, \SiIII$\lambda$4638.28 \\
11 &   4641.85 & 2& 10-11 & \OII\ $\lambda$4641.81, 4643.39, \NII$\lambda$4643.08 \\
12 &   5320.82 &21& 22-30 & \OII\ $\lambda$5322.53 \\
13 &   5327.18 &21& 22-30 & -  \\
14 &   6445.34 &14& 19-25 & - \\
15 &   6450.79 &14& 19-25 & \CIV\ $\lambda6449.90$\\
16 &   6454.08 &14& 19-25 & \OII\ $\lambda$6457.05, \NII$\lambda$6457.68\\
17 &   6467.02 &14& 19-25 & - \\
\hline
\end{tabular}
\tablefoot{Line numbers refer to the transitions
indicated in Fig.~\ref{niii-doublet-quartet}, `M\#' are the multiplet
numbers according to \citet{Moore75}, and `low-up' refer to the
corresponding (packed) lower and upper levels as
provided in Table~\ref{atom_lev_niii}.} 
\end{table}

A detailed comparison between the various \NII\ (for the later
subtypes) and \NIII\ lines from {\sc fastwind} and {\sc cmfgen} is
provided in Appendix~\ref{apen_cmfgen}. Overall, the following trends
and problems have been identified.

For models {\tt d10v}, {\tt d8v}, and {\tt s10a}
(Figs.~\ref{d10v_nii}, \ref{d8v_nii}, and \ref{s10a_nii},
respectively), the agreement of the \NII\ lines is almost perfect. 
For model {\tt s8a} (Fig.~\ref{s8a_nii}), on the other hand, big
discrepancies are found. Most of our lines are much stronger than
those from {\sc cmfgen}, because of the following reason. 
Within the line-forming region, the electron density as
calculated by {\sc fastwind} is a factor of $\approx 8$ higher than
the one as calculated by {\sc cmfgen}, thus enforcing higher
recombination rates from \NIII\ to \NII, more \NII\ and thus stronger
lines. This is the only model with such a large discrepancy in the
electron density (e.g., the electron densities from models {\tt s10a}
and {\tt s6a} agree very well), and the reason for this discrepancy
needs to be identified in future work. Nevertheless, this discrepancy
would not lead to erroneous Nitrogen abundances: When analyzing the
observations, a prime diagnostic tool are the wings of the Hydrogen
Balmer lines that react, via Stark-broadening, sensitively on the
electron density. After a fit of these wings has been obtained, one
can be sure that the electron stratification of the model is
reasonable (though the derived gravity might be erroneous then). With
a `correct' electron density, however, the abundance determination via
\NII\ should be no further problem.

Let us concentrate now on the \NIII\ lines, beginning with the triplet
\trip\ and accounting for the fact that at cooler temperatures the
$\lambda$4642 component is dominated by an overlapping \OII\ line (see
Table~\ref{tab_lines_niii}).
Except for the dwarf model {\tt d6v} and the supergiant models {\tt
s8a} and {\tt s6a}, both codes predict very similar absorption (for
the cooler models) and emission lines, thus underpinning the results
from our various tests performed in Sect.~\ref{test-dr}.
Interestingly, our emission lines for the hottest dwarfs ({\tt d4v}
and {\tt d2v}, Figs.~\ref{d4v} and \ref{d2v}) are slightly stronger than
those predicted by {\sc cmfgen}, a fact that is not too worrisome
accounting for the subtleties involved in the formation process.

What {\it is} worrisome, however, is the deviation for models {\tt d6v},
{\tt s8a} and {\tt s6a} (Figs.~\ref{d6v}, \ref{s8a} and \ref{s6a}). 
Whereas {\sc fastwind} predicts only slightly refilled absorption
profiles for all three models, {\sc cmfgen} predicts almost completely
refilled (i.e., EW $\approx$ 0) profiles for {\tt d6v} and {\tt s8a},
and well developed emission for {\tt s6a}. 

We investigated a number of possible reasons for this discrepancy. At
first, the difference in electron density for model {\tt s8a} (see
above) could not explain the strong deviations. At least for our {\sc
fastwind} models, dielectronic recombination does not play any role,
and we checked that this is also the case for the {\sc cmfgen} models,
by switching off DR. As it finally turned out, the physical driver
that overpopulates the 3d level in {\sc cmfgen} in the considered
parameter range is a coupling with the \OIII\ resonance line at
374~\AA, which is discussed in the next section.

With respect to \NIII$\lambda4097$, we find a certain trend in the
deviation between {\sc fastwind} and {\sc cmfgen} absorption profiles.
Both for dwarfs and supergiants, our line is weaker at cooler
temperatures, similar at intermediate ones and stronger at hotter
temperatures. 

For \NIII$\lambda4379$, the {\sc cmfgen} models predict more absorption for
models with $\Teff \leq 35,000$~K, because of an overlapping \OII\ and 
\CIII\ line. For hotter models, the presence of \OII\ is no longer
important, but there is a still a difference for $\lambda4379$, going into
emission at the hottest temperatures. 

For the quartet multiplet at \qua, we find a similar trend as for 
$\lambda$4097. For dwarfs, the lines are weaker at cooler
temperatures, similar at {\tt d6v} and stronger in absorption at {\tt
d4v}. Likewise then, the emission at {\tt d2v} is weaker than in {\sc
cmfgen}. For supergiants, the situation is analogous, at least for all
models but the hottest one
where both codes predict identical emission.

Finally, lines $\lambda4003$ and $\lambda4195$ (note the blueward
\SiIII\ blend at hotter temperatures) show quite a good agreement for
both dwarfs and supergiants, with somewhat larger discrepancies only
for models {\tt d4v} and {\tt s4a}. 


In summary, the overall agreement between \NII\ and \NIII\ profiles as
predicted by {\sc fastwind} and {\sc cmfgen} is satisfactory, and for most
lines and models the differences are not cumbersome. Because of the involved
systematics, however, abundance analyses might become slightly biased as a
function of temperature, if performed either via {\sc fastwind} or via {\sc
cmfgen}. Additionally, we have identified also some strong deviations,
namely with respect to \NII\ line-strengths for supergiants around \Teff
$\approx$ 30~kK,\footnote{rooted within a different stratification of
electron density.} and regarding the emission strengths of the \NIII\
triplet, for models {\tt d6v}, {\tt s8a}, and {\tt s6a}. 

\section{Coupling with \OIII} 
\label{Ocoup} 
After numerous tests we were finally able to identify the origin of
the latter discrepancy. It is the overlap of two resonance lines from
\OIII\ and \NIII\ around 374 \AA\ (for details, see
Table~\ref{tab_ocoup}) that is responsible for the stronger emission
in {\sc cmfgen} models compared to the {\sc fastwind} results. Whereas
this process is accounted for in {\sc cmfgen}, our present {\sc
fastwind} models cannot do so, because Oxygen is treated only as a
background element, and no exact line transfer (including the overlap)
is performed. We note that this is not exactly the \cite{bowen35}
mechanism as mentioned in Sect.~\ref{intro-niii}, since this mechanism
involves also the overlap between another \OIII\ resonance line and
the \HeII\ Ly-$\alpha$ line at 303~\AA, which does not play a strong
role in our {\sc cmfgen} models as we have convinced
ourselves.\footnote{Insofar, the \OIII\ lines at $\lambda\lambda$3340,
3444, 3759 mentioned in Sect.~\ref{intro-niii} remain in absorption.}
Here, only the coupling between \NIII\ and \OIII\ at 374~\AA\ is
decisive. 

\begin{table}
\caption{Overlapping \NIII\ and \OIII\ resonance lines around 374 \AA.}
\label{tab_ocoup}
\tabcolsep1.mm
\begin{center}
\begin{tabular}{l|clclccc} 
\hline
\hline
\multicolumn{1}{l}{transition(s)}
&\multicolumn{1}{c}{$\lambda$}
&\multicolumn{1}{l}{low}
&\multicolumn{1}{c}{$g_{\rm l}$}
&\multicolumn{1}{l}{up}
&\multicolumn{1}{c}{$g_{\rm u}$}
&\multicolumn{1}{c}{$gf$}
&\multicolumn{1}{c}{ratio}\\
\multicolumn{1}{l}{}
&\multicolumn{1}{c}{(\AA)}
&\multicolumn{1}{l}{}
&\multicolumn{1}{c}{}
&\multicolumn{1}{l}{}
&\multicolumn{1}{c}{}
&\multicolumn{1}{c}{}
&\multicolumn{1}{c}{}\\
\hline
\NIII\ (packed)      &         &2p $^2$P$^0$        & 6 & 3d $^2$D       & 10  & 2.5 & \\  
\NIII\ coupl. comp.  & 374.434 & 2p $^2$P$^0_{3/2}$ & 4 & 3d $^2$D$_{5/2}$ & 6  & 1.5 & 0.60\\  
\hline
\OIII\ (packed)      &         & 2p$^2$ $^3$P     & 9  & 3s $^3$P$^0$   & 9 & 0.72 &\\  
\OIII\ coupl. comp.  & 374.432 & 2p$^2$ $^3$P$_2$  & 5  & 3s $^3$P$^0_1$ & 3 & 0.06 & 0.14 \\  
\hline
\end{tabular}
\end{center}
\tablefoot{
Packed levels/transitions and overlapping (`coupled') ones. `low' = lower
term/level, `up' = upper term/level, $g_{\rm l}$ and $g_{\rm u}$ corresponding
statistical weights. `Ratio' indicates the opacity ratio between overlapping
and packed lines. Individual components at 374.20{\ldots}374.44 \AA\
for \NIII\ (3 components) and at 373.80{\ldots}374.43 \AA\ for \OIII\ (6
components). Note that 0.01\AA\ correspond to 8\,\kms. Data from NIST database.
}
\end{table}

\begin{figure}
\resizebox{\hsize}{!}
  {\includegraphics[angle=90]{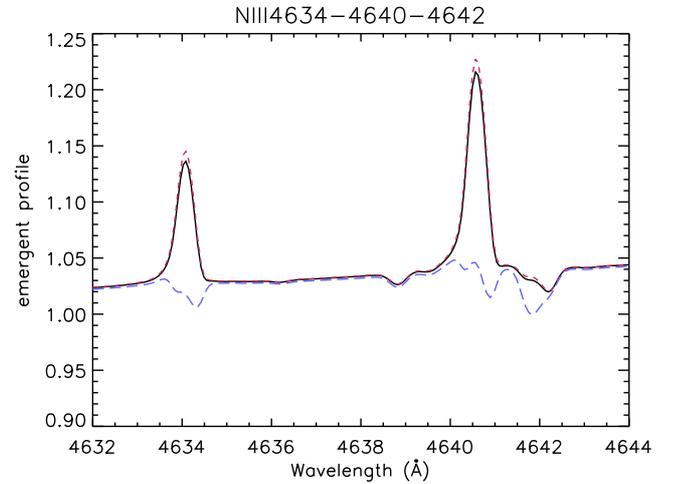}}
\caption{\NIII\ \trip\ for three versions of model {\tt s6a}, as
calculated by {\sc cmfgen}. Solid black: standard version; Dotted/red: 
photo cross-sections for level 3d without resonances (i.e., no DR);
dashed/blue: \NIII\ resonance line at 374 \AA\ forbidden to be affected by
\OIII. For the latter model, the optical triplet lines remain in absorption!}
\label{ocoup_emlines}
\end{figure}

In the following we discuss some details of the process, by means of
the {\sc cmfgen} model {\sc s6a} that displays the largest difference to
the {\sc fastwind} predictions. For a further investigation, we `decoupled'
\NIII\ from \OIII\ by setting the corresponding oscillator strength of the
transition \OIII\ 2p$^2$ $^3$P$_2$ $\rightarrow$ 3s $^3$P$^0_1$ to a very
low value, and compared the results with those from the
`coupled' standard model. Figure~\ref{ocoup_emlines} shows that the well
developed emission lines from the standard models switch into absorption,
and become very similar to the corresponding {\sc fastwind} profiles
(Fig.~\ref{s6a}). On the other hand, discarding DR 
by neglecting the resonances in the photo cross-sections from level 3d had
almost no effect on the profiles, in accordance with our previous arguments.
(Actually, without resonances there is even more emission than before, which
shows that for this model the ionization via doubly excited levels outweighs
DR.)

\begin{figure}
\resizebox{\hsize}{!}
  {\includegraphics[angle=90]{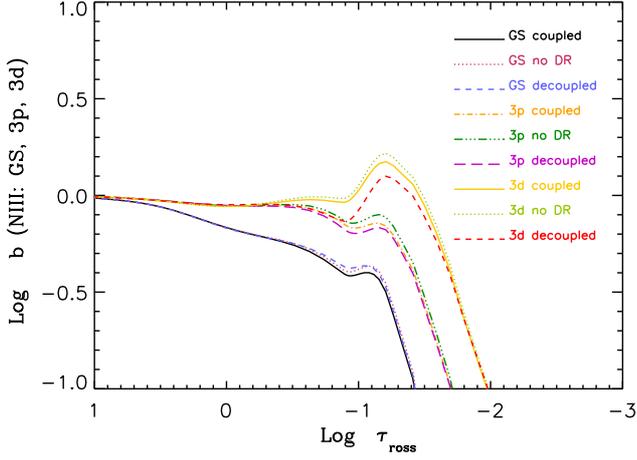}}
\caption{NLTE departure coefficients for three versions of 
model {\tt s6a}, as calculated by {\sc cmfgen}. Lower group of curves: \NIII\
ground state; intermediate group: level 3p, upper group: level 3d.
See text.}
\label{ocoup_departures}
\end{figure}

Figure~\ref{ocoup_departures} displays the corresponding NLTE departure 
coefficients for the \NIII\ ground state, level 3p and level 3d.
Because of the superlevel approach, there is no
difference\footnote{assumed to be balanced by collisions.} between
level 3d $^2$D$_{5/2}$ (coupled to \OIII) and 3d $^2$D$_{3/2}$ (not
coupled). Whilst the ground state and the 3p level remain similar
for all three models (standard, `decoupled' and `coupled' with
resonance-free photo cross-section for 3d), level 3d becomes much
stronger overpopulated in the two `coupled' models, compared to the
`decoupled' one.

\begin{figure}
\resizebox{\hsize}{!}
  {\includegraphics{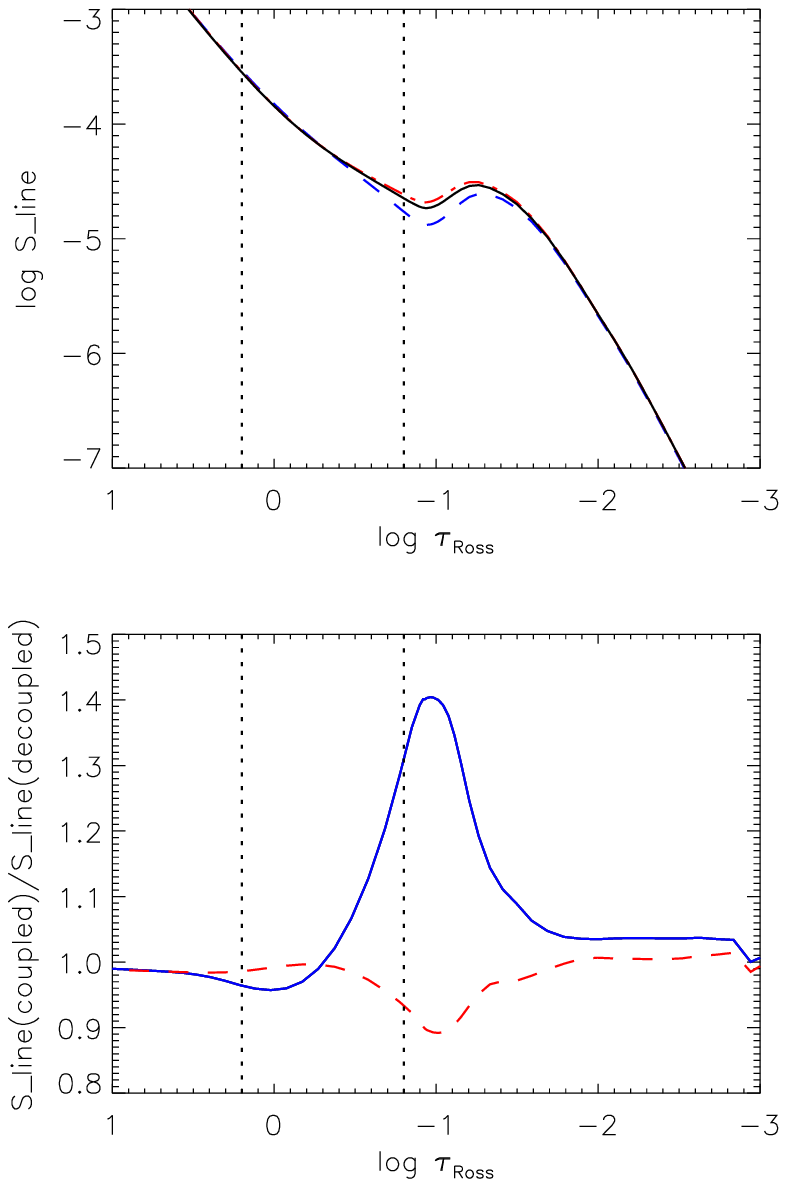}}
\caption{Line source functions for the \NIII/\OIII\ resonance lines at
374~\AA, for two versions of model {\tt s6a}, as calculated by {\sc cmfgen}.
{\it Upper panel:} \NIII\ `decoupled' (dashed/blue); \OIII\ `decoupled'
(dashed-dotted/red); \NIII\ and \OIII\ `coupled' (solid, identical).
{\it Lower panel:} Source function ratios. Solid: 
\NIII\ (`coupled')/\NIII\ (`decoupled'); dashed:
\OIII\ (`coupled')/\OIII\ (`decoupled'). The optical triplet lines 
are formed in the range indicated by the dotted lines.}
\label{ocoup_sline}
\end{figure}

The origin of this stronger overpopulation becomes obvious from
Fig.~\ref{ocoup_sline}. The upper panel displays the source functions of the
\NIII/\OIII\ resonance lines for the `decoupled' and `coupled' case (again,
no difference for individual components, due to the superlevel approach).
For the `decoupled' case, the \NIII\ source function is significantly lower
than the \OIII\ one, whereas for the `coupled' standard model both source
functions are identical, at a level very close to the `decoupled' \OIII\
case. These changes are visualized in the lower panel: because of the line
overlap, the \NIII\ source function increases by a factor up to 1.35 (in the
line forming region of the optical triplet lines), whereas \OIII\ decreases
only marginally.

The source function equality itself results from the strong coupling of the
two resonance lines and the fact that both of them are optically thick from
the lower wind on, with similar opacities (proportional to the product of
cross-section, ionization fraction and abundance).\footnote{In the formation
region of the optical lines, the opacity ratio between the \NIII\ and
\OIII\ resonance lines ranges from roughly 5 for the coolest models to 0.5
for the hottest ones.} It can be shown that the magnitude of the
coupled source function mostly depends on those processes that determine
the individual source functions in addition to the mean line intensity. 
This can be easily seen by using the Sobolev approximation,
\beq
\Jbar = (1-\beta) S + \beta_\mathrm{c} I_\mathrm{c},
\label{jbarsob}
\eeq
with scattering integral $\Jbar$, local and core escape probabilities
$\beta$ and $\beta_{\rm c}$, and core intensity $I_{\rm c}$
\citep{Sobo60}. In case of overlapping lines, the source function $S$
and the opacities need to account for all components with appropriate
weights.

We now assume that the individual source functions (for the packed levels) 
can be approximated by  
\beq
S_\mathrm{N} \approx \Jbar + \delta_\mathrm{N}, \qquad S_\mathrm{O} \approx
\Jbar + \delta_\mathrm{O},
\label{slsob}
\eeq
where $\delta_{\rm N}$ and $\delta_{\rm O}$ correspond to the additional source
terms (mostly from cascades to the upper levels), and the well-known
factor\footnote{corresponding to (1-$\varepsilon$) in a two-level
atom, with $\varepsilon$ roughly the ratio of collisional to
spontaneous radiative rate coefficient, for the downward transition.}
in front of $\Jbar$~ has been approximated by unity.

$\Jbar$ includes the
contributions from all transitions between the fine structure components, 
$\Jbar$ = $\sum gf_i \Jbar_i/\sum gf_i$, exploiting the fact that the reduced
occupation numbers for the fine structure levels are equal, $n_i/g_i$ =
$n_j/g_j$.

Let us first consider the case of no fine structure splitting, i.e., the
overlapping resonance lines should be the only ones connecting the upper and
lower levels. Without line overlap, we obtain the well-known result
(solving for Eqs.~\ref{jbarsob} and \ref{slsob} in parallel)
\beq
S_\mathrm{N}^{\rm decoup} = \frac{\beta_{\rm c,N} I_\mathrm{c} +
\delta_\mathrm{N}}{\beta_\mathrm{N}}, \qquad 
S_\mathrm{O}^{\rm decoup} = \frac{\beta_{\rm c,O} I_\mathrm{c} +
\delta_\mathrm{O}}{\beta_\mathrm{O}}.
\eeq
Likewise, but accounting for the line overlap and assuming $\beta~\ll~1$ for
both components, we find
\beqa 
S_\mathrm{N}^{\rm coup} & \approx & S_\mathrm{O}^{\rm coup} \approx
\frac{\beta_\mathrm{c} I_\mathrm{c}}{\beta} +
\frac{\delta_\mathrm{N}}{\beta_\mathrm{N}}
+\frac{\delta_\mathrm{O}}{\beta_\mathrm{O}} \approx  \nonumber \\ 
& \approx & S_\mathrm{N}^{\rm decoup} +
\frac{\delta_\mathrm{O}}{\beta_\mathrm{O}}
 \approx    S_\mathrm{O}^{\rm decoup} +
\frac{\delta_\mathrm{N}}{\beta_\mathrm{N}},
\eeqa
Thus, the coupled source functions are larger than in the decoupled case,
but (almost) independent on the ratio of their contribution. (In the above
equation, the first term of the sum does not depend on any specific opacity
as long as the lines -- coupled or decoupled -- remain optically thick). 

For the case of two multiplet lines from both \NIII\ and \OIII, with one
of each overlapping, the corresponding result for the packed levels reads
\beq
S_\mathrm{N}^{\rm coup} \approx  S_\mathrm{O}^{\rm coup} \approx
\frac{\beta_\mathrm{c} I_\mathrm{c}}{\beta} + 
\frac{1}{3}\bigl(\frac{\delta_\mathrm{N}}{\beta_\mathrm{N}}
+\frac{\delta_\mathrm{O}}{\beta_\mathrm{O}}\bigr),
\label{S_twocomp_coup}
\eeq
where the escape probabilities $\beta_{\rm N}$ and $\beta_{\rm O}$
refer to the total opacity arising from both multiplet lines, and the
factor `1/3' can be traced down to the fact that three (optically thick)
lines participate in total, two single lines from \NIII\ and \OIII, and
one coupled line complex. 

The corresponding source functions for the decoupled case with packed levels
would read
\beq
S_\mathrm{N}^{\rm decoup} = \frac{\beta_\mathrm{c} I_\mathrm{c}}{\beta} +
\frac{1}{2}
\frac{\delta_\mathrm{N}}{\beta_\mathrm{N}}, \qquad
S_\mathrm{O}^{\rm decoup} = \frac{\beta_\mathrm{c} I_\mathrm{c}}{\beta} +
\frac{1}{2}
\frac{\delta_\mathrm{O}}{\beta_\mathrm{O}}, \qquad
\label{S_twocomp_decoup}
\eeq 
(the factor of two arising from two participating multiplet lines), and a
comparison with Eq.~\ref{S_twocomp_coup} shows that in most cases the
coupled source function would lie in between the corresponding ones that
neglect the line overlap. Again, our result is (almost) independent of the
opacity ratio between the overlapping \NIII\ and \OIII\ lines but also
independent of the weights of the individual lines within each
multiplet.\footnote{as long as $\beta \ll 1$.} 

Generalization to more multiplet lines is straightforward, and our analytic
result compares well to the actual case where the \NIII\ and \OIII\ source
functions for the coupled case are identical and lie in between the values
for the decoupled situation, see Fig.~\ref{ocoup_sline}, upper panel.

\begin{figure*}
\resizebox{\hsize}{!} {\includegraphics{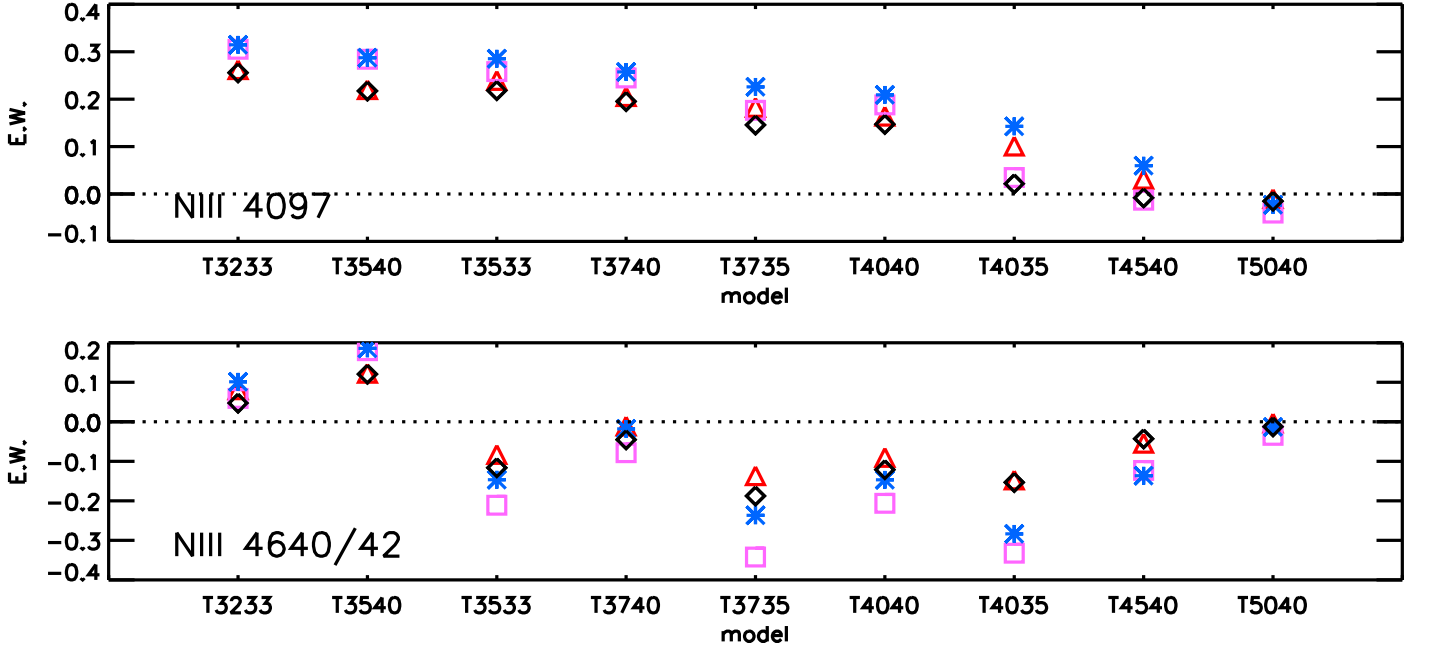}}
\caption{ Comparison of equivalent widths (EW) for \NIII\ $\lambda4097$ and
$\lambda\lambda4640/42$ for models with different Nitrogen abundances and
wind-strengths. Black diamonds: [N] = 7.78, mass-loss rates as in
Table~\ref{models-mih}. Red triangles: as black diamonds, but with half the
mass-loss rate. Purple squares: [N] = 8.18, mass-loss rates as in
Table~\ref{models-mih}. Blue asterisks: as purple squares, but with half the
mass-loss rate. All background abundances are solar \citep{asplund05}.} 
\label{grid-abundmdot}
\end{figure*}

Thus, whenever the source function of the \OIII$\lambda$374 line is
significantly larger than the one from \NIII, a decent effect on the
emission strength of the optical triplet lines is to be expected. This
situation is particularly met around \Teff\ = 30 to 33~kK, since in this
region the upper level of the \OIII\ line is significantly populated by
cascades from higher levels. 

First test calculations with a simplified Oxygen model atom performed
by {\sc fastwind} confirm the general effect, but realistic results
cannot be provided before a detailed model atom has been constructed.
Let us note, however, that most of the other optical \NIII\ lines are
barely affected by the resonance line coupling, and that these lines
can be used for diagnostic purposes already now.

In Paper~II we derive Nitrogen abundances for LMC O-stars
from the VLT-FLAMES survey of massive stars. Though there are only few
objects in the critical temperature range, at least for one object,
N11-029 (O9.7Ib), we encountered the problem that the observed,
refilled \NIII\ triplet (EW $\approx$ 0) could not be reproduced by
{\sc fastwind}, though with {\sc cmfgen}. We interpret this problem as
due to the resonance line overlap, but stress also the fact that other
diagnostic lines enable a satisfactory abundance analysis.

\section{Influence of various parameters}
\label{varpar}
Let us finally investigate the influence of important parameters on the
strength of the optical emission lines. We stress that the following results 
have an only qualitative, differential character, as long as the coupling with
\OIII\ has not been accounted for, at least in the range \Teff\ $\la$~35kK.

\paragraph{Nitrogen abundance.} Figure~\ref{grid-abundmdot} displays
the reaction on a variation of Nitrogen abundance and mass-loss rate.
All models have been calculated with background abundances following
\citet{asplund05}. The (purple) squares correspond (almost, except for
the background) to our previous results for [N] = 8.18 (0.4 dex larger
than the solar value), and mass-loss rates according to
Table~\ref{models-mih}. Reducing the Nitrogen abundance to solar
values, [N]~=~7.78, results in considerably less emission (black
diamonds), which is a consequence of the fact that the relative
overpopulation of the 3d level decreases significantly when the
formation depth proceeds inwards.  The influence on
\NIII$\lambda$4097, on the other hand, is less extreme, and follows
the standard trend that a higher abundance results in stronger
absorption lines. 
\paragraph{Mass-loss rates.} Also in Fig.~\ref{grid-abundmdot}, asterisks (for
[N] = 8.18) and triangles (for [N] = 7.78) correspond to a situation where the
mass-loss rates have been decreased, by a factor of two compared to
Table~\ref{models-mih}. Whereas the effect for the enriched models is
significant, the models with solar nitrogen abundance are much less affected, 
though the predicted emission is still much stronger than for models without
a wind at all, cf. Fig.~\ref{grid-wind} (triangles). 

\begin{figure*}
\resizebox{\hsize}{!}
  {\includegraphics{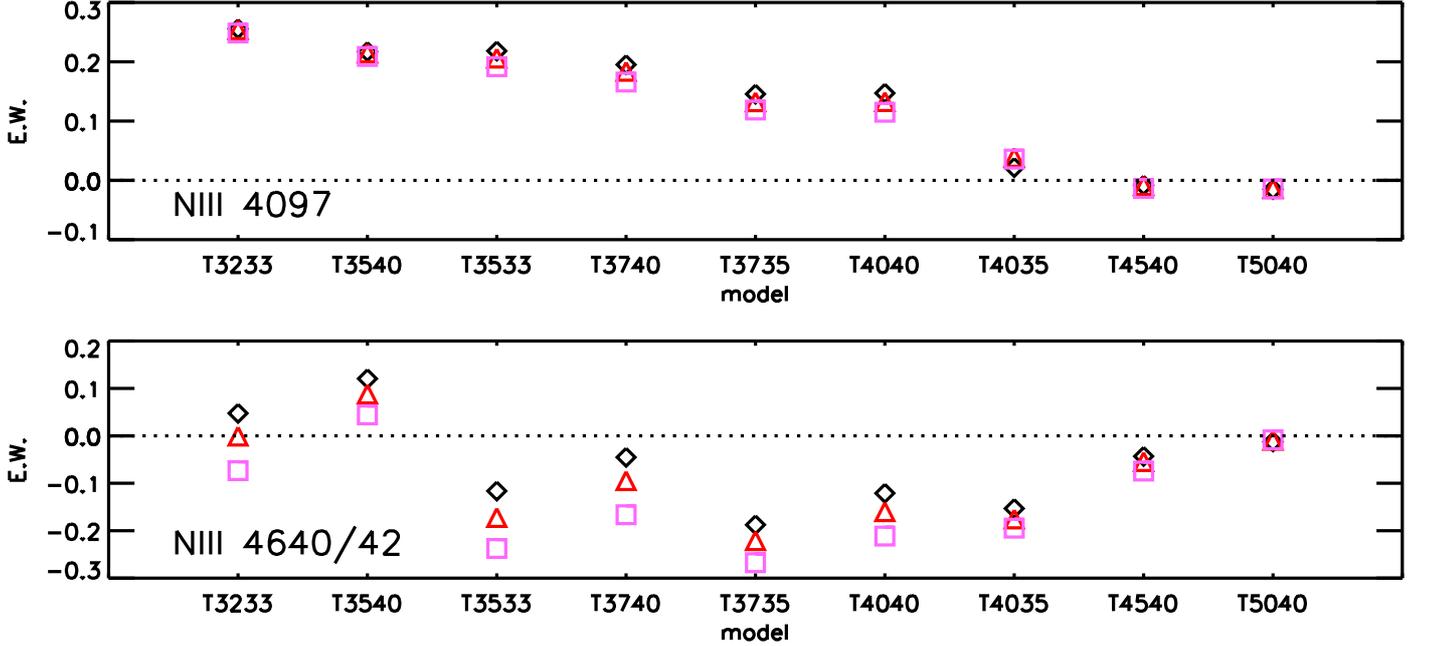}}
\caption{
Comparison of equivalent widths (EW) for \NIII\ $\lambda4097$ 
and $\lambda\lambda4640/42$ for models with different background
abundances, [N] = 7.78 and mass-loss rates as in Table~\ref{models-mih}. 
Black diamonds: $Z/Z_\odot = 1.0$; red triangles: $Z/Z_\odot = 0.5$; purple
squares: $Z/Z_\odot = 0.2$. Background abundances scaled to solar
abundance pattern from \citet{asplund05}.} 
\label{grid-bgabund}
\end{figure*}

The origin of this different behaviour is, again, rooted in the
specific run of the relative overpopulation of level 3d. For a large
Nitrogen abundance, the formation takes place in a region where the
overpopulation increases quickly with increasing mass-loss rate,
whereas for a lower abundance (deeper formation) it remains rather
constant over a certain range of formation depths, such that a
moderate change in \mdot\ has much less effect.
Figures~\ref{ocoup_departures} and \ref{ocoup_sline} (upper panel),
with a formation region corresponding to [N] = 7.92, visualize the
general situation (irrespective of whether coupling with \OIII\ is
important or not): an increase of the Nitrogen abundance shifts the
formation depth to the right (towards lower \taur), where the
departure coefficient of level 3d and the source function display a
`bump', due to the large velocity gradient, inducing very strong
pumping by the resonance line(s). A decrease of the wind-strength, on
the other hand, shifts the position of the bump to the right, while
(almost) preserving the formation depth with respect to $\taur$. Thus,
in case of a large Nitrogen abundance the strong overpopulation is
lost when \mdot\ decreases, leading to a corresponding loss of
emission strength.  For a lower abundance, however, the formation
takes place in a region where the overpopulation displays a `plateau',
and a reduction of \mdot, i.e., a shift of the bump to the right, has
a much weaker effect. 

\paragraph{Wind clumping.} Since the \NIII\ emission lines of O-stars are
formed in the middle or outer photosphere,\footnote{only for a very large
Nitrogen abundance or mass-loss rate, (`slash stars' or Wolf-Rayets) the
formation takes place in the wind.} wind clumping has no direct
effect on their strength, as long as the photosphere remains unclumped
(which is most likely the case, e.g., \citealt{Puls06, Sundqvist11,
Najarro11}). Only if the transition region wind/photosphere were clumped,
such direct effects are to be expected, due to increased recombination and
optically thick clumps affecting the resonance lines \citep{Sundqvist11}.
Indirect effects, however, can be large, since clumpy winds have a lower
mass-loss rate than corresponding smooth ones. Since typical \mdot\
reductions are expected to be on the order of 2{\ldots}3 when
accounting for micro- and macro-clumping (\citealt{Sundqvist11} and
references therein), Figure~\ref{grid-abundmdot} (asterisks vs. squares and
triangles vs. diamonds) gives also an impression on the expected effects
when clumping were in- or excluded in the spectrum synthesis.

\paragraph{Background abundances.} Figure~\ref{grid-bgabund} displays
the effects when only the background abundances are varied, while, 
inconsistently, the Nitrogen abundance is kept at its solar value and
the wind-strengths correspond to the Galactic WLR. Diamonds refer to
solar composition, triangles to $Z/Z_\odot$ = 0.5 (roughly LMC) and
squares to $Z/Z_\odot$ = 0.2 (roughly SMC). In line with our reasoning
from Sect.~\ref{full-blank}, the emission increases significantly when
the background abundances decrease, due to reduced line-blocking. Note
that for objects with unprocessed Nitrogen this effect might be
compensated by a lower Nitrogen abundance, and, even more, because of
lower wind-strengths (\mdot\ $\propto (Z/Z_\odot)^{0.7}$,
\citealt{mokiem07b}). Note also that the correlation between triplet
emission and $\lambda$4097 absorption strength has become rather weak,
due to the influence of the resonance line connected to level 3s (see
Sect.~\ref{wind-effects}).  Let us finally stress that because of the
strong dependence of emission line strength on line-blocking, a
sensible assumption on the atmospheric iron abundance is required, due
to its dominating effect on the EUV fluxes. 

\section{Summary and conclusions}
\label{conclusions}
In this paper we described the implementation of a Nitrogen~III
model ion into the NLTE atmosphere/spectrum synthesis code {\sc
fastwind}, to allow for future analyses of Nitrogen abundances in
O-stars.\footnote{Corresponding \NII/\NIV/\NV\ model ions will be
presented in Paper~II.} In particular, we concentrated on a
re-investigation of the \NIII\ \trip\ emission line formation.
Previous, seminal work by MH has suggested that the formation
mechanism of these emission lines in O((f)) and O(f) stars is
dominated by two processes: the overpopulation of the 3d level via
dielectronic recombination, and the strong drain of the 3p level
(towards 2p$^2$) by means of the `two electron transitions'. 

To account for the dielectronic recombination process in our new
\NIII\ model, 
we tested two different implementations. Within the implicit method,
the contribution of dielectronic recombination is included in the
photoionization cross-sections (provided from OPACITY project data),
whereas the explicit method uses resonance-free photoionization
cross-sections and directly accounts for the different stabilizing
transitions from autoionizing levels. Both methods have been
implemented into {\sc fastwind}, and the equivalence of both methods
has been shown.

First tests were able to reproduce the results by MH quite
well, after adapting our new atomic data to values as used by them. This
result was achieved by applying similar assumptions, i.e., wind-free
atmospheres and the same approximate treatment of line-blanketing.
The triplet emission becomes even increased when using our new
atomic data, mostly because of higher oscillator strengths for the `two
electron transitions'.

After switching to fully line-blanketed models (still wind-free),
however, the situation changes dramatically. Then, the strong emission
present in the `pseudo line-blanketed' models is lost (consistent with
results from the {\sc tlusty} OSTAR2002 grid).  We have investigated
this mechanism in detail, and shown that it is rooted in the overall
lower ionizing EUV-fluxes (not only until the \NIII\ edge as assumed
by MH), either directly or indirectly, and the strong collisional
coupling of the ground state and the 2p$^2$ levels. In any case, these
levels become much less underpopulated in fully line-blanketed models,
thus preventing an effective drain of the 3p level. We also
questioned the role of dielectronic recombination regarding the
overpopulation of the 3d level. After setting the DR contribution to
the 3d level to zero, almost no reaction on the triplet lines was
found throughout the whole model grid. Both results lead to the
conclusion that, under realistic conditions of
line-blocking/-blanketing with solar background abundances, both
the drain and the dielectronic recombination lose their key role as
assigned to them by MH. 

This key role is now played by the stellar wind. Already MH suggested
that the emission lines of Of stars (contrasted to O((f)) and O(f)
stars) might be formed due to the \citet{swings48} mechanism, an
overpopulation of the 3d level by pumping through the corresponding
resonance line in an ``extended atmosphere''.
This suggestion could not be proven though, since MH had not the tools to
model such atmospheres. Nowadays, this is no longer a
problem, and when we include the wind in our calculations (with
mass-loss rates following the `unclumped' Galactic WLR), it turns out
that we obtain almost the identical emission as present in the
original MH simulations (performed without wind and without realistic
line-blanketing). By inspection of the net rates into and from the
3d level, we noticed how the wind induces the overpopulation of the
3d level via pumping from the ground state rather than by
dielectronic recombination. A prerequisite of this process is that the
wind-strength is large enough to display a significantly accelerating
velocity field already in the photospheric formation region of the
triplet lines, to allow the resonance lines to leave detailed balance. 

The most important implication of our study is that under Galactic
conditions DR plays only a secondary role, both for O-stars with
``compact'' and with ``extended'' atmospheres, whereas the key process
is the overpopulation due to the resonance line {\it in the presence
of an accelerating velocity field}. Note that without such a velocity
field the resonance line loses its impact though. Consequently, it is
to be expected that hydrostatic NLTE codes, such as {\sc tlusty} and
{\sc detail/surface}, will not be suited to quantitatively 
synthesize the \NIII\ triplet lines (and, to a lesser extent, also
\NIII$\lambda$4097, because of their interrelation), unless the
wind-strength is significantly below the Galactic WLR or the
background metallicity, $Z$, is much lower than the Galactic one 
(see \citealt{Heap06} who performed Nitrogen diagnostics for SMC O-stars
by means of {\sc tlusty}).
Note also that our results have been derived for O-star
conditions only, and should be valid as long as the \NIII\ emission
lines, to a major extent, are formed in the photosphere or in the
transition region. For objects with significantly denser winds, e.g.,
WN-stars, our analysis would certainly need to be extended, since
additional effects might be present or might even dominate.

In order to check our new model atom and the predictive power of {\sc
fastwind}, we performed first comparisons with results from the
alternative model atmosphere code {\sc cmfgen}, for a small grid of
O-type dwarfs and supergiants. For this objective we used a set of
important \NII\ and \NIII\ lines in the blue part of the visual
spectrum. The overall agreement between both codes is mostly
satisfactory, though some systematic deviations demand a further
clarification in terms of a comparison with observations. 

Within the range 30,000~K $\leq \Teff\ \leq$ 35,000~K, however, some
major discrepancies have been found. Here, {\sc cmfgen} triggers the
emission at \trip\ earlier, i.e., at cooler temperatures, than
calculations from {\sc fastwind}. This effect could be traced down to
the overlap of the \NIII\ resonance line (actually, one of its
fine-structure components, at 374.43 \AA) with a resonance line from
\OIII, which is of similar strength throughout the O-star domain. We
studied this effect by means of the corresponding {\sc cmfgen}
calculations (since {\sc fastwind} lacks a detailed \OIII\ model), and
also by analytic considerations.

As long as both resonance lines are optically thick, source function
equality is achieved. Under most conditions, the `coupled' source
function lies in between the individual, `decoupled' ones from \NIII\
and \OIII. In the critical temperature range now, the decoupled \OIII\
source function is predicted to be much larger than the \NIII\ one
(due to substantial cascades from upper levels), and leads, after
being coupled with \NIII, to significant values for the combined
source function, beyond the decoupled \NIII\ case. Thus, the important
3d level becomes more pumped, and the triplet emission occurs at
cooler temperatures than without coupling. When the coupling is
neglected in {\sc cmfgen}, the predicted triplet emission vanishes and
the profiles become similar to those from {\sc fastwind}. 

It might be suspected that the impact of the \OIII\ resonance line
overlap introduces an additional parameter to be known when the
Nitrogen abundance shall be determined via the triplet lines, namely
the Oxygen abundance. As we have shown by our analytic considerations,
however, this parameter remains rather unimportant\footnote{except for
certain differences in the EUV-fluxes etc.} as long as both resonance
lines are optically thick, which is true in the interesting parameter
range. We have tested this prediction by lowering the \OIII\
oscillator strength in {\sc cmfgen} by a large factor (50), and found
no difference in the coupled source function and triplet emission.

Summarizing, not only the \NIII\ resonance line itself is responsible
for the (strong) triplet emission, but also the overlapping \OIII\
resonance line, at least at later O-types. Indeed, for one
corresponding object studied in Paper~II, N11-029 (O9.7Ib), we 
encountered the problem that the observed \NIII\ triplet
could not be reproduced by {\sc fastwind}, though with {\sc cmfgen},
and we interpreted this problem as due to the resonance line overlap.
Insofar, we need to incorporate this process into {\sc fastwind} if we
aim at deriving Nitrogen abundances at cooler temperatures 
from the triplet lines alone.
Fortunately, the contamination of the other optical lines by this
process remains weak, so that these lines can be used already now, and
the Nitrogen abundance determination by {\sc fastwind} is not hindered.

\smallskip
\noindent
Our study implies two important consequences that need to be tested in future 
comparisons with observations.
\begin{itemize}
\item[(i)] Since the efficiency of DR and `two electron' drain is
strongly dependent on the degree of line-blanketing/-blocking, we predict 
that in a metal-poor environment (e.g., in the Small Magellanic Cloud with
$Z/Z_\odot \approx 0.2$) the emission becomes stronger again, due to
less EUV line-blocking. On the other hand, in such a low-$Z$
environment also the wind-strengths and the base-line Nitrogen abundance
become lower, and the combined effects need to be investigated in detail. 

\item[(ii)] As outlined above, the triplet lines from O-type stars are of photospheric
origin and depend, via the \NIII\ (and \OIII) resonance lines, on the
actual wind-strength (independent of clumping and X-ray
properties), determining the onset of the accelerating velocity
field. Thus, their emission strengths might be used to constrain the
stellar mass-loss rate (after the line-formation has been shown to
work reliably), if the Nitrogen abundance can be derived independently
from other lines. In particular, `weak-winded' stars (e.g.,
\citealt{bouret03, martins05b, puls08b, Marcolino09, Najarro11}) have
a rather weak wind for their luminosity, being a factor of 10 to 100
thinner than predicted/observed for their counterparts with `normal'
winds. Thus, it might be
suspected that weak-winded stars display much less emission than stars
with `normal' winds of similar type. This requires, that a) there are
weak-winded stars also at intermediate O-types, and b) that the winds
of their `normal' counterparts are not strongly clumped, which would
diminish also their emission because of lower-than-thought mass-loss
rates.

Using the \NIII\ triplet as an independent mass-loss diagnostics would
be somewhat similar to the corresponding application of the NIR
\Bra-line (\citealt{Najarro11}, see also \citealt{puls08b}), but
with the advantage that the emission strength of the triplet is a
rather monotonic function of \mdot, whereas \Bra\ changes its
behaviour from weak to normal winds considerably.
\end{itemize}

\begin{acknowledgements}
{We thank our anonymous referee for valuable comments and suggestions.
J.G.R.G. gratefully acknowledges financial support from the German
DFG, under grant 418 SPA 112/1/08 (agreement between the DFG and the
Instituto de Astrofisica de Canarias), and J.P and F.N. acknowledge
financial support from the Spanish Ministerio de Ciencia e
Innovaci\'on under projects AYA2008-06166-C03-02 and
AYA2010-21697-C05-01. 
Many thanks to John Hillier for providing the {\sc cmfgen} code, and
particularly to Keith Butler for his advice and help on the Nitrogen
atomic data and useful suggestions on the manuscript.}
\end{acknowledgements}

\bibliographystyle{aa}
\bibliography{bib_niii}

\begin{thebibliography}{70}
\expandafter\ifx\csname natexlab\endcsname\relax\def\natexlab#1{#1}\fi

\bibitem[{{Allen}(1973)}]{allen73}
{Allen}, C.~W. 1973, {Astrophysical quantities} (London: University of London,
  Athlone Press, 1973, 3rd ed.)

\bibitem[{{Asplund} {et~al.}(2005){Asplund}, {Grevesse}, \&
  {Sauval}}]{asplund05}
{Asplund}, M., {Grevesse}, N., \& {Sauval}, A.~J. 2005, in Astronomical Society
  of the Pacific Conference Series, Vol. 336, Cosmic Abundances as Records of
  Stellar Evolution and Nucleosynthesis, ed. {T.~G.~Barnes III \& F.~N.~Bash},
  25

\bibitem[{{Asplund} {et~al.}(2009){Asplund}, {Grevesse}, {Sauval}, \&
  {Scott}}]{asplund09}
{Asplund}, M., {Grevesse}, N., {Sauval}, A.~J., \& {Scott}, P. 2009, \araa, 47,
  481

\bibitem[{{Bautista} {et~al.}(1998){Bautista}, {Romano}, \&
  {Pradhan}}]{Bautista98}
{Bautista}, M.~A., {Romano}, P., \& {Pradhan}, A.~K. 1998, \apjs, 118, 259

\bibitem[{{Bell} {et~al.}(1995){Bell}, {Hibbert}, {Stafford}, \&
  {Brage}}]{Bell95}
{Bell}, K.~L., {Hibbert}, A., {Stafford}, R.~P., \& {Brage}, T. 1995, \mnras,
  272, 909

\bibitem[{{Berrington} {et~al.}(1987){Berrington}, {Burke}, {Butler}, {Seaton},
  {Storey}, {Taylor}, \& {Yan}}]{berrington87}
{Berrington}, K.~A., {Burke}, P.~G., {Butler}, K., {et~al.} 1987, Journal of
  Physics B Atomic Molecular Physics, 20, 6379

\bibitem[{{Bouret} {et~al.}(2003){Bouret}, {Lanz}, {Hillier}, {Heap}, {Hubeny},
  {Lennon}, {Smith}, \& {Evans}}]{bouret03}
{Bouret}, J.-C., {Lanz}, T., {Hillier}, D.~J., {et~al.} 2003, \apj, 595, 1182

\bibitem[{{Bowen}(1935)}]{bowen35}
{Bowen}, I.~S. 1935, \apj, 81, 1

\bibitem[{{Brott} {et~al.}(2011{\natexlab{a}}){Brott}, {de Mink}, {Cantiello},
  {Langer}, {de Koter}, {Evans}, {Hunter}, {Trundle}, \& {Vink}}]{Brott11a}
{Brott}, I., {de Mink}, S.~E., {Cantiello}, M., {et~al.} 2011{\natexlab{a}},
  \aap, 530, A115

\bibitem[{{Brott} {et~al.}(2011{\natexlab{b}}){Brott}, {Evans}, {Hunter}, {de
  Koter}, {Langer}, {Dufton}, {Cantiello}, {Trundle}, {Lennon}, {de Mink},
  {Yoon}, \& {Anders}}]{Brott11b}
{Brott}, I., {Evans}, C.~J., {Hunter}, I., {et~al.} 2011{\natexlab{b}}, \aap,
  530, A116

\bibitem[{{Bruccato} \& {Mihalas}(1971)}]{brucato71}
{Bruccato}, R.~J. \& {Mihalas}, D. 1971, \mnras, 154, 491

\bibitem[{{Cunto} \& {Mendoza}(1992)}]{cunto92}
{Cunto}, W. \& {Mendoza}, C. 1992, Revista Mexicana de Astronomia y
  Astrofisica, vol.~23, 23, 107

\bibitem[{Eissner(1991)}]{eissner91}
Eissner, W. 1991, J. Phys. IV (France), 1, C1

\bibitem[{Eissner \& Nussbaumer(1969)}]{eissner69}
Eissner, W. \& Nussbaumer, H. 1969, in Premiere Reunion de l'Association
  Europeene de Spectroscopie Atomique No.~42 (Paris-Orsay: Faculte des
  Sciences)

\bibitem[{{Evans} {et~al.}(2006){Evans}, {Lennon}, {Smartt}, \&
  {Trundle}}]{evans06}
{Evans}, C.~J., {Lennon}, D.~J., {Smartt}, S.~J., \& {Trundle}, C. 2006, \aap,
  456, 623

\bibitem[{{Evans} {et~al.}(2011){Evans}, {Taylor}, {H{\'e}nault-Brunet},
  {Sana}, {de Koter}, {Sim{\'o}n-D{\'{\i}}az}, {Carraro}, {Bagnoli}, {Bastian},
  {Bestenlehner}, {Bonanos}, {Bressert}, {Brott}, {Campbell}, {Cantiello},
  {Clark}, {Costa}, {Crowther}, {de Mink}, {Doran}, {Dufton}, {Dunstall},
  {Friedrich}, {Garcia}, {Gieles}, {Gr{\"a}fener}, {Herrero}, {Howarth},
  {Izzard}, {Langer}, {Lennon}, {Ma{\'{\i}}z Apell{\'a}niz}, {Markova},
  {Najarro}, {Puls}, {Ramirez}, {Sab{\'{\i}}n-Sanjuli{\'a}n}, {Smartt},
  {Stroud}, {van Loon}, {Vink}, \& {Walborn}}]{Evans11}
{Evans}, C.~J., {Taylor}, W.~D., {H{\'e}nault-Brunet}, V., {et~al.} 2011, \aap,
  530, A108

\bibitem[{{Fernley} {et~al.}(1999){Fernley}, {Hibbert}, {Kingston}, \&
  {Seaton}}]{Fernley99}
{Fernley}, J.~A., {Hibbert}, A., {Kingston}, A.~E., \& {Seaton}, M.~J. 1999,
  Journal of Physics B Atomic Molecular Physics, 32, 5507

\bibitem[{{Grevesse} \& {Sauval}(1998)}]{grevesse98}
{Grevesse}, N. \& {Sauval}, A.~J. 1998, Space Science Reviews, 85, 161

\bibitem[{{Heap} {et~al.}(2006){Heap}, {Lanz}, \& {Hubeny}}]{Heap06}
{Heap}, S.~R., {Lanz}, T., \& {Hubeny}, I. 2006, \apj, 638, 409

\bibitem[{{Heger} \& {Langer}(2000)}]{Heger00}
{Heger}, A. \& {Langer}, N. 2000, \apj, 544, 1016

\bibitem[{{Hillier} {et~al.}(2003){Hillier}, {Lanz}, {Heap}, {Hubeny}, {Smith},
  {Evans}, {Lennon}, \& {Bouret}}]{Hillier03}
{Hillier}, D.~J., {Lanz}, T., {Heap}, S.~R., {et~al.} 2003, \apj, 588, 1039

\bibitem[{{Hillier} \& {Miller}(1998)}]{hilliermiller98}
{Hillier}, D.~J. \& {Miller}, D.~L. 1998, \apj, 496, 407

\bibitem[{{Howarth} {et~al.}(1997){Howarth}, {Siebert}, {Hussain}, \&
  {Prinja}}]{Howarth97}
{Howarth}, I.~D., {Siebert}, K.~W., {Hussain}, G.~A.~J., \& {Prinja}, R.~K.
  1997, \mnras, 284, 265

\bibitem[{{Hubeny}(1988)}]{hubeny88}
{Hubeny}, I. 1988, Computer Physics Communications, 52, 103

\bibitem[{{Hubeny} \& {Lanz}(1995)}]{hubeny95}
{Hubeny}, I. \& {Lanz}, T. 1995, \apj, 439, 875

\bibitem[{{Hunter} {et~al.}(2009{\natexlab{a}}){Hunter}, {Brott}, {Langer},
  {Lennon}, {Dufton}, {Howarth}, {Ryans}, {Trundle}, {Evans}, {de Koter}, \&
  {Smartt}}]{hunter09b}
{Hunter}, I., {Brott}, I., {Langer}, N., {et~al.} 2009{\natexlab{a}}, \aap,
  496, 841

\bibitem[{{Hunter} {et~al.}(2008){Hunter}, {Brott}, {Lennon}, {Langer},
  {Dufton}, {Trundle}, {Smartt}, {de Koter}, {Evans}, \& {Ryans}}]{hunter08}
{Hunter}, I., {Brott}, I., {Lennon}, D.~J., {et~al.} 2008, \apjl, 676, L29

\bibitem[{{Hunter} {et~al.}(2007){Hunter}, {Dufton}, {Smartt}, {Ryans},
  {Evans}, {Lennon}, {Trundle}, {Hubeny}, \& {Lanz}}]{hunter07}
{Hunter}, I., {Dufton}, P.~L., {Smartt}, S.~J., {et~al.} 2007, \aap, 466, 277

\bibitem[{{Hunter} {et~al.}(2009{\natexlab{b}}){Hunter}, {Lennon}, {Dufton},
  {Trundle}, {Sim{\'o}n-D{\'{\i}}az}, {Smartt}, {Ryans}, \& {Evans}}]{hunter09}
{Hunter}, I., {Lennon}, D.~J., {Dufton}, P.~L., {et~al.} 2009{\natexlab{b}},
  \aap, 504, 211

\bibitem[{{Kelleher} {et~al.}(1999){Kelleher}, {Mohr}, {Martin}, {Wiese},
  {Sugar}, {Fuhr}, {Olsen}, {Musgrove}, {Reader}, {Sansonetti}, \&
  {Dalton}}]{nist}
{Kelleher}, D.~E., {Mohr}, P.~J., {Martin}, W.~C., {et~al.} 1999, in Society of
  Photo-Optical Instrumentation Engineers (SPIE) Conference Series, Vol. 3818,
  Society of Photo-Optical Instrumentation Engineers (SPIE) Conference Series,
  ed. {G.~R.~Carruthers \& K.~F.~Dymond}, 170

\bibitem[{{Lanz} \& {Hubeny}(2003)}]{LanzHubeny03}
{Lanz}, T. \& {Hubeny}, I. 2003, \apjs, 146, 417

\bibitem[{{Lenorzer} {et~al.}(2004){Lenorzer}, {Mokiem}, {de Koter}, \&
  {Puls}}]{lenorzer04}
{Lenorzer}, A., {Mokiem}, M.~R., {de Koter}, A., \& {Puls}, J. 2004, \aap, 422,
  275

\bibitem[{{Marcolino} {et~al.}(2009){Marcolino}, {Bouret}, {Martins},
  {Hillier}, {Lanz}, \& {Escolano}}]{Marcolino09}
{Marcolino}, W.~L.~F., {Bouret}, J., {Martins}, F., {et~al.} 2009, \aap, 498,
  837

\bibitem[{{Markova} \& {Puls}(2008)}]{MP08}
{Markova}, N. \& {Puls}, J. 2008, \aap, 478, 823

\bibitem[{{Martins} {et~al.}(2005{\natexlab{a}}){Martins}, {Schaerer}, \&
  {Hillier}}]{martins05a}
{Martins}, F., {Schaerer}, D., \& {Hillier}, D.~J. 2005{\natexlab{a}}, \aap,
  436, 1049

\bibitem[{{Martins} {et~al.}(2005{\natexlab{b}}){Martins}, {Schaerer},
  {Hillier}, {Meynadier}, {Heydari-Malayeri}, \& {Walborn}}]{martins05b}
{Martins}, F., {Schaerer}, D., {Hillier}, D.~J., {et~al.} 2005{\natexlab{b}},
  \aap, 441, 735

\bibitem[{{Meynet} \& {Maeder}(2000)}]{MeynetMaeder00}
{Meynet}, G. \& {Maeder}, A. 2000, \aap, 361, 101

\bibitem[{{Mihalas}(1971)}]{mihalas71}
{Mihalas}, D. 1971, \apj, 170, 541

\bibitem[{{Mihalas}(1978)}]{mihalasbook78}
{Mihalas}, D. 1978, {Stellar atmospheres (2nd edition)} (San Francisco:
  W.~H.~Freeman and Co., 1978)

\bibitem[{{Mihalas} \& {Hummer}(1973)}]{mihalas73}
{Mihalas}, D. \& {Hummer}, D.~G. 1973, \apj, 179, 827

\bibitem[{{Mokiem} {et~al.}(2007){Mokiem}, {de Koter}, {Vink}, {Puls}, {Evans},
  {Smartt}, {Crowther}, {Herrero}, {Langer}, {Lennon}, {Najarro}, \&
  {Villamariz}}]{mokiem07b}
{Mokiem}, M.~R., {de Koter}, A., {Vink}, J.~S., {et~al.} 2007, \aap, 473, 603

\bibitem[{{Moore}(1975)}]{Moore75}
{Moore}, C.~E. 1975, {Selected tables of atomic spectra - A: Atomic energy
  levels - Second edition - B: Multiplet table; N I, N II, N III. Data derived
  from the analyses of optical spectra}, ed. {Moore, C.~E.}

\bibitem[{{Najarro} {et~al.}(2011){Najarro}, {Hanson}, \& {Puls}}]{Najarro11}
{Najarro}, F., {Hanson}, M.~M., \& {Puls}, J. 2011, ArXiv e-prints

\bibitem[{{Najarro} {et~al.}(1996){Najarro}, {Kudritzki}, {Cassinelli},
  {Stahl}, \& {Hillier}}]{najarro96}
{Najarro}, F., {Kudritzki}, R.~P., {Cassinelli}, J.~P., {Stahl}, O., \&
  {Hillier}, D.~J. 1996, \aap, 306, 892

\bibitem[{{Nikitin} \& {Yakubovskii}(1963)}]{nikitin63}
{Nikitin}, A.~A. \& {Yakubovskii}, O.~A. 1963, Soviet Astronomy, 7, 189

\bibitem[{{Nussbaumer} \& {Storey}(1983)}]{nussbaumer83}
{Nussbaumer}, H. \& {Storey}, P.~J. 1983, \aap, 126, 75

\bibitem[{{Oke}(1954)}]{oke54}
{Oke}, J.~B. 1954, \apj, 120, 22

\bibitem[{{Pauldrach} {et~al.}(2001){Pauldrach}, {Hoffmann}, \&
  {Lennon}}]{pauldrach01}
{Pauldrach}, A.~W.~A., {Hoffmann}, T.~L., \& {Lennon}, M. 2001, \aap, 375, 161

\bibitem[{{Pauldrach} {et~al.}(1994){Pauldrach}, {Kudritzki}, {Puls}, {Butler},
  \& {Hunsinger}}]{pauldrach94c}
{Pauldrach}, A.~W.~A., {Kudritzki}, R.~P., {Puls}, J., {Butler}, K., \&
  {Hunsinger}, J. 1994, \aap, 283, 525

\bibitem[{{Pauldrach} \& {Puls}(1990)}]{PP90}
{Pauldrach}, A.~W.~A. \& {Puls}, J. 1990, \aap, 237, 409

\bibitem[{{Puls} {et~al.}(2006){Puls}, {Markova}, {Scuderi}, {Stanghellini},
  {Taranova}, {Burnley}, \& {Howarth}}]{Puls06}
{Puls}, J., {Markova}, N., {Scuderi}, S., {et~al.} 2006, \aap, 454, 625

\bibitem[{{Puls} {et~al.}(2010){Puls}, {Sundqvist}, \& {Rivero
  Gonz{\'a}lez}}]{Puls10}
{Puls}, J., {Sundqvist}, J.~O., \& {Rivero Gonz{\'a}lez}, J.~G. 2010, in Active
  OB stars: structure, evolution, mass-loss \& critical limits, ed. {C.~Neiner,
  G.~Wade, G.~Meynet, \& G.~Peters} (Cambridge Univ. Press), {IAU Symp., 272}
  [arXiv:1009.0364]

\bibitem[{{Puls} {et~al.}(2005){Puls}, {Urbaneja}, {Venero}, {Repolust},
  {Springmann}, {Jokuthy}, \& {Mokiem}}]{puls05}
{Puls}, J., {Urbaneja}, M.~A., {Venero}, R., {et~al.} 2005, \aap, 435, 669

\bibitem[{{Puls} {et~al.}(2008){Puls}, {Vink}, \& {Najarro}}]{puls08b}
{Puls}, J., {Vink}, J.~S., \& {Najarro}, F. 2008, \aapr, 16, 209

\bibitem[{{Repolust} {et~al.}(2005){Repolust}, {Puls}, {Hanson}, {Kudritzki},
  \& {Mokiem}}]{Repo05}
{Repolust}, T., {Puls}, J., {Hanson}, M.~M., {Kudritzki}, R.-P., \& {Mokiem},
  M.~R. 2005, \aap, 440, 261

\bibitem[{{Repolust} {et~al.}(2004){Repolust}, {Puls}, \&
  {Herrero}}]{repolust04}
{Repolust}, T., {Puls}, J., \& {Herrero}, A. 2004, \aap, 415, 349

\bibitem[{{Seaton}(1958)}]{seaton58}
{Seaton}, M.~J. 1958, \mnras, 118, 504

\bibitem[{{Seaton}(1962)}]{seaton62}
{Seaton}, M.~J. 1962, in Atomic and Molecular Processes, ed. {D.~R.~Bates} (New
  York, Academic Press), 375

\bibitem[{{Sobolev}(1960)}]{Sobo60}
{Sobolev}, V.~V. 1960, {Moving envelopes of stars} (Cambridge: Harvard
  University Press, 1960)

\bibitem[{{Sota} {et~al.}(2011){Sota}, {Ma{\'{\i}}z Apell{\'a}niz}, {Walborn},
  {Alfaro}, {Barb{\'a}}, {Morrell}, {Gamen}, \& {Arias}}]{sota11}
{Sota}, A., {Ma{\'{\i}}z Apell{\'a}niz}, J., {Walborn}, N.~R., {et~al.} 2011,
  \apjs, 193, 24

\bibitem[{{Stafford} {et~al.}(1994){Stafford}, {Bell}, \&
  {Hibbert}}]{stafford94}
{Stafford}, R.~P., {Bell}, K.~L., \& {Hibbert}, A. 1994, \mnras, 266, 715

\bibitem[{{Sundqvist} {et~al.}(2011){Sundqvist}, {Puls}, {Feldmeier}, \&
  {Owocki}}]{Sundqvist11}
{Sundqvist}, J.~O., {Puls}, J., {Feldmeier}, A., \& {Owocki}, S.~P. 2011, \aap,
  528, A64

\bibitem[{{Swings}(1948)}]{swings48}
{Swings}, P. 1948, Annales d'Astrophysique, 11, 228

\bibitem[{{Swings} \& {Struve}(1940)}]{swings40}
{Swings}, P. \& {Struve}, O. 1940, \apj, 91, 546

\bibitem[{{van Regemorter}(1962)}]{vanregemorter62}
{van Regemorter}, H. 1962, \apj, 136, 906

\bibitem[{{Vink} {et~al.}(2010){Vink}, {Brott}, {Gr{\"a}fener}, {Langer}, {de
  Koter}, \& {Lennon}}]{Vink10}
{Vink}, J.~S., {Brott}, I., {Gr{\"a}fener}, G., {et~al.} 2010, \aap, 512, L7

\bibitem[{{Vink} {et~al.}(2000){Vink}, {de Koter}, \& {Lamers}}]{vink00}
{Vink}, J.~S., {de Koter}, A., \& {Lamers}, H.~J.~G.~L.~M. 2000, \aap, 362, 295

\bibitem[{{Walborn}(1971)}]{walborn71b}
{Walborn}, N.~R. 1971, \apjs, 23, 257

\bibitem[{{Walborn} {et~al.}(2004){Walborn}, {Morrell}, {Howarth}, {Crowther},
  {Lennon}, {Massey}, \& {Arias}}]{walborn04}
{Walborn}, N.~R., {Morrell}, N.~I., {Howarth}, I.~D., {et~al.} 2004, \apj, 608,
  1028

\bibitem[{{Yan} \& {Seaton}(1987)}]{yu87}
{Yan}, Y. \& {Seaton}, M.~J. 1987, Journal of Physics B Atomic Molecular
  Physics, 20, 6409

\end{thebibliography}

\Online
\appendix
\section{Dielectronic Recombination: Implementation to {\sc fastwind}}
\label{theory-dr}
In the present work we implemented dielectronic recombination into {\sc
fastwind}, which, so far, could not (or only approximately) deal with this
process.  To this end, new rates (both for the dielectronic recombination
and for the inverse process) had to be inserted into the system of the rate
equations.

\subsection{Explicit method}
\label{exp-dr} 

To calculate these rates for the `explicit method' (see Sect.~\ref{fw-dr}),
we follow the formulation as provided by \cite{nussbaumer83}. In compact
notation, the dielectronic recombination for an element X and charge $m+1$
proceeds via
\beq
X^{m+1}_p + e^- \rarrow X_a^m \rarrow X_b^m + h\nu
\label{dr_nuss}
\eeq
where $p$ is a parent state of the $m+1$ times ionized element X, $a$ is an
autoionizing state and $b$ is a bound state. We denote the initial state of
expression~\ref{dr_nuss}, composed of the recombining ion and the free
electron, as a continuum state $c$. We refer to the first process as
dielectronic capture and to its inverse as autoionization. In general,
dielectronic captures and autoionizations link state $a$ to a large number
of continuum states $c$. 

As a final result,\footnote{See also \citealt{mihalasbook78} for a
simplified derivation.} \cite{nussbaumer83} could express the dielectronic
recombination coefficient between autoionizing state $a$ and bound state
$j$, $\alpha_{\rm DR}(aj)$, as
\beq
n_{\rm e} N_k^{m+1} \alpha_{\rm DR}(aj) = b_a (n_a^m)^{*} A_{aj}^{\rm R}, 
\label{dr-rate-3}
\eeq
with $A_{aj}^{\rm R}$ the corresponding radiative transition
probability for the stabilizing transition (Einstein coefficient for
spontaneous emission, corrections for induced emission will be applied
below), total ion density $N_k^{m+1}$ (element $k$), $(n_a^m)^{*}$ the
LTE population of state $a$ with respect to the actual (NLTE)
ground-state population of the next higher ion and the actual electron
density, and $b_a$ the related NLTE departure coefficient. 

The last quantity can be expressed in terms of (i) the autoionization
coefficient(s), $A_{ac}^{\rm a}$, between state $a$ and all possible
compound-states $c$ that can form $a$, (ii) the radiative transition
probabilities, $A_{ai}^{\rm R}$, between state $a$ and all possible bound and
autoionizing states with lower energy $i$ to which state $a$ can decay, 
and (iii) the departure coefficients of the
contributing parent levels, $b_p$ (here with respect to the ground-state of
the same ion, $n_1^{m+1}$)
\beq
b_a = \frac{n_a^m}{{n_a^m}^*}= \frac{\sum_c{b_p {A_{ac}^{\rm a}}}}
{\sum_c{A_{ac}^{\rm a}} + \sum_i{A_{ai}^{\rm R}}}.
\eeq
Usually, the autoionizing probabilities for state $a$ are much larger than
the radiative probabilities for decay, and often there is only one parent
level, namely the ground-state of ion $m+1$, $n_p = n_1^{m+1}$, i.e., $b_p =
1$. Under these conditions (which are similar for excited parent levels
assumed to be in LTE with respect to the ground level), $b_a \rightarrow 1$, and
{\it the dielectronic rate depends only on the LTE population of state $a$ and
the radiative transition probability $A_{aj}^{\rm R}$}. All dependencies on the
autoionization probabilities have `vanished', and we need only the value of
$(n_a^m)^{*}$ that can be derived from the Saha-equation and the
ground-state population of ion $m+1$, without including the autoionizing
levels into the rate equations. 

In stellar atmospheres, one needs (in addition to the spontaneous emission
to level $j$) to account for stimulated emission as well, i.e.,
\beq
A_{aj}^{\rm R} \rightarrow A_{aj}^{\rm R} + B_{aj} \bar{J} =
A_{aj}^{\rm R} 
\left(\displaystyle\ 1 + \frac{c^2}{2h \nu^3} \bar{J}
\right)
\label{einstein}
\eeq
with $B_{aj}$ the Einstein coefficient for stimulated emission and 
$\bar{J}$ the scattering integral (profile weighted, frequency integrated
mean intensity) for the stabilizing
transition. Since the important resonances are broad, the scattering integrals 
might be replaced by the mean intensities, $J_\nu$, of
the pseudo-continuum (i.e., including all background opacities/emissivities)
at the frequency of the stabilizing line.

Finally, we can define the total dielectronic rate to level $j$ from any
possible autoionizing state $a^i$,
\beqa
\lefteqn{n_{\rm e} N_k^{m+1} \sum_i{\alpha_{\rm DR}(a^i j)} \approx
n_{\rm e} n_1^{m+1} \frac{1}{g_k} C_{\rm i} T^{-3/2} \times} \\ \nonumber 
& &\times \sum_i{g(a^i) \exp(-E_{a^i}/kT) A_{a^i j}^{\rm R} \
\Bigl( 1 + \frac{c^2}{2h \nu_{ij}^3} \bar{J}_{ij}
\Bigr)},
\label{dr-rate-sum}
\eeqa
which might need to be augmented by departure coefficients $b_{a^i}$
inside the rhs sum if the parent levels are not the ground-state or
not in LTE with respect to the ground-state. The inverse upward rate
involves only the line process(es),
\beq
n_j^m \sum_i {B_{ja^i} \bar{J}_{ij}} = n_j^m \sum_i{\frac{c^2}{2h\nu_{ij}^3} 
A_{a^i j}^{\rm R} \frac{g(a^i)}{g_j}\bar{J}_{ij}},
\label{upward-rate}
\eeq
with $B_{ji}$ the Einstein-coefficient(s) for absorption. It is easy to show
that for LTE conditions ($n_j^m = (n_j^m)^{*}$ and Planckian radiation in the
lines) the upward and downward rates cancel each other exactly, as
required for thermalization.

Though we have followed here the derivation by \cite{nussbaumer83}, our
final results for the downward and upward rates are identical with the
formulation as used by \cite{mihalas73} in their investigation. 

These rates have been implemented into {\sc fastwind} and are used 
whenever the `explicit method' is applied. The only input parameters
that need to be specified in the atomic data input file are the
transition frequencies and the oscillator-strengths for the
stabilizing lines, $f_{ja}$, which relate to the pro\-ducts $g_a
A_{aj}^{\rm R}$ via
\beq
g_a A_{aj}^{\rm R} = \frac{8 \pi^2 e^2}{m_{\rm e} c^3} \nu_{aj}^2 g_j f_{ja}.
\eeq
\smallskip
\noindent
For convenience and for consistency with the formulation of `normal'
recombination rates (see Sect.~\ref{imp-dr}), the quantity $(n_a^m)^{*}$ is
expressed in terms of the LTE-population (with respect to actual ionization
conditions) of the lower, stabilizing level $(n_j^m)^{*}$
\beq
(n_a^m)^{*} = (n_j^m)^{*} \frac{g_a}{g_j} \exp(-h\nu_{aj}/kT),
\label{a_via_j}
\eeq
so that the downward rate (for a specific autoionizing level $a$) can be 
expressed as
\beqa
\label{dr-rate-final}
n_{\rm e} N_k^{m+1} \alpha_{\rm DR}(aj) & = & [b_a] (n_j^m)^{*} 
\frac{8 \pi^2 e^2}{m_{\rm e} c^3} \nu_{aj}^2  f_{ja}  \exp(-h\nu_{aj}/kT) \times 
\nonumber \\
& \times & \Bigl( 1 + \frac{c^2}{2h \nu_{aj}^3} \bar{J}_{aj} \Bigr). 
\eeqa
Summation over all contributing autoionizing states yields the final rate.

In this explicit method, the rates for dielectronic recombinations and
inverse processes are calculated in a separate step and then added to the
rates involving resonance-free photoionization cross-sections
alone. In our model ion (Sect.~\ref{atom niii}), we use such
cross-sections defined in terms of the Seaton approximation
(Eq.~\ref{Seaton_cross}), which, together with the data for the stabilizing
transitions, have been taken from the {\sc wm}-basic atomic
database. Note that we consider processes both from/to
ground-states as well as from/to excited states within ion $m+1$, so far on the
assumption that the autoionizing levels are in LTE (i.e., without including
these levels into the model atom and rate equations).

\subsection{Implicit method}
\label{imp-dr} Within the implicit method, the DR contribution is
already contained within the `conventional'
recombination rates, $(n_j/n_k)^*R_{kj}$, to yield a total number of
recombinations
\beq
n_k\left(\displaystyle\frac{n_j}{n_k}\right)^* R_{kj} := {n_j}^* R_{kj}.
\eeq
As usual, $n_k$ is the actual population of the recombining ion in state $k$
and $(n_j/n_k)^*$ the LTE population ratio of the considered level to which
the process recombines (either directly or via the intermediate doubly
excited state). $R_{kj}$ is defined as
\beq
R_{kj} = 4 \pi \int^{\infty}_{\nu_{0}}
\frac{\alpha_{jk}(\nu)}{h\nu}\left(\displaystyle\ \frac{2h\nu^3}{c^2} +
J_\nu\right) e^{-h\nu/kT}\ \dd \nu
\label{recomb_rate}
\eeq
with mean intensity $J_\nu$ and total photoionization
cross-section (including resonances), $\alpha_{jk}(\nu)$. 

In the following, we show that this formulation is consistent with the rates
derived in Sect.~\ref{exp-dr}. We split the cross-section into a
resonance-free contribution, and a contribution from the
resonances,
\beq
\alpha_{jk}(\nu) = \alpha_{jk}^{\rm no\_res}(\nu) + \alpha_{jk}^{\rm res} (\nu). 
\eeq
The total recombination rate is then
the sum of direct and dielectronic recombination,
\beq
{n_j}^* R_{kj} = {n_j}^* (R_{kj}^{\rm dir} +R_{kj}^{\rm DR}),
\eeq
with
\beq
{n_j}^* R_{kj}^{\rm DR} = {n_j}^*  4 \pi \ \int_{\rm res}
\frac{\alpha_{jk}^{\rm res}(\nu)}{h\nu}\left(\displaystyle\ \frac{2h\nu^3}{c^2}
+
J_\nu\right) e^{-h\nu/kT}\ \dd \nu.
\eeq
The resonances are narrow enough so that most of the frequency
dependent quantities can be drawn in front of the integral, evaluated 
at the average position of the resonances $i$,
\beqa
\lefteqn{{n_j}^* R_{kj}^{\rm DR} \approx} \\ 
& & {n_j}^* \sum_{\rm i} \frac{4\pi}{h\nu_{ij}} 
\frac{2h\nu_{ij}^3}{c^2}  e^{-h\nu_{ij}/kT} \
\Bigl(\int_{\rm res(i)} \alpha_{ji}^{\rm res}(\nu) \
\bigl( 1 + \frac{c^2}{2h \nu_{ij}^3} J_\nu \bigr) \ \dd \nu \Bigr).  \nonumber
\eeqa
The integral over the cross-sections of the resonances corresponds to the
cross-section of the stabilizing transitions,
\beq
\frac{4\pi}{h\nu_{ij}} \ \int_{\rm res(i)} \alpha_{ji}^{\rm res} \dd \nu   
= B_{ji} = \frac{4\pi}{h\nu_{ij}} \frac{\pi e^2}{m_{\rm e} c} f_{ji},
\label{int_stab}
\eeq
with Einstein-coefficient $B_{ji}$ and oscillator-strength $f_{ji}$. Like\-wise,
\[
\int_{\rm res(i)} \alpha_{ji}^{\rm res} J_\nu \dd \nu \propto f_{ji} \bar{J}_{ij}.
\]
Then, indeed, we recover the result from Eq.~\ref{dr-rate-final} (explicit
formulation),
\beq
n_j^* R_{kj}^{\rm DR} = n_j^* \ \sum_{\rm i}
\frac{8 \pi^2 e^2}{m_{\rm e} c^3} \nu_{ij}^2  f_{ji}  \exp(-h\nu_{ij}/kT) 
\left(\displaystyle\ 1 + \frac{c^2}{2h \nu_{ij}^3} \bar{J}_{ij} \right),
\eeq
if, as outlined in Sect.~\ref{exp-dr}, the autoionizing levels are in or
close to LTE. Note that this is a necessary condition for the implicit
method to yield reliable results,\footnote{which has been used to
calculate the total cross-sections as provided by, e.g., the OPACITY
Project.} otherwise one has to use exclusively the explicit approach and to
include the autoionizing levels into the model atom and rate equations.

The proof that the rates for the inverse photoionization process, calculated
either by the implicit method (using total photoionization cross-sections)
or via rates from resonance-free cross-sections plus rates involving the
excitation of the second electron, are consistent, is analogous and omitted
here.

\begin{figure}[b]
\resizebox{\hsize}{!}
  {\includegraphics{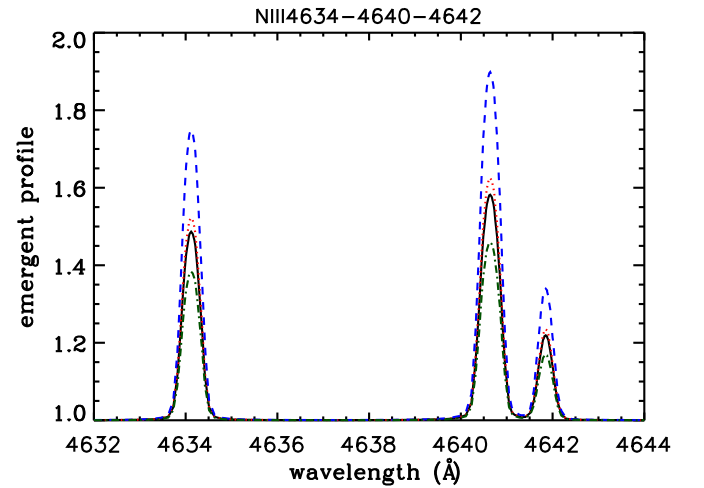}}
\caption{Consistency of implicit and explicit method. Comparison of \NIII\
\trip\ profiles from model `T3740', for calculations using a different
treatment of DR to the 3d level. Implicit method (solid/black), explicit
method (dotted/red), explicit method with oscillator strength for the
stabilizing transition increased (dashed/blue) and decreased
(dashed-dotted/black) by a factor of two.}
\label{op-vs-exp}
\end{figure}

\subsection{Consistency check}
\label{consistency}

To check the consistency of our implementation of implicit and
explicit DR treatment, we carried out the following test. First, to
ensure consistent {\it data}, we derived the oscillator strength
corresponding to the wide PEC resonance in the total photoionization
cross-section (Fig.~\ref{smooth-op}) by integrating over the
cross-section and applying Eq.~\ref{int_stab}.\footnote{Of course, the
underlying contribution responsible for direct ionization needs to be
subtracted.} The resulting value ($f = 0.45$, which is somewhat
smaller than the data provided by the {\sc wm}-basic database, $f =
0.60$) was then used within the explicit method, at the original
wavelength (which coincides with the position of the resonance).
As displayed in Fig.~\ref{op-vs-exp} for the case of model `T3740'
with `pseudo line-blanketing' (see Sect.~\ref{mih-hum}), both methods
result in very similar line profiles for the \NIII\ emission triplet, 
proving their consistency.
Figure~\ref{op-vs-exp} also displays the strong reaction of the \NIII\
triplet when the oscillator strength is either increased or diminished by a
factor of two.

\section{Details on the \NIII\ model ion}

\begin{figure*}
\begin{minipage}{9cm}
\resizebox{\hsize}{!} {\includegraphics{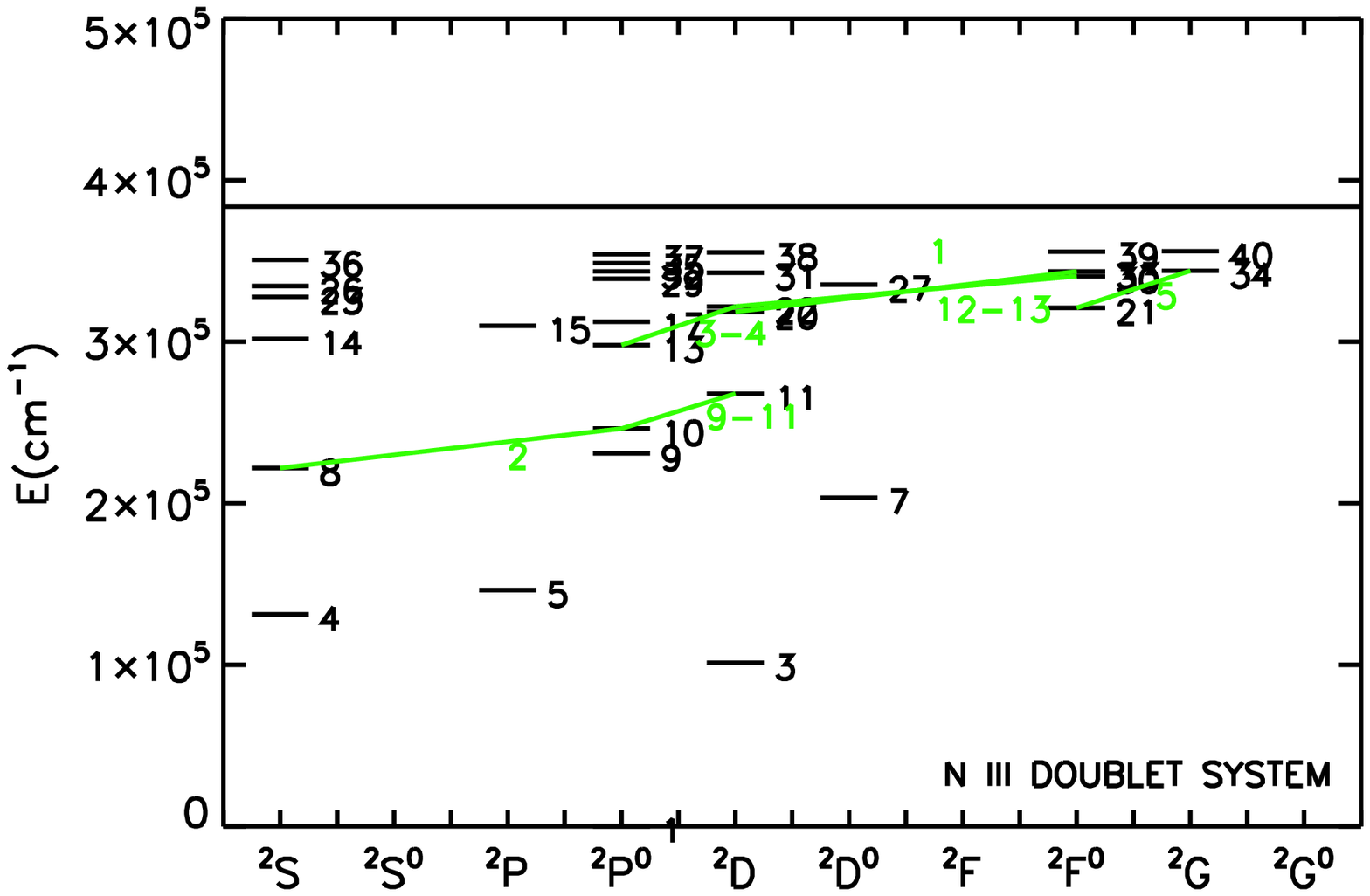}}
\end{minipage}
\begin{minipage}{9cm}
\resizebox{\hsize}{!}
 {\includegraphics{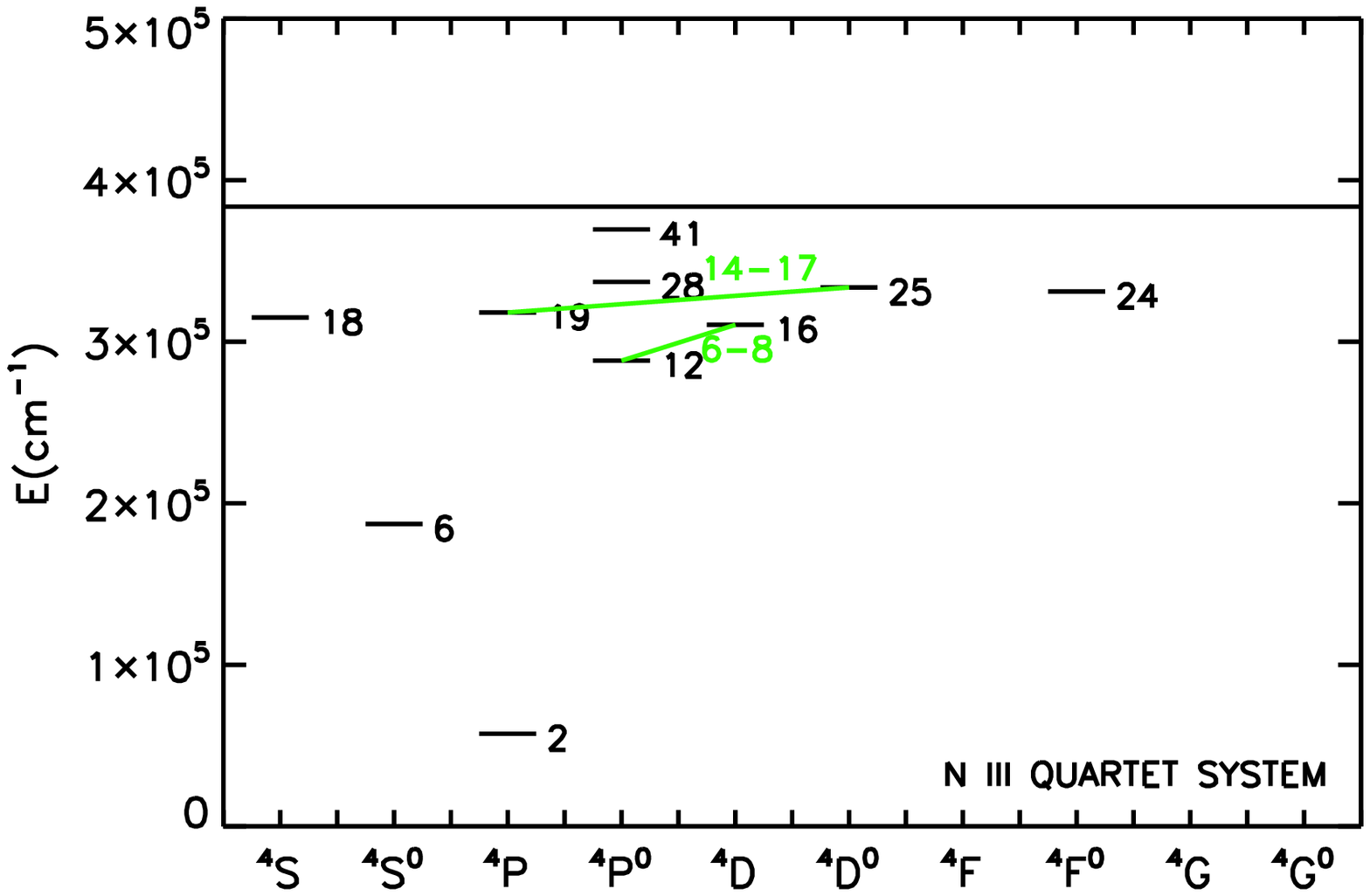}}
\end{minipage}
\caption{Grotrian diagrams for the \NIII\ doublet (left) and quartet (right) 
system. Level numbers
refer to Table~\ref{atom_lev_niii}. Levels \# 8, 10, 11 refer to the 3s,
3p and 3d levels involved in the formation of \NIII\ \trip, and
\# 3, 4, 5 refer to levels 2p$^2$ ($^2$D, $^2$S, $^2$P) that drain level 3p via
`two electron transitions'. Important optical transitions are indicated by
green lines and numbers referring to Table~\ref{tab_lines_niii} in the main
section.}
\label{niii-doublet-quartet}
\end{figure*}

Table~\ref{atom_lev_niii} displays the electronic configurations and term
designations of our \NIII\ ionic model, and Fig.~\ref{niii-doublet-quartet}
the corresponding Grotrian diagrams for the doublet and the quartet systems,
indicating important diagnostic transitions in the optical (cf.
Sect.~\ref{comp-cmfgen}).

\begin{table}[!h]
\caption{Electronic configurations and term designations of our \NIII\ model
ion. The level numbers correspond to the entries in the Grotrian diagrams,
Fig.~\ref{niii-doublet-quartet}.}
\label{atom_lev_niii}
\tabcolsep1.4mm
\begin{tabular}{rll | rll}
\hline 
\hline
\# & Configuration & Desig. & \#  & Configuration  & Desig. \\
\hline
  1      & 1s$^2$ 2s$^2$($^1$S) 2p           &    2p $^2$P$^0$    & 
  22     & 1s$^2$ 2s   2p($^3$P$^0$) 3p    &    3p'   $^2$D        \\
  2      & 1s$^2$ 2s   2p$^2$              &    2p$^2$ $^4$P    &      
 23      & 1s$^2$ 2s   2p($^3$P$^0$) 3p    &    3p'   $^2$S        \\
  3      & 1s$^2$ 2s   2p$^2$              &    2p$^2$ $^2$D    &      
 24      & 1s$^2$ 2s   2p($^3$P$^0$) 3d    &    3d'   $^4$F$^0$        \\
  4      & 1s$^2$ 2s   2p$^2$              &    2p$^2$ $^2$S  &       
 25      & 1s$^2$ 2s   2p($^3$P$^0$) 3d    &    3d'   $^4$D$^0$        \\
  5      & 1s$^2$ 2s   2p$^2$              &    2p$^2$ $^2$P  &      
 26      & 1s$^2$ 2s$^2$($^1$S) 5s           &    5s   $^2$S        \\
  6      & 1s$^2$ 2p$^3$                     &    2p$^3$ $^4$S$^0$  &      
 27      & 1s$^2$ 2s   2p($^3$P$^0$) 3d    &    3d'   $^2$D$^0$        \\
  7      & 1s$^2$ 2p$^3$                     &    2p$^3$ $^2$D$^0$  &      
 28      & 1s$^2$ 2s   2p($^3$P$^0$) 3d    &    3d'   $^4$P$^0$        \\
  8      & 1s$^2$ 2s$^2$($^1$S) 3s           &    3s   $^2$S  &      
 29      & 1s$^2$ 2s$^2$($^1$S) 5p           &    5p   $^2$P$^0$        \\
  9      & 1s$^2$ 2p$^3$                     &    2p$^3$ $^2$P$^0$  &      
 30      & 1s$^2$ 2s   2p($^3$P$^0$) 3d    &    3d'   $^2$F$^0$        \\
 10      & 1s$^2$ 2s$^2$($^1$S) 3p           &    3p   $^2$P$^0$  &      
 31      & 1s$^2$ 2s$^2$($^1$S) 5d           &    5d   $^2$D        \\
 11      & 1s$^2$ 2s$^2$($^1$S) 3d           &    3d   $^2$D    &    
 32      & 1s$^2$ 2s   2p($^3$P$^0$) 3d    &    3d'   $^2$P$^0$        \\
 12      & 1s$^2$ 2s   2p($^3$P$^0$) 3s    &    3s'   $^4$P$^0$    &    
 33      & 1s$^2$ 2s$^2$($^1$S) 5f           &    5f   $^2$F$^0$        \\
 13      & 1s$^2$ 2s   2p($^3$P$^0$) 3s    &    3s'   $^2$P$^0$  &       
 34      & 1s$^2$ 2s$^2$($^1$S) 5g           &    5g   $^2$G        \\
 14      & 1s$^2$ 2s$^2$($^1$S) 4s           &    4s   $^2$S    &    
 35      & 1s$^2$ 2s$^2$($^1$S) 6p           &    6p   $^2$P$^0$        \\
 15      & 1s$^2$ 2s   2p($^3$P$^0$) 3p    &    3p'   $^2$P    &    
 36      & 1s$^2$ 2s$^2$($^1$S) 6s           &    6s   $^2$S        \\
 16      & 1s$^2$ 2s   2p($^3$P$^0$) 3p    &    3p'   $^4$D    &    
 37      & 1s$^2$ 2s   2p($^1$P$^0$) 3s    &    3s''   $^2$P$^0$        \\
 17      & 1s$^2$ 2s$^2$($^1$S) 4p           &    4p   $^2$P$^0$    &    
 38      & 1s$^2$ 2s$^2$($^1$S) 6d           &    6d   $^2$D        \\
 18      & 1s$^2$ 2s   2p($^3$P$^0$) 3p    &    3p'   $^4$S  &    
 39      & 1s$^2$ 2s$^2$($^1$S) 6f           &    6f   $^2$F$^0$        \\
 19      & 1s$^2$ 2s   2p($^3$P$^0$) 3p    &    3p'   $^4$P  &    
 40      & 1s$^2$ 2s$^2$($^1$S) 6g           &    6g   $^2$G        \\
 20      & 1s$^2$ 2s$^2$($^1$S) 4d           &    4d   $^2$D  &    
 41      & 1s$^2$ 2s   2p($^3$P$^0$) 4s    &    4s'   $^4$P$^0$        \\
 21      & 1s$^2$ 2s$^2$($^1$S) 4f           &    4f   $^2$F$^0$       & \\
\hline
\end{tabular}
\end{table}

\section{Comparison of line profiles with results from {\sc cmfgen}} 
\label{apen_cmfgen} 

In this appendix, we display a comparison of \NIII\ (and partly \NII) line
profiles from {\sc fastwind} (black) and {\sc cmfgen} (green), for all
models from the grid presented in Table~\ref{grid-cmfgen}. For a discussion,
see Sect.~\ref{comp-cmfgen}.

\clearpage
\begin{figure}
\resizebox{\hsize}{!}{\includegraphics{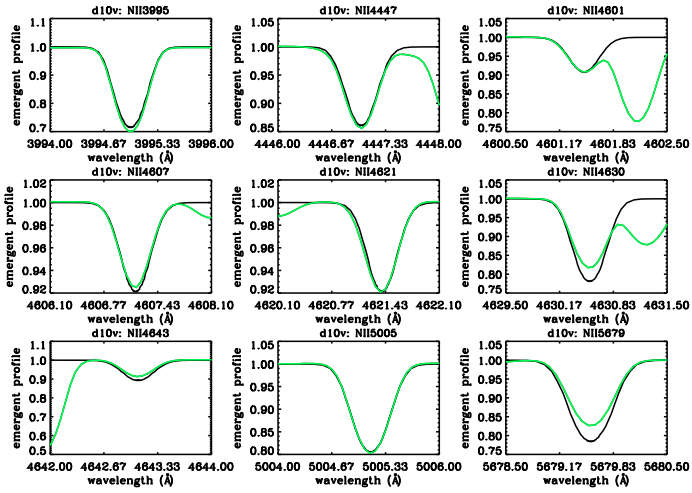}}
\caption{Comparison of \NII\ line profiles from present work (black) and
{\sc cmfgen} (green), for model {\tt d10v} (for parameters, see
Table~\ref{grid-cmfgen}).} 
\label{d10v_nii}
\end{figure}

\begin{figure}
\resizebox{\hsize}{!}{\includegraphics{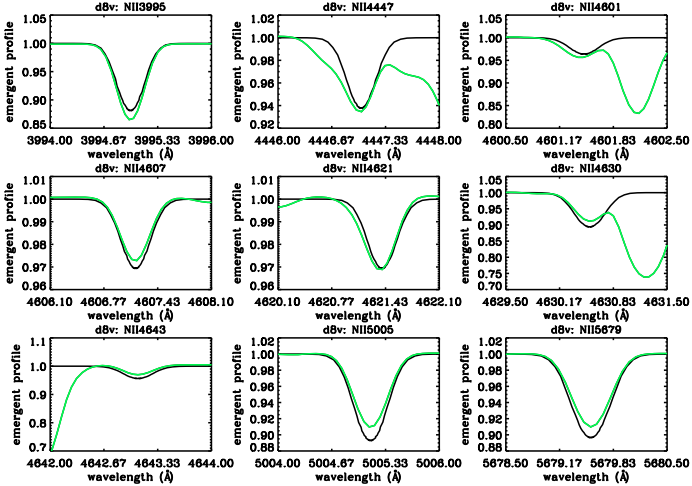}}
\caption{As Fig.~\ref{d10v_nii}, but for model {\tt d8v}.}
\label{d8v_nii}
\end{figure}

\begin{figure}
\resizebox{\hsize}{!}{\includegraphics{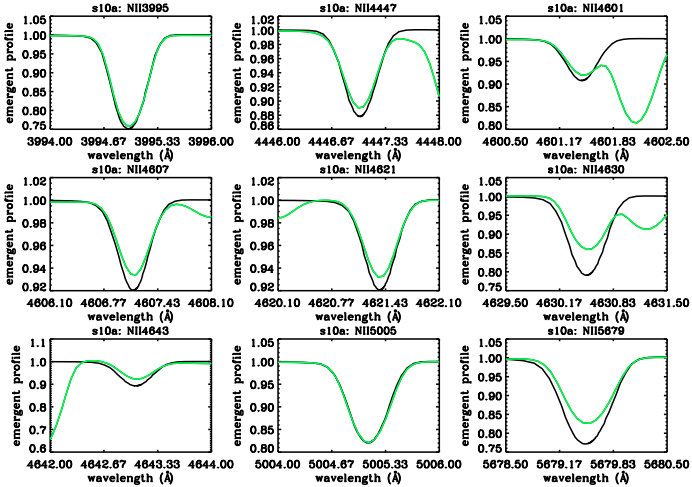}}
\caption{As Fig.~\ref{d10v_nii}, but for model {\tt s10a}. \vspace{0.3cm}} 
\label{s10a_nii}
\end{figure}

\begin{figure}
\resizebox{\hsize}{!}{\includegraphics{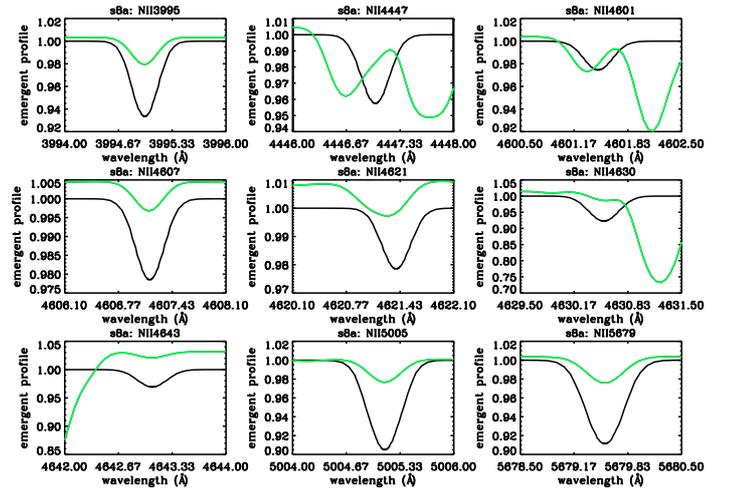}}
\caption{As Fig.~\ref{d10v_nii}, but for model {\tt s8a}.}
\label{s8a_nii}
\end{figure}

\begin{figure}
\resizebox{\hsize}{!}{\includegraphics{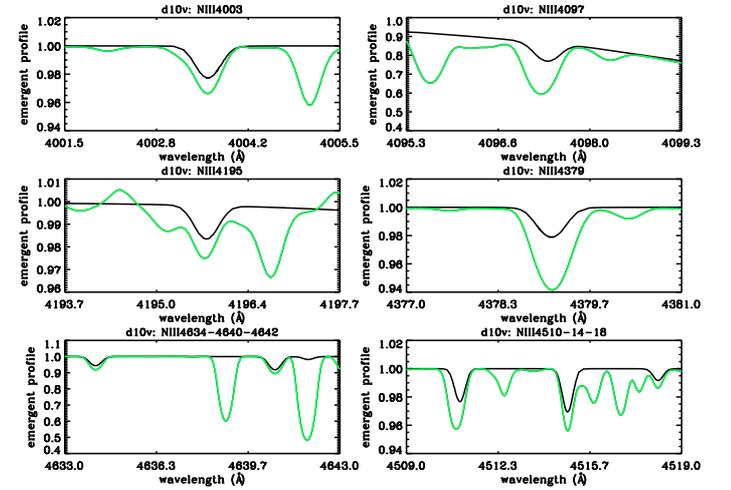}}
\caption{Comparison of \NIII\ line profiles from present work (black) and
{\sc cmfgen} (green), for model {\tt d10v}.} 
\label{d10v}
\end{figure}

\begin{figure}
\resizebox{\hsize}{!}{\includegraphics{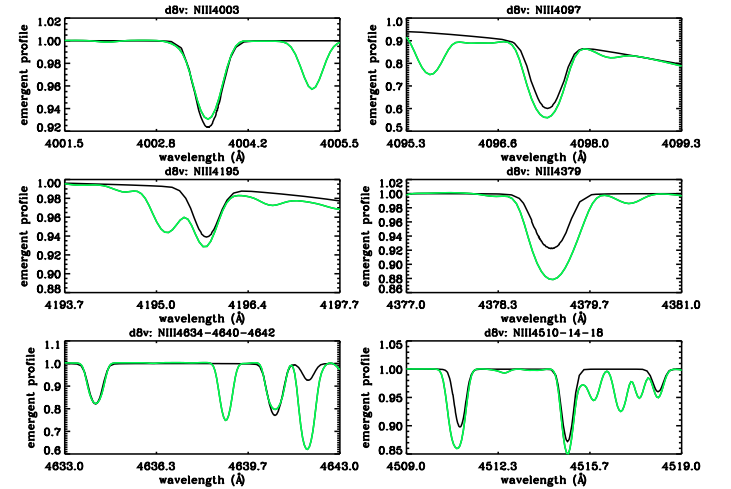}}
\caption{As Fig.~\ref{d10v}, but for model {\tt d8v}.}
\label{d8v}
\end{figure}

\begin{figure}
\resizebox{\hsize}{!}{\includegraphics{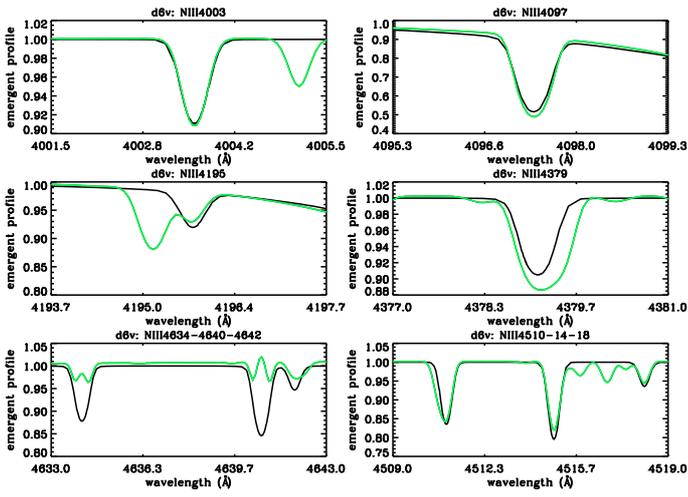}}
\caption{As Fig.~\ref{d10v}, but for model {\tt d6v}.}
\label{d6v}
\end{figure}

\begin{figure}
\resizebox{\hsize}{!}{\includegraphics{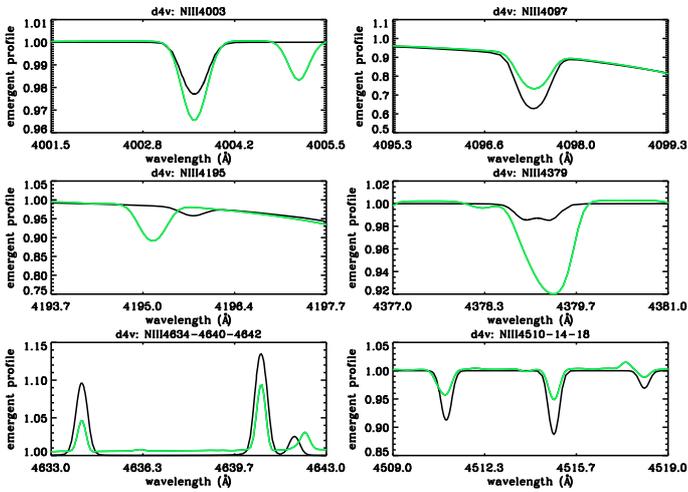}}
\caption{As Fig.~\ref{d10v}, but for model {\tt d4v}.\vspace{0.3cm}}
\label{d4v}
\end{figure}

\begin{figure}
\resizebox{\hsize}{!}{\includegraphics{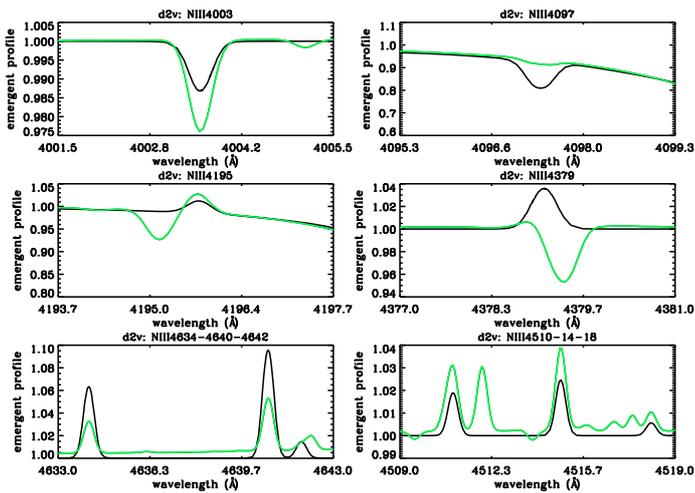}}
\caption{As Fig.~\ref{d10v}, but for model {\tt d2v}.}
\label{d2v}
\end{figure}

\begin{figure}
\resizebox{\hsize}{!}{\includegraphics{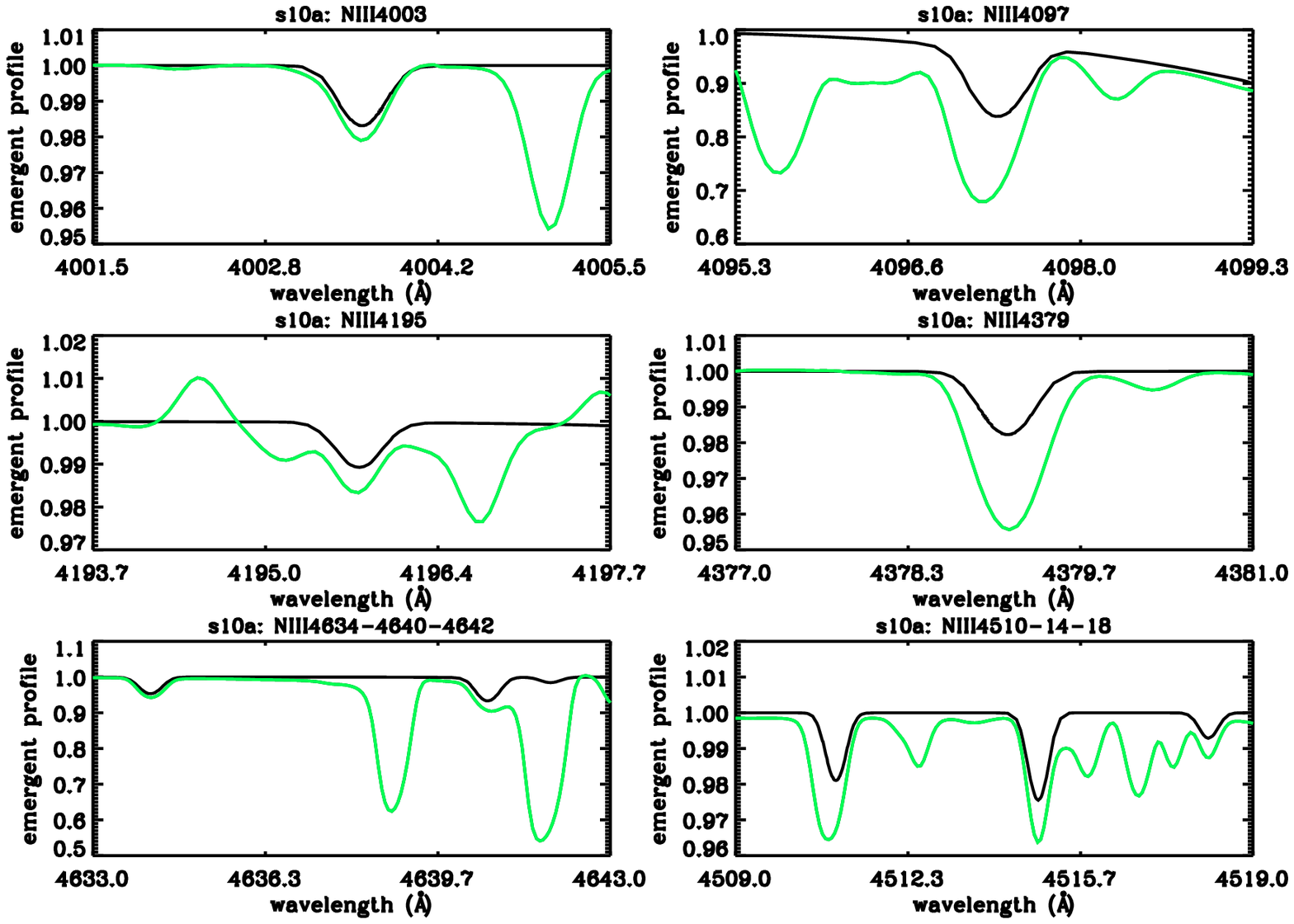}}
\caption{As Fig.~\ref{d10v}, but for model {\tt s10a}.}
\label{s10a}
\end{figure}

\begin{figure}
\resizebox{\hsize}{!}{\includegraphics{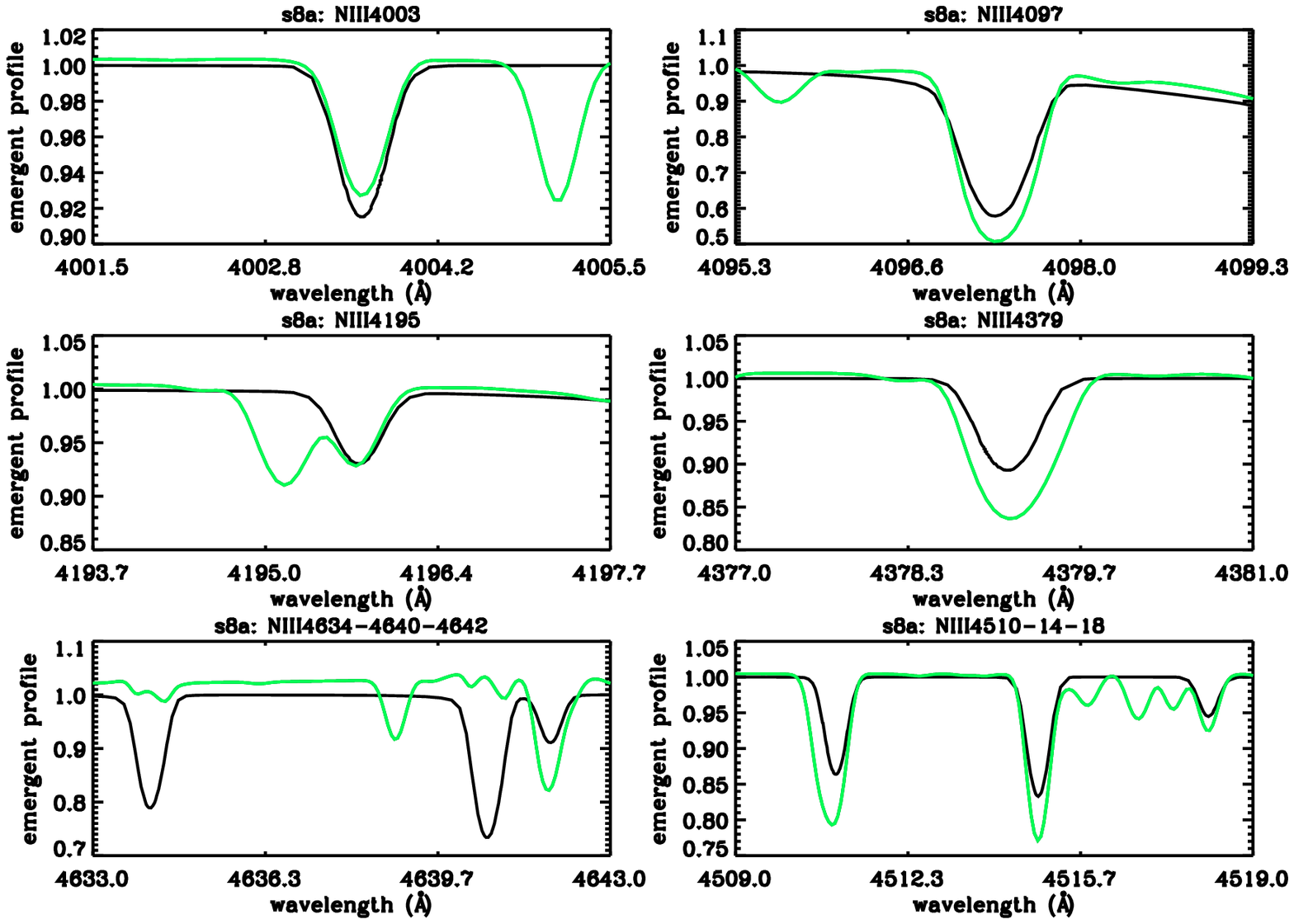}}
\caption{As Fig.~\ref{d10v}, but for model {\tt s8a}.}
\label{s8a}
\end{figure}

\begin{figure}
\resizebox{\hsize}{!}{\includegraphics{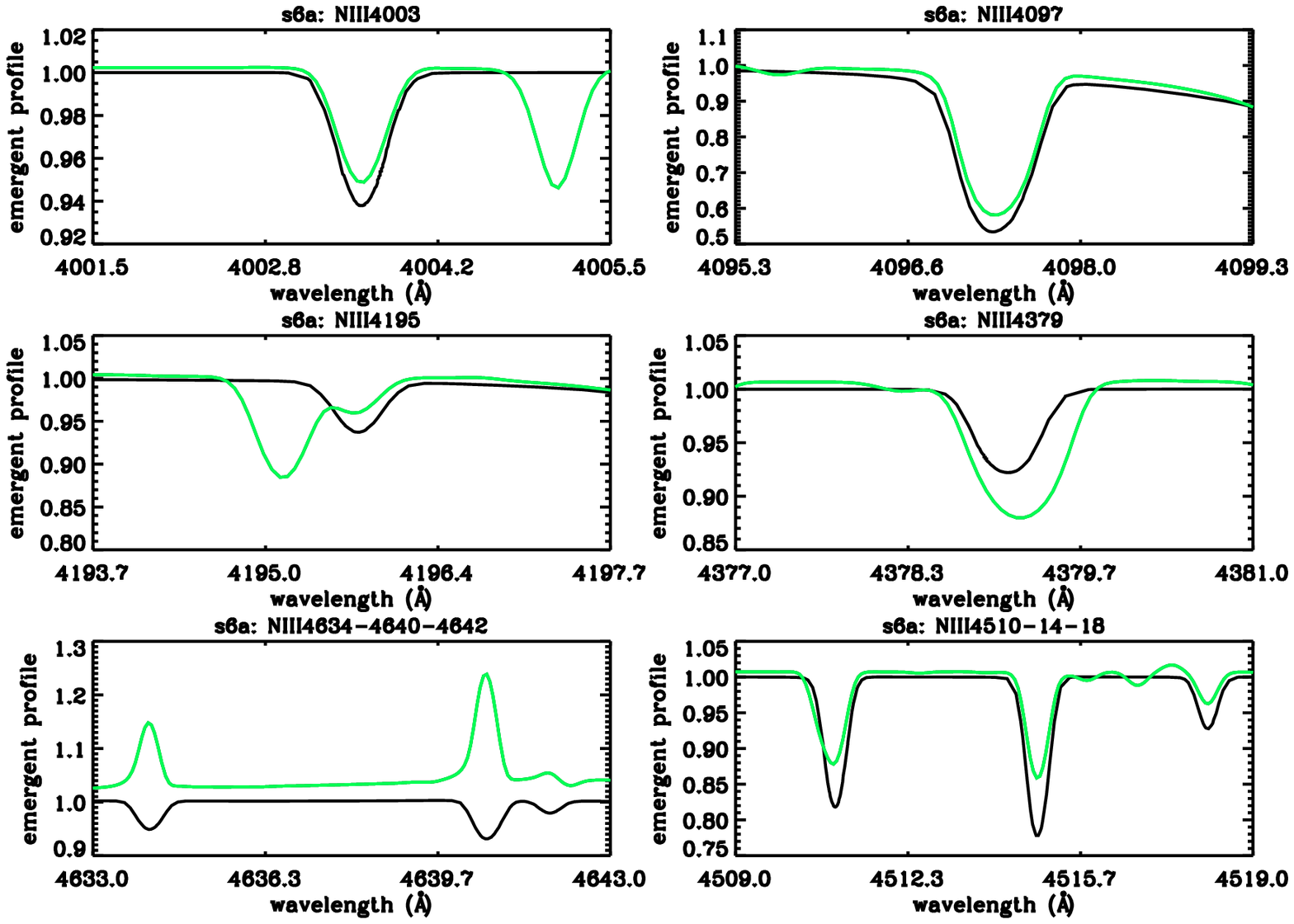}}
\caption{As Fig.~\ref{d10v}, but for model {\tt s6a}.}
\label{s6a}
\end{figure}

\begin{figure}
\resizebox{\hsize}{!}{\includegraphics{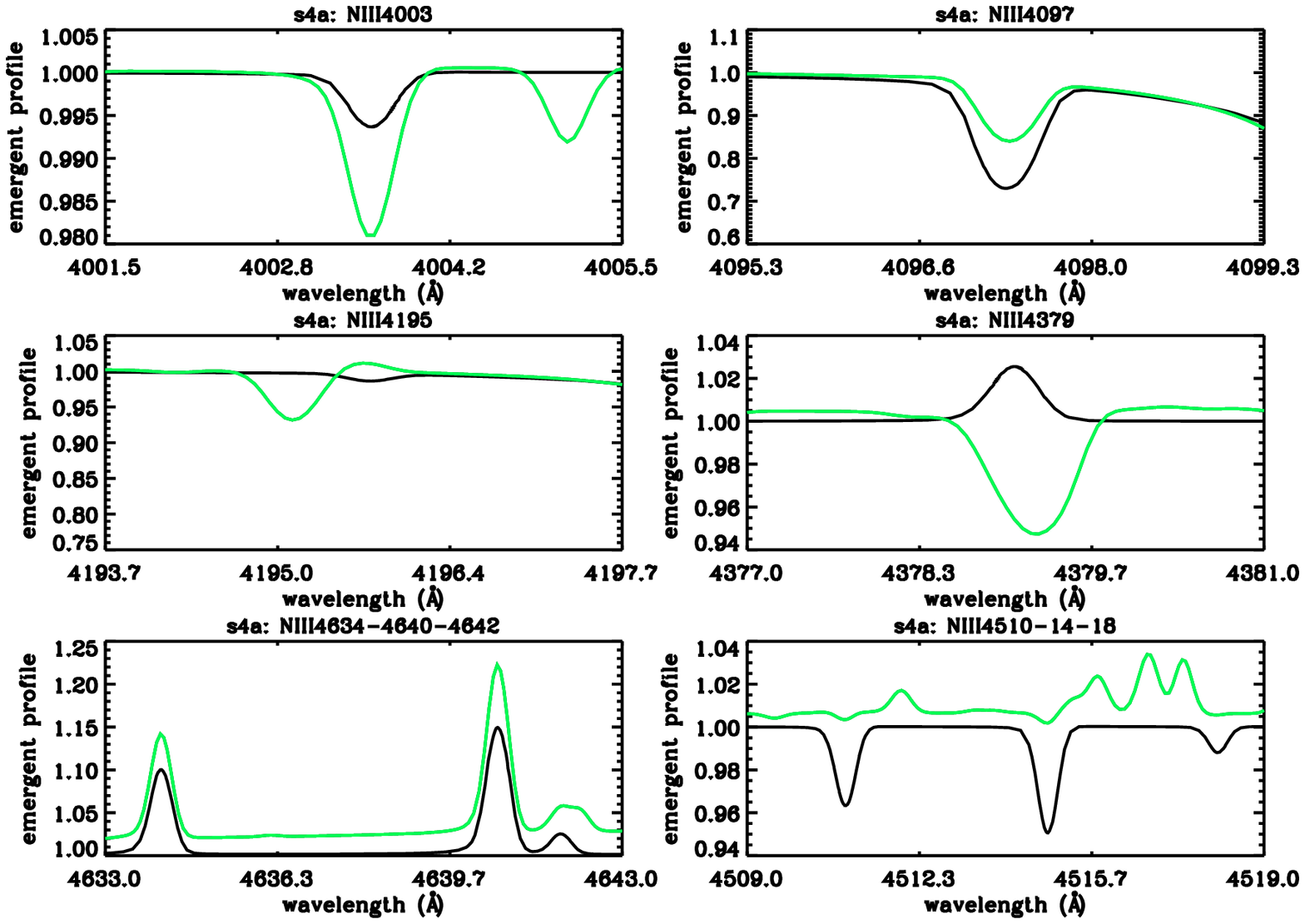}}
\caption{As Fig.~\ref{d10v}, but for model {\tt s4a}.}
\label{s4a}
\end{figure}

\begin{figure}
\resizebox{\hsize}{!}{\includegraphics{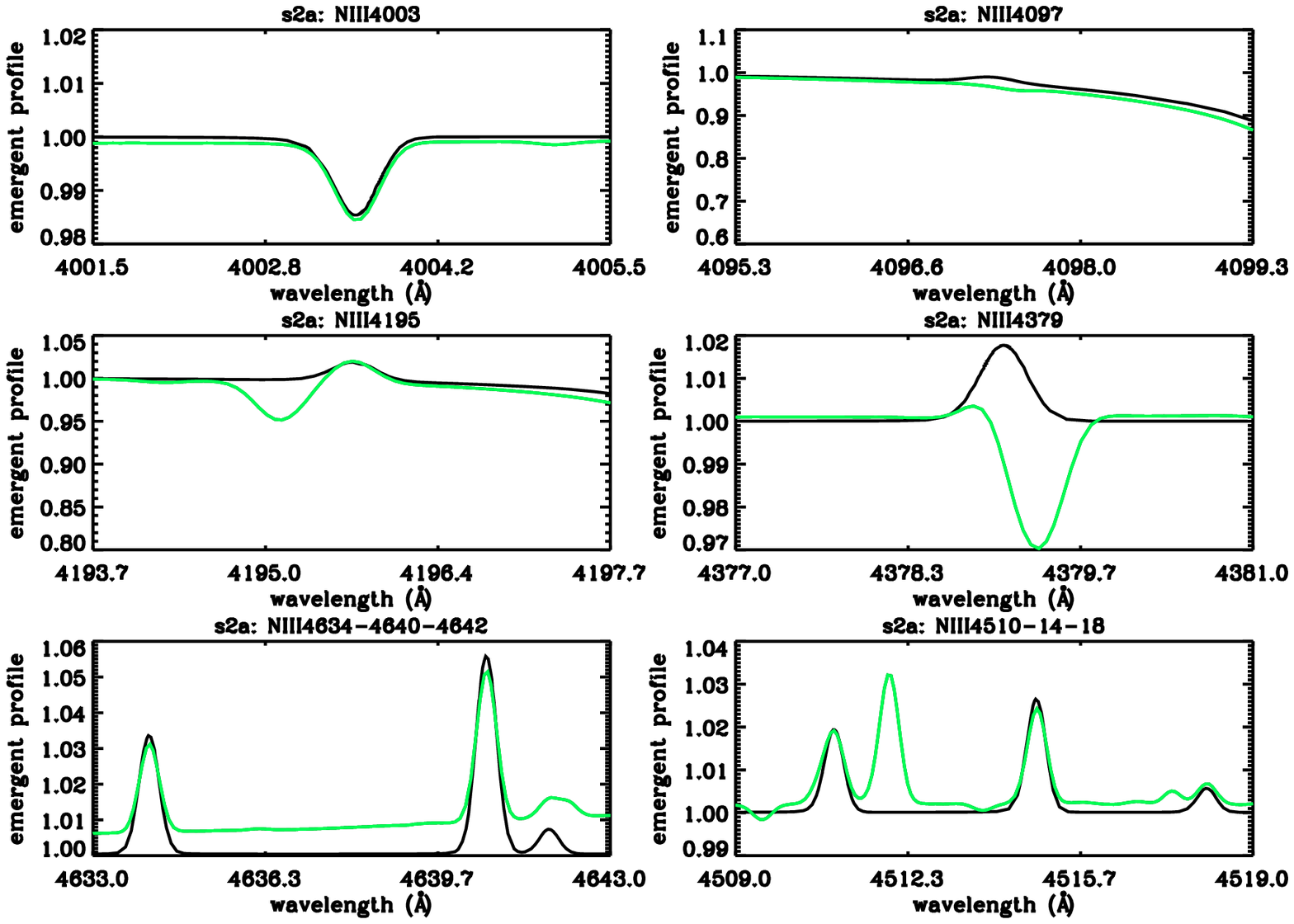}}
\caption{As Fig.~\ref{d10v}, but for model {\tt s2a}.}
\label{s2a}
\end{figure}

\end{document}